\documentclass[11pt]{article}
\usepackage{graphicx}
\usepackage{amssymb,amsthm,amsmath,float}
\usepackage{epstopdf}
\newcommand{\Eq}[1]{Eq.~(\ref{#1})}
\newcommand{\Th}[1]{Theorem~\ref{#1}}
\newcommand{\Lem}[1]{Lemma~\ref{#1}}
\newcommand{\Con}[1]{Conjecture~\ref{#1}}
\newcommand{\Def}[1]{Def.~\ref{#1}}
\newcommand{\Sec}[1]{\S \ref{#1}}
\newcommand{\Fig}[1]{Fig.~\ref{#1}}
\newcommand{\Tbl}[1]{Table~\ref{#1}}
\newcommand{\App}[1]{Appendix~\ref{#1}}

\newcommand{\InsertFig}[4]
{\begin{figure}[ht]
       \centerline{
         \includegraphics[width=#4]{#1}
       }
       \caption{{\footnotesize  #2}
       \label{#3}}
\end{figure}}

\newcommand{\InsertFigTwo}[5] {
\begin{figure}[ht]
       \centerline{
         \includegraphics[width=#5]{#1}
         \hskip 0.5in
         \includegraphics[width=#5]{#2}
       }
       \caption{{\footnotesize  #3}
       \label{#4}}
\end{figure}}

\newcommand{\R}{{\mathbb{ R}}}

\newcommand{\T}{{\mathbb{ T}}}
\newcommand{\Z}{{\mathbb{ Z}}}

\newcommand{\bS}{{\mathbb{ S}}}
\newcommand{\cO}{{\cal O}}
\newcommand{\pt}{\mbox{{\huge .}}}
\newcommand{\ba}{{\bf a}}
\newcommand{\bb}{{\bf b}}

\newcommand{\bs}{{\bf s}}

\newcommand{\sgn}{\mathop{\rm sgn}}
\newcommand{\fix}[1]{{\cal #1}}
\newcommand{\Tr}{\mathop{\rm Tr}}
\newcommand{\per}[1]{({#1})^{\infty}}
\newcommand{\Int}[1]{\mathop{\rm Int}{#1}}

\newcommand{\floor}[1]{{\lfloor{#1}\rfloor}}
\newcommand{\ceil}[1] {{\lceil{#1}\rceil}}
\newcommand{\frc}[1]{\{ {#1} \}}

\newcommand{\hen} {H\'enon }
\newtheorem{teo}{Theorem}
\newtheorem{lem}[teo]{Lemma}
\newtheorem{cor}[teo]{Corollary}
\newtheorem{con}{Conjecture}
\newtheorem{defn}{Definition}
\floatstyle{ruled}
\newfloat{Algorithm}{thp}{loa}[teo]

\title{Symbolic Codes for Rotational Orbits}
\author{
     H.~R.~Dullin\thanks
     {
      H.R.Dullin@lboro.ac.uk. Supported in part by DFG grant Du 302/2,
      and EPSRC grant GR/R44911/01. } \\
     Department of Mathematical Sciences,\\
     Loughborough University \\
     Loughborough LE11 3TU, UK\\
\and
        J.~D.~Meiss and D.~G.~Sterling\thanks
      {
        James.Meiss@Colorado.EDU, DSterling@somalogic.com.
        JDM was supported in part by NSF grant DMS-0202032,
        DGS was supported by an NRC postdoctoral fellowship.
      }\\
     Department of Applied Mathematics\\
     University of Colorado \\
     Boulder, CO 80309-0526 \\
}
\date{\today}
\begin{document}
\maketitle

\begin{abstract}
\noindent
Symbolic codes for rotational orbits and ``islands-around-islands" are constructed for the quadratic, area-preserving \hen map. The codes are based upon continuation from an anti-integrable limit, or alternatively from the horseshoe. Given any sequence of rotation numbers we obtain symbolic sequences for the corresponding elliptic and hyperbolic rotational orbits. These are shown to be consistent with numerical evidence. The resulting symbolic partition of the phase space consists of wedges constructed from images of the symmetry lines of the map.

\vspace*{1ex}
\noindent
\end{abstract}

\section{Introduction}

Symbolic dynamics has been profitably used in the
study of many dynamical systems since its invention by Hadamard in 
1898 and naming by Morse and Hedlund in 1938 \cite{Lind95, Kitchens98, Hao98}.
Among its many uses, symbolic dynamics can provide useful information about topological
invariants such as the enumeration of periodic orbits and the
entropy. 
It can also lead to numerical methods for finding
periodic, homoclinic and chaotic orbits, and facilitate
characterization of transport. Symbolic dynamics applies most directly to systems that are
hyperbolic; indeed it was invented to describe geodesics on surfaces
with negative curvature and famously applies to hyperbolic toral
automorphisms and the Smale horseshoe. 

In this paper we continue the study, began in \cite{SDM99}, of the coding of orbits of H\'enon's quadratic, area-preserving mapping \cite{Henon69}. Previously we used the concept of an anti-integrable (AI) limit \cite{Aubry90} to define codes by continuation (see \Sec{sec:AILimit}) and studied the bifurcations of homoclinic orbits that destroy the horseshoe of this map.  In the current paper we study the codes of orbits that are born in rotational bifurcations of the elliptic fixed point.  The rule that we obtain identifies the subset of orbits in the horseshoe that become rotational orbits, encircling the elliptic fixed point of the \hen map.

The extension of symbolic dynamics to systems with stable orbits has proved difficult, except for the case of one-dimensional maps where classical results apply \cite{Sarkovskii64, Metropolis73}. The dissipative \hen mapping \cite{Henon76} is a natural system to attempt the generalization of these results to multidimensional, nonhyperbolic systems, especially as it reduces (when $b = 0$) to the one-dimensional logistic map. Just as the symbols for the logistic map are based on the itinerary of an orbit relative to the critical point, Grassberger and Kantz \cite{Grassberger85} proposed that the symbols for the \hen map could be determined by partitioning the plane using the stable and unstable foliation and their ``primary homoclinic tangencies," that is, points at which the local curvature of the unstable manifold diverges. This led to the concept of a ``primary pruning front" in the symbol plane as determining the allowed sequences for a particular map \cite{Cvitanovic88}. Symbol sequences defined in this way can exhibit monodromy, that is, morph into new sequences on paths that encircle codimension-two bifurcations, such as a cusp \cite{Hansen92};  along these paths, the primary tangencies may exhibit discontinuities \cite{Giovannini92}. As far as we know, this method has not been applied to the area-preserving case ($b=\pm 1$), though it has been applied to other dissipative systems, such as the cubic, generalized \hen map \cite{Fang94}.

Several other methods have also been proposed to obtain symbolic codes for orbits of the \hen map. Biham and Wenzel defined the codes as the signature of a pseudo-gradient method for finding periodic orbits \cite{Biham89}. Sterling and Meiss proved that this technique works sufficiently close to an anti-integrable limit \cite{Sterling98}. Unfortunately, it does not always converge to fixed points, and sometimes gives two codes for the same orbit \cite{Grassberger89}. Hansen and Cvitanovic defined codes by approximating the two-dimensional map near $b=0$ by a sequence of one-dimensional unimodal maps \cite{Hansen98}. The code defining an orbit can change if it crosses the critical point of one of the approximating maps. 

Primary homoclinic tangencies have also been used to construct codes for the area-preserving standard map. For parameter values where elliptic periodic orbits and their associated islands are small (large $k$), Christiansen and Politi have shown that a primary set of homoclinic tangencies can be identified by their proximity to the dominant fold lines in the map \cite{Christiansen95}. Gaps between these points can be connected with symmetry lines associated with the reversibility of the map to form a curve that creates a symbol partition \cite{Christiansen97}. The symbol boundary can be modified to include elliptic islands, \cite{Christiansen96, Christiansen97}, by choosing appropriate images of the symmetry lines. This method does not explain why symmetry lines are important, nor does it give a recipe for selecting the proper lines.

In \Sec{sec:RotationalOrbits} we give a rule for constructing the symbolic codes for orbits of the \hen map with any given rotation number. These codes are consistent with those obtained by numerical continuation from the AI limit. We also show that these codes have definite symmetry properties, and that the symbolic partition in an elliptic island has the form of a wedge with apex at the elliptic fixed point and whose boundaries are constructed from specific symmetry lines. These wedges are similar to those found by Christiansen and Politi for the standard map.

Our rotational codes are closely related to those for maps with a natural angle variable, for example for circle maps \cite{Veerman86, Zheng91} and cat maps \cite{Percival87a}. We review these ideas in the Appendices.

We also develop a systematic rule for obtaining the symbolic codes of rotational orbits with ``higher class" \cite{Meiss86}, that is to ``islands-around-islands," in \Sec{sec:classTwo}-\ref{sec:classCCode}. These correspond, for example, to orbits that rotate around orbits that rotate around the elliptic fixed point; thus they are defined by a sequence of rotation numbers. Our construction provides codes for the elliptic and hyperbolic orbits for each sequence of rotation numbers. Again the codes are shown to be consistent with the numerical evidence. The resulting symbolic partition is constructed from a sequence of wedges defined by symmetry lines.

A different method for constructing symbolic codes for islands-around-islands was given previously \cite{Aizawa84,Afraimovich00}; however in these cases the entire set of orbits in an island was assigned the same sequence and the motivation was to study the transport implications for chaotic orbits outside the islands \cite{Meiss86, Meiss86b}.

\section{Symbolic codes for the \hen map by continuation}\label{sec:AILimit}

The area-preserving \hen diffeomorphism can be written as\footnote
{
   We do not use H\'{e}non's original form, $(\xi,\eta) \rightarrow (1-a\xi^2+\eta,b\xi)$,
   since it becomes linear at $a=0$, and is orientation-reversing for $b>0$.
   Our form can be obtained by setting $b = -1$, and
   defining $k = a$, $x = k\eta$, and $x'=-k\xi$.
}
\begin{equation}\label{eq:Henon}
      (x,x') \mapsto  f(x,x') = (x', -x -k + x'^{2 }) \;.
\end{equation}
An orbit of this map is written as a bi-infinite sequence $\cO =
(\ldots x_{-1}, x_0, x_{1}\ldots)$ with $(x_{t+1}, x_{t+2}) = f(x_t, x_{t+1})$.
The dynamics of this map exhibit the full range
of behavior of typical area-preserving maps, including infinite
cascades of period-doubling bifurcations, invariant circles, cantori,
transport, twistless bifurcations, etc.  \cite{MacKay93, Meiss92,
Dullin99}.  As $k$ increases, the dynamics become increasingly
chaotic, and beyond a critical value, $k = k_H$, the set of bounded
orbits forms a Smale horseshoe (this set has measure zero---almost all
orbits escape to infinity).  It was proved by Devaney and Nitecki that
there is indeed a hyperbolic horseshoe when $k > 5+2\sqrt{5} > k_H$
\cite{Devaney79, Sterling98}.  Our numerical studies indicate that
\cite{Sterling98,SDM99}
\begin{equation}\label{eq:kH}
      k_H \approx 5.699310786700 \;.
\end{equation}

The dynamics of the bounded orbits in the hyperbolic horseshoe is conjugate to the full shift on two symbols. A simple way of constructing these symbols for the \hen map is to define
\begin{equation}\label{eq:sDefine}
s_{t} = \sgn x_{t}
\end{equation}
because the bounded invariant set is divided by the coordinate axes, see \Fig{fig:horseshoe}.

     \InsertFig{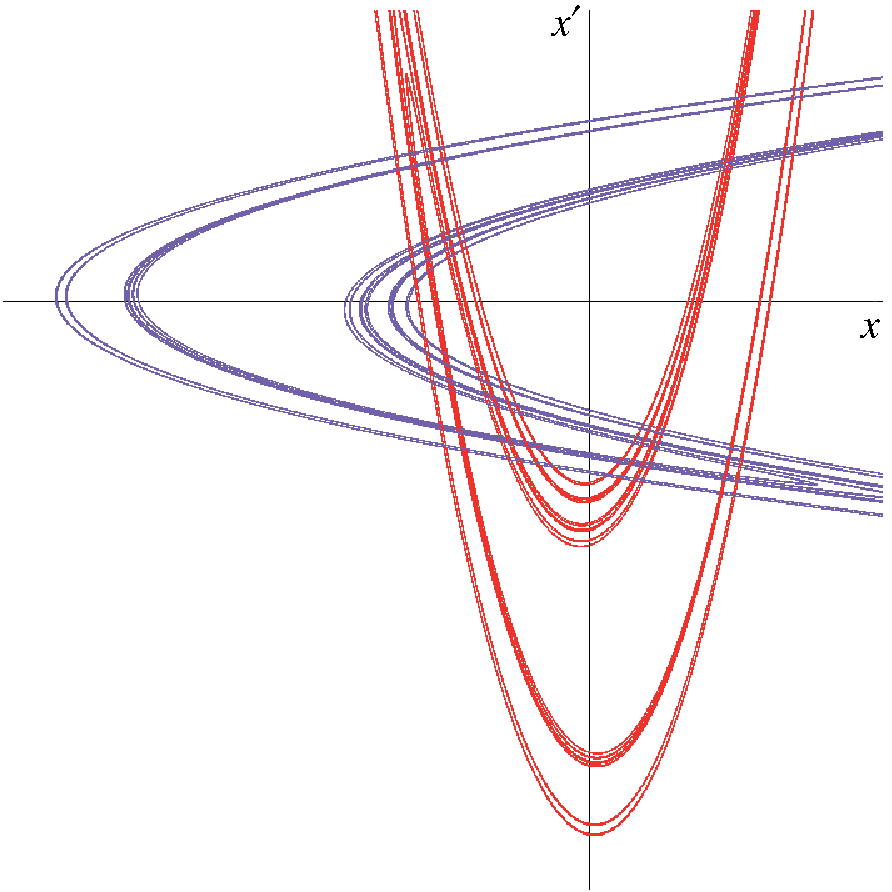} {Stable (blue) and unstable (red) manifolds of the
     hyperbolic fixed point for the \hen map when $k=6$. The closure of the
     intersection of these manifolds is the horseshoe of
     bounded orbits.}
     {fig:horseshoe}{4in}

An alternative method for obtaining these codes is to view the \hen
map as arising by continuation from an {\em anti-integrable} (AI) limit
\cite{Aubry90, Aubry92}; for \Eq{eq:Henon} we define
\begin{equation} \label{eq:zDefine}
     z = \epsilon x \;, \; \mbox{with} \; \epsilon \equiv \frac{1}{\sqrt{k+1}} \;,
\end{equation}
to rescale the map as
\[
\epsilon (z_{t+1} + z_{t-1}-\epsilon) = z_t^2-1 \;.
\]
To define $\epsilon$, we assumed that $k > -1$, but this is not much
of a restriction since there are no bounded orbits for the \hen map
when $k < -1$.  The case $\epsilon \rightarrow 0$ is the AI limit
\cite{Sterling98, SDM99}.  In this limit orbits reduce to sequences
$z_t = \pm 1$, where the choice of sign is arbitrary---the map is
equivalent to the full shift on the symbols $\pm$.  Every such
sequence continues away from the limit to an orbit of the \hen map.
Conversely, when there is a horseshoe these symbols agree with
\Eq{eq:sDefine}, and every bounded orbit of the \hen horseshoe
continues to an orbit at the AI limit.

Each point in the horseshoe is represented by a bi-infinite sequence of signs,
together with the  binary point representing the current position:
\[
        (x_t,x_{t+1}) \cong \bs \equiv (\ldots s_{t-2}\,s_{t-1} \pt s_t \, s_{t+1}\ldots)
\]
We denote the right-shift by $\sigma$, so that
\begin{equation}\label{eq:shiftMap}
     \sigma \bs = (\ldots s_{t-1}\,s_t \pt  s_{t+1}\ldots)
\end{equation}
When there is a hyperbolic horseshoe for $k > k_H$, the \hen map $f$, restricted to 
the set of bounded orbits, is conjugate to the shift
map $\sigma$ acting on the sequence $\bs$ that is the code for $x$.

We use continuation away from the AI limit to find orbits with a given
code as $k$ varies \cite{Sterling98}.  
Such a ``global coding'' is complete if every smooth one-parameter 
family in the extended space $\R^2 \times \R$ (phase space times parameter)
connects to the AI limit. Because of twistless bifurcations 
it is not enough to simply take the map parameter $k$ as 
a family parameter \cite{Dullin99,SDM99}. We do not know
any counterexample to this ``smooth no-bubble conjecture'', 
and adopt it as our working hypothesis.

We use the signs $+$ and $-$ to denote the symbols $s_t$.  For
example the hyperbolic fixed point, at $x_t = 1+\sqrt{1+k}$, has code
$s_t = +$; we denote the bi-infinite sequence for this orbit by $\bs =
\per{+}$.  In general a period-$n$ orbit is given by the bi-infinite
concatenation of a sequence of $n$ symbols; we represent this with a
superscript $\infty$: $\per{s_0\,s_1\,\ldots\,s_{n-1}}$.  We
often denote repeated symbols with a superscript, thus $\per{-+^7}$ is
a period-$8$ orbit.  The {\em parity} of a finite symbol sequence is
defined to be the product of its symbols:
\begin{equation}\label{eq:parity}
    \pi(s_0s_1\ldots s_{n-1}) = \prod_{t=0}^{n-1} s_t  \;.
\end{equation}
Equivalently the parity is {\em even} or {\em odd} if the number of
minus signs is as well.  A parity can be assigned to any periodic
orbit by computing the parity of its fundamental sequence; for example,
$\pi(\per{-+++}) = -1$.

Using the continuation method, we can follow orbits from the AI limit
to visualize the relation between the symbolic codes and the positions
of the corresponding points in phase space.  If we distinguish points solely by a single
symbol, $s_0$, this gives a partition of the phase space into regions
corresponding to the $+$ and the $-$ code.  Using the conjugacy
between the shift $\sigma$ and the \hen map on these orbits, we could
reconstruct the full symbol sequence for each orbit by following the
sequence of visits to the two elements of the partition.

     \InsertFigTwo {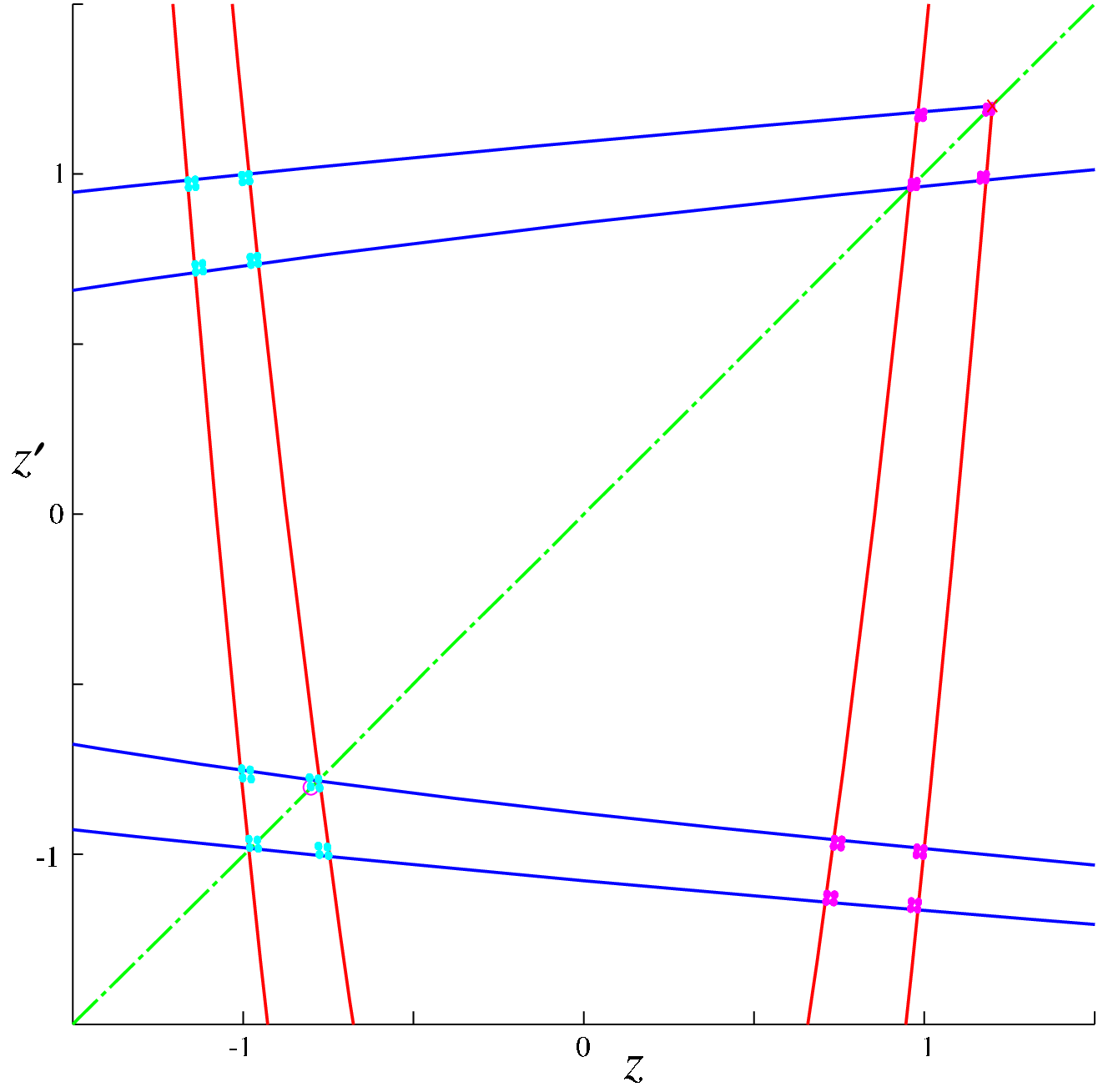}{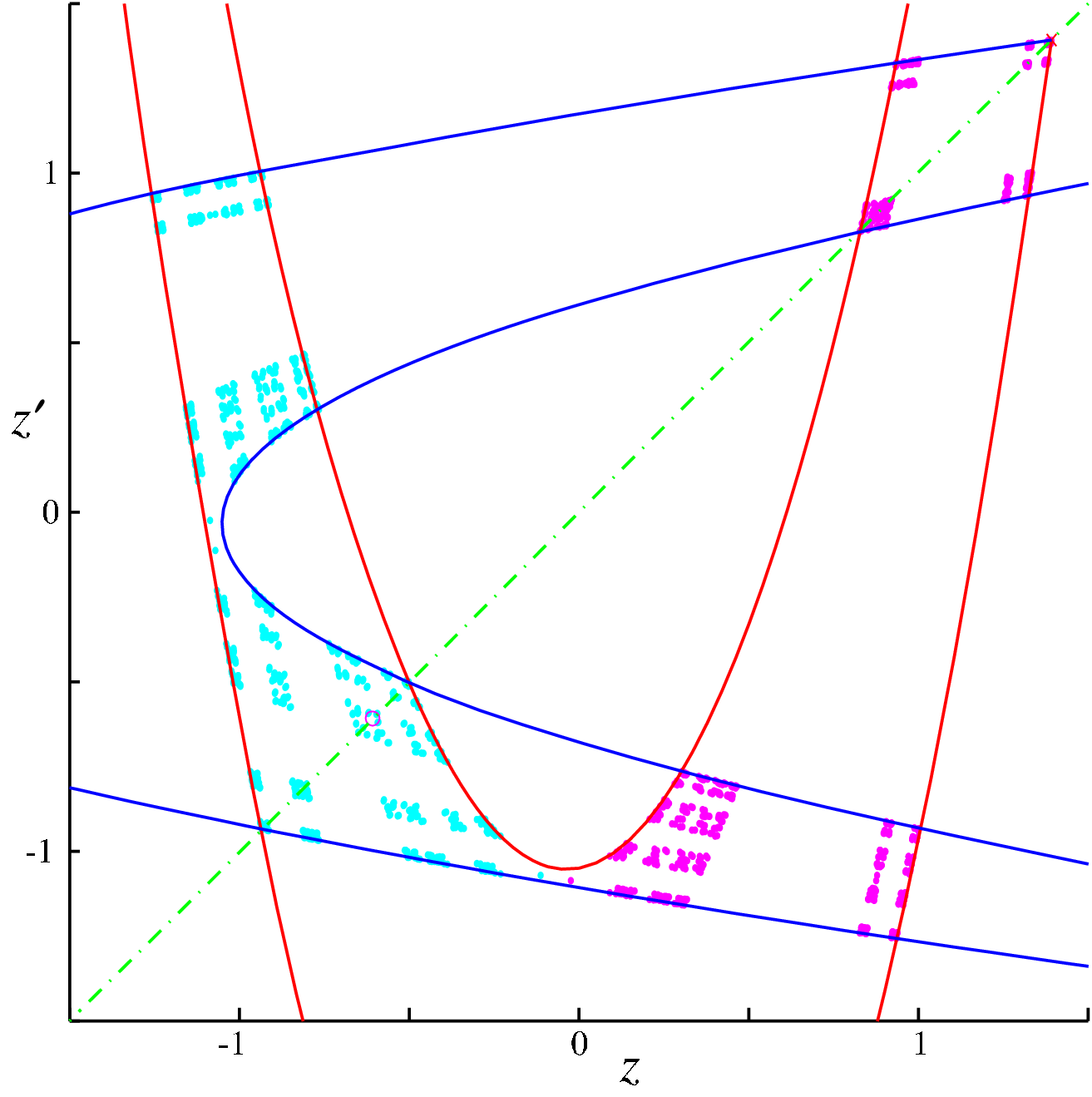} {Points on the $99$ periodic
     orbits of the \hen map with periods up to $10$ for two different parameter
     values. In the left panel $k = 24$ ($\epsilon = 0.2$) and in the right
     $k = 5.5$ ($\epsilon \approx 0.392$). The blue (red) curves are the
     initial lobes of the stable (unstable) manifolds for the hyperbolic
     fixed point, $\per{+}$, located in the upper right corner.}
     {fig:AI-boundary} {3in}

In \Fig{fig:AI-boundary}, a point is colored cyan if $s_{0} = -$, and
magenta if $s_0 = +$.  We use the scaled coordinates $(z,z')$, defined
in \Eq{eq:zDefine} for the plots, so that as $k \rightarrow \infty$
all bounded orbits converge to the four AI states $(z,z') = (\pm 1,
\pm 1)$.  As $k$ decreases the trajectories move away from these
points as shown in the left frame of \Fig{fig:AI-boundary}.  For this
value of $\epsilon$ it can be proven that all of the trajectories
reside within the union of four small squares with sides $M_\infty
\approx 0.568$ that are centered on the four AI states
\cite{Sterling98}---actually we observe that they are contained in
smaller rectangles bounded by segments of the stable and unstable
manifolds of the hyperbolic fixed point.  These segments intersect at
points on the ``type one" homoclinic orbits with symbol sequences
$+^{\infty}\pt -(-)-+^{\infty}$ and $+^{\infty}\pt -(+)-+^{\infty}$
\cite{SDM99}.  As $\epsilon$ increases the bounded orbits move further
from the AI states, and the stable and unstable manifolds of $\per{+}$
approach their first tangency.  Just below $k = k_H$ (right panel of
\Fig{fig:AI-boundary}) the two type-one homoclinic orbits have been
destroyed in a homoclinic bifurcation and only a subset of symbol
sequences are now allowed.  Nevertheless, the symbol boundary is still
very simple (at least up to this period), and is delineated by the
near tangency between the first lobes of the stable and unstable
manifolds.

One phenomenon special to area-preserving maps is the existence of
elliptic periodic orbits.  For example, the second fixed point,
$\per{-}$ (at $x_t = 1-\sqrt{1+k}$) becomes elliptic at $k = 3$.  More
generally this orbit has positive {\em residue} for any $k > -1$.
Recall that Greene's residue, $r$, is a convenient encoding for the
linear stability of an orbit of an area-preserving map: given a point
$(x,x')$ on a period-$n$ orbit of $f$, then
\begin{equation}\label{eq:residue}
       r \equiv \frac14 ( 2- \Tr(Df^n(x,x'))
\end{equation}
Orbits are elliptic when $0<r<1$, hyperbolic when $r<0$ and reflection hyperbolic when $r>1$.
For example, the fixed point $\per{-}$  is elliptic for $-1<k<3$ and at $k=3$ undergoes a
period-doubling bifurcation becoming hyperbolic with reflection. More generally \cite{SDM99}
for a period $n$ orbit, as $k \rightarrow \infty$,
\begin{equation}\label{eq:resLimit}
    r \sim - \pi(\bs) \frac14 (4k)^{n/2} \;.
\end{equation}
Thus the parity of a symbol sequence determines the character of the
orbit for large $k$.  Often, we refer to orbits that are born
with positive residue as ``elliptic" even though they typically
undergo bifurcations that destroy their stability and may even
smoothly continue to orbits with negative residue.

Our goal for this paper is to investigate the structure of the symbol
partition when there are elliptic periodic orbits and their associated
rotational orbits.  In particular, \Fig{fig:k20boundary} shows that
the symbol boundary near an elliptic orbit exhibits a structure
radically different from the horseshoe shown in \Fig{fig:AI-boundary}.
In the following sections, we will explain the wedge shaped symbol
boundary that eminates from the elliptic fixed point in
\Fig{fig:k20boundary}.  To do this, we must construct the codes for
rotational orbits.

     \InsertFig {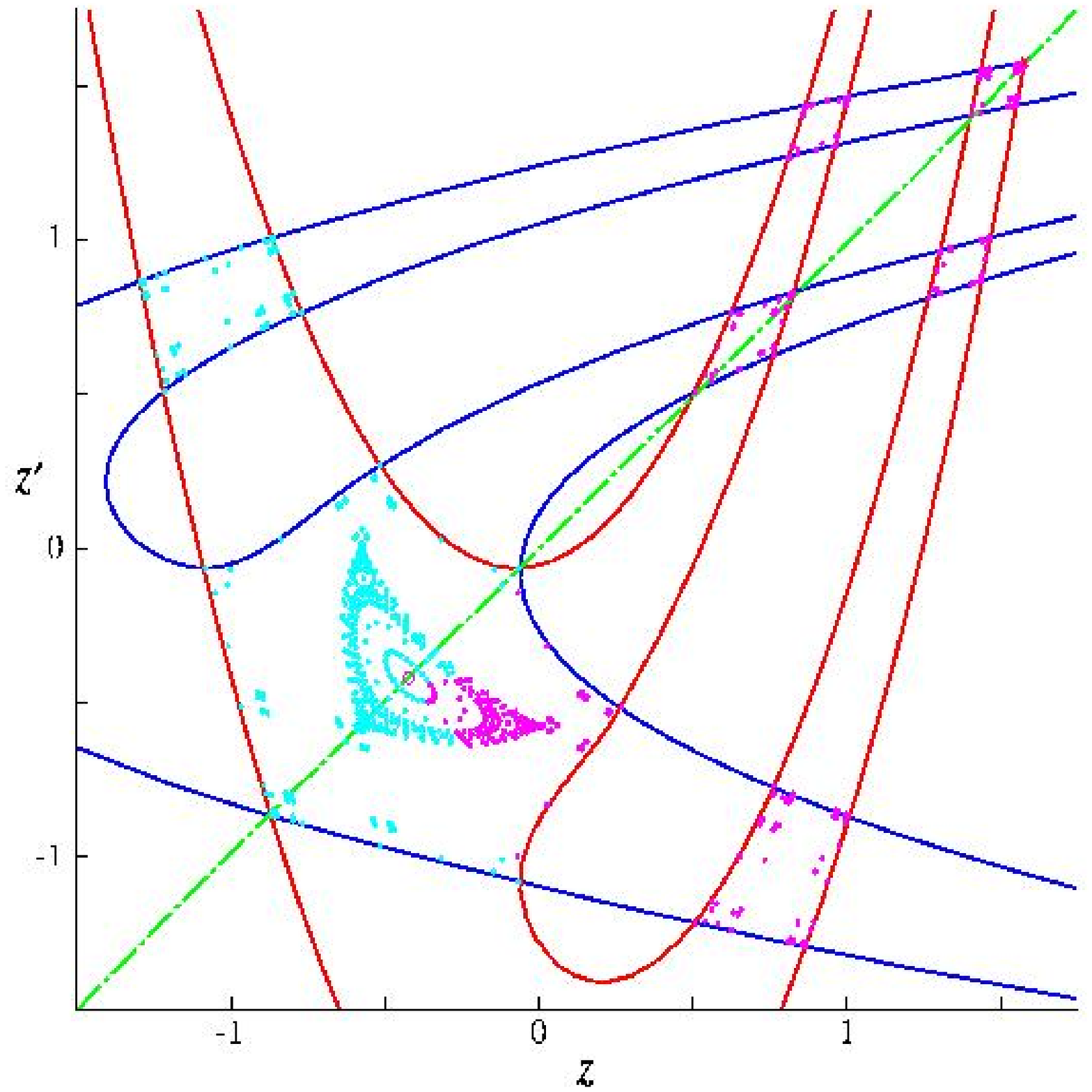} {Periodic orbits of the \hen map for $k
     = 2.0$.  Plotted are points on all periodic orbits up to period
     $10$ that still exist, as well as all rotational (class-one)
     orbits with periods up to $31$ that exist at this parameter
     value.  Points with $s_0 = +$ ($-$) are colored magenta (cyan).
     It is apparent that the first few lobes of the stable and
     unstable manifolds no longer delineate the symbol boundary in the
     vicinity of the elliptic fixed point.} {fig:k20boundary} {4in}

\section{Rotational (Class-One) Orbits}\label{sec:RotationalOrbits}

One special class of orbits for the \hen map are the {\em rotational} orbits---those
that are created by a rotational bifurcation of the elliptic fixed point $\per{-}$.
Generally, a rotational bifurcation occurs when the linearized rotation number at an elliptic fixed
point passes through a given rotation number $0 \le \omega \le \frac12$, i.e., when
\begin{equation}\label{eq:rotBifRes}
       r = \frac14 (2 - e^{2\pi i \omega} - e^{-2\pi i \omega}) =  \sin^2(\pi \omega)
\end{equation}
Since the residue of the elliptic fixed point is $r(\per{-}) = \frac12\sqrt{1+k}$, we can solve
 \Eq{eq:rotBifRes} for $k$ or $\omega$ to obtain
\begin{align}\label{eq:komega}
    k(\omega) &= \cos(2\pi\omega) ( \cos(2\pi\omega) - 2) \nonumber\\
    \omega(k) &= \frac{1}{\pi} \arcsin\left(\frac{(1+k)^{1/4}}{\sqrt{2}}\right) \;.
\end{align}
Since the residue of a period-$n$ orbits grows like $k^{n/2}$ as $k \rightarrow \infty$ 
it is more convenient for numerical work to use the scaled residue
\begin{equation}\label{eq:scaledresidue}
   \rho \equiv \frac{1}{\sqrt{k+1}} \sgn(r) \left[ 4 |r| \right]^{1/n} \,.
\end{equation}
Notice  $\rho=0 \Leftrightarrow r=0$, and as $k \rightarrow \infty$, \Eq{eq:resLimit} implies that $\rho \rightarrow \pm2$.

When $\omega(k)$ is rational, a rotational bifurcation typically\footnote
{
  When the period is larger than four. The tripling and quadrupling
  bifurcations are special, see e.g. \cite{MeyerHall92}.
}
creates a pair of orbits, one elliptic and one hyperbolic, we call
these {\em class-one} rotational orbits following Meiss \cite{Meiss86}
(the two fixed points are therefore class-zero orbits). 
A theorem of Franks implies that there are class-one period $q$ orbits 
for each rational rotation number $p/q$ whenever $0 < p/q < \omega(k)$ \cite{Franks90}.
Each elliptic class-one periodic orbit is typically encircled by rotational orbits
of class-two, giving rise to ``island chains.'' We will discuss
these higher class orbits in \Sec{sec:classTwo}.  When $\omega(k)$
is irrational and Diophantine, Moser's twist theorem implies that the rotational bifurcation gives birth to an invariant circle \cite{Moser94}. If $\omega(k)$ is not sufficiently irrational, we expect that a cantorus is born (as indicated by Aubry-Mather theory, though we know of no proof of this for the \hen map).

We begin this section with some observations from our numerical
experiments.  We then construct a rule that generalizes these
observations and appears to be consistent with all class-one orbits.
This construction has a natural geometric interpretation in terms of
circle maps.  The code that we obtain is related to that for
rotational orbits in other systems, for example the codes obtained for
minimizing rotational orbits in twist maps by Aubry~\cite{Aubry83},
for cat maps by Percival and Vivaldi~\cite{Percival87a}, billiards by
B\"{a}cker and Dullin~\cite{Backer97}, and for circle maps by
Lin~\cite{Hao98}.  We discuss some of these relationships in the
Appendices.

\subsection{Preliminary Numerical Observations}\label{sec:Numerical}
It was not obvious to us which of the possible period-$n$ orbits at
the AI limit should be identified as the class-one rotational orbits.
Thus we began by searching for periodic codes that continue to orbits
that collide with $\per{-}$ at $k(\omega)$ defined by \Eq{eq:komega}.
A brute force way to systematically search for rotational orbits,
would be to choose a period and continue {\em all} periodic codes of
that length from the AI limit until the corresponding orbits
bifurcate.  For example, at period $4$ there are three distinct
periodic codes: $\per{---+}, \per{--++}$, and $\per{-+++}$.  Upon
continuation, we discover that the orbits corresponding to
$\per{--++}$ and $\per{-+++}$ collide when $k = k(\frac14) = 0$ at the
position of the elliptic fixed point.  These two orbits have negative
and positive residues, respectively (in agreement with
\Eq{eq:resLimit}) during their entire lifetimes, and thus they
represent the hyperbolic and elliptic $\frac14$ orbits.  In our
previous numerical experiments, we followed each of the 1.47 million
orbits of the \hen map with periods up to 24 in order to obtain a
numerical estimate of the parameter at which the \hen map horseshoe is
destroyed~\cite{SDM99}.

We can actually considerably reduce the number of sequences that need
to be checked by computing the rotation number of the symbol sequence
using the ``self-rotation number" \cite{DSM00}.  The self-rotation
number of symbol sequence is the same as that of the orbit, and is
unchanged as $k$ varies.  As shown in Table 2 of \cite{DSM00} the
self-rotation number of $\per{---+}$ is $\frac12$, thus it cannot be
one of the rotational period-four orbits---indeed it arises in a
period-doubling bifurcation of the period-two orbit, $\per{-+}$.

Proceeding in this way we identified all of the class-one rotational
orbits with periods less than $20$~\cite{SterlingThesis}.  There are
118 such orbits; the first few of these are shown in
\Tbl{tbl:classOne}.  In the table we chose a canonical permutation to
order these sequences (we will discuss the canonical ordering in
\Sec{sec:RotationalCodes}).

Our numerical observations indicate that all of the class-one
rotational orbits smoothly continue to the AI limit; we know of no
proof of this observation.  We observe that the codes for the negative
and positive residue orbits differ only in one symbol, the second:
indeed, this symbol is simply the sign of the residue, indicated by 
a $*$ in the symbol sequence.

At this point the pattern for the rotational codes that we will
describe in the next section became clear.

\begin{table}[tbp]
 \centering
     \begin{tabular}{|l|c|c|c|} \hline
    $\omega$ &Code  &  $k(\omega)$ & PD \\
       \hline
         $\tfrac01$  &$\per{*}$     & -1         & 3 \\
       \hline
         $\tfrac12$  &$\per{-+}$    &  3         & 4 \\
       \hline
         $\tfrac13$  &$\per{-*+}$   &  $\tfrac54$    & $\tfrac54$  \\
       \hline
         $\tfrac14$   &$\per{-*++}$  &  0         &0.2174036214$^\dagger$ \\
       \hline
         $\tfrac15$  &$\per{-*+++}$ & $\tfrac{7-5\sqrt5}{8}$  &-0.2404626622$^\star$ \\
         $\tfrac25$  &$\per{-*--+}$ & $\tfrac{7+5\sqrt5}{8}$  &2.822983929$^\star$ \\
       \hline
         $\tfrac16$&$\per{-*+^{4}}$ & $-\tfrac34$            &-0.4766507416 \\
       \hline
         $\tfrac17$&$\per{-*+^{5}}$  &-0.8582400707          &-0.6124008240 \\
         $\tfrac27$&$\per{-*+--++}$  & 0.4945574340          & 0.7226142786 \\
         $\tfrac37$&$\per{-*-^{4}+}$ & 2.613682637           & 3.099045238  \\
       \hline
         $\tfrac18$&$\per{-*+^{6}}$  &$\tfrac12 - \sqrt2$    &-0.6974167690 \\
         $\tfrac38$&$\per{-*--+--+}$ &$\tfrac12 +\sqrt2$     & 2.4246586398 \\
       \hline
    $\tfrac19$&$\per{-*+^{7}}$       &-0.9452647974          &-0.7546238304 \\
    $\tfrac29$&$\per{-*++--+++}$     &-0.3171426657          &-0.0739574299 \\
    $\tfrac49$&$\per{-*-^{6}+}$      & 2.762407463           & 3.187557989  \\
       \hline
    $\tfrac{1}{10}$&$\per{-*+^{8}}$    &$\tfrac{-1-3\sqrt5}{8}$&-0.7954212145 \\
    $\tfrac{3}{10}$&$\per{-*+--+--++}$ &$\tfrac{-1+3\sqrt5}{8}$& 0.8673114431 \\
       \hline
    $\tfrac{1}{11}$&$\per{-*+^{9}}$     &-0.9747995591        &-0.8258655309 \\
    $\tfrac{2}{11}$&$\per{-*+++--+^{4}}$&-0.6582603930        &-0.4217831728 \\
    $\tfrac{3}{11}$&$\per{-*+--++--++}$ & 0.3048831897        & 0.5506136016 \\
    $\tfrac{4}{11}$&$\per{-*--+--+--+}$ & 1.738564049         & 2.225483132  \\
    $\tfrac{5}{11}$&$\per{-*-^{8}+}$    & 2.839612714         & 3.215096092  \\
        \hline
    $\tfrac{1}{12}$& $\per{-*+^{10}}$      &$\tfrac34-\sqrt3$   &-0.8493921692 \\
    $\tfrac{5}{12}$& $\per{-*-^{4}+-^{4}+}$&$\tfrac34+\sqrt3$   & 2.908515654  \\
       \hline
       \multicolumn{4}{l}{\footnotesize{$^\dagger  \, \text{The smaller real root of}\,
               16 k^4-64 k^3-8 k^2+1$}}\\
       \multicolumn{4}{l}{\footnotesize{$^\star  \,\text{The two real roots of}\,
                     4096 k^6-13312 k^5+512 k^4+17344 k^3-7520 k^2-16276 k-3251$}}\\
     \end{tabular}
     \caption{\footnotesize
     Codes for class-one rotational orbits. The $*$ represents $\pm$ for positive and
     negative residue orbits, respectively. The third column is the value \Eq{eq:komega}
     at which the orbits collide with $\per{-}$ in a rotational bifurcation.
     The column ``PD" is the value of $k$ for which the positive
     residue orbits undergo period-doubling. The elliptic $\tfrac13$ and $\tfrac{3}{10}$ orbits actually
     are created earlier in saddle-center bifurcations at $k=1.0$ and
     $k = 0.7063926$, respectively.
      \label{tbl:classOne}}
\end{table}

\subsection{Rotational Codes}\label{sec:RotationalCodes}
Since the dynamics on a smooth invariant circle with irrational rotation number $\omega$
is conjugate to a rigid rotation with this rotation number, it is natural to
construct a symbolic code for rotational orbits based this conjugacy. Letting $\theta$ denote
a point on the circle with unit circumference, the rigid rotation is
\begin{equation}\label{eq:circleMap}
   \theta \mapsto \theta + \omega  \mod 1 \quad \Rightarrow \quad
    \theta_{t}  = \frc{\omega t + \alpha} \;,
\end{equation}
where $\alpha$ is the initial phase. Here $\frc{x} \equiv x - \floor{x}$ denotes the fractional
part of $x$, and $\floor{x} \equiv \max_{p\in \Z}\{p : p \le x\}$ is the $\mathop{floor}$ function.
To agree with our sign choice in \Eq{eq:Henon}, we will depict this as a clockwise rotation,
see \Fig{fig:s-code}.

To obtain the codes for the circle map,
divide the circle into two wedges
\begin{equation}\label{eq:wedgeDefine}
    \begin{split}
        W_{-}(\omega) &= (-\omega,\omega) \\
        W_{+}(\omega) &= [\omega,1-\omega] \;,
    \end{split}
\end{equation}
see \Fig{fig:s-code}.
Note that $W_{-}(\omega)$ is open and $W_{+}(\omega)$ is closed,
and that these wedges divide the circle into two parts for
any $0 < \omega \le \tfrac12$.

     \InsertFig{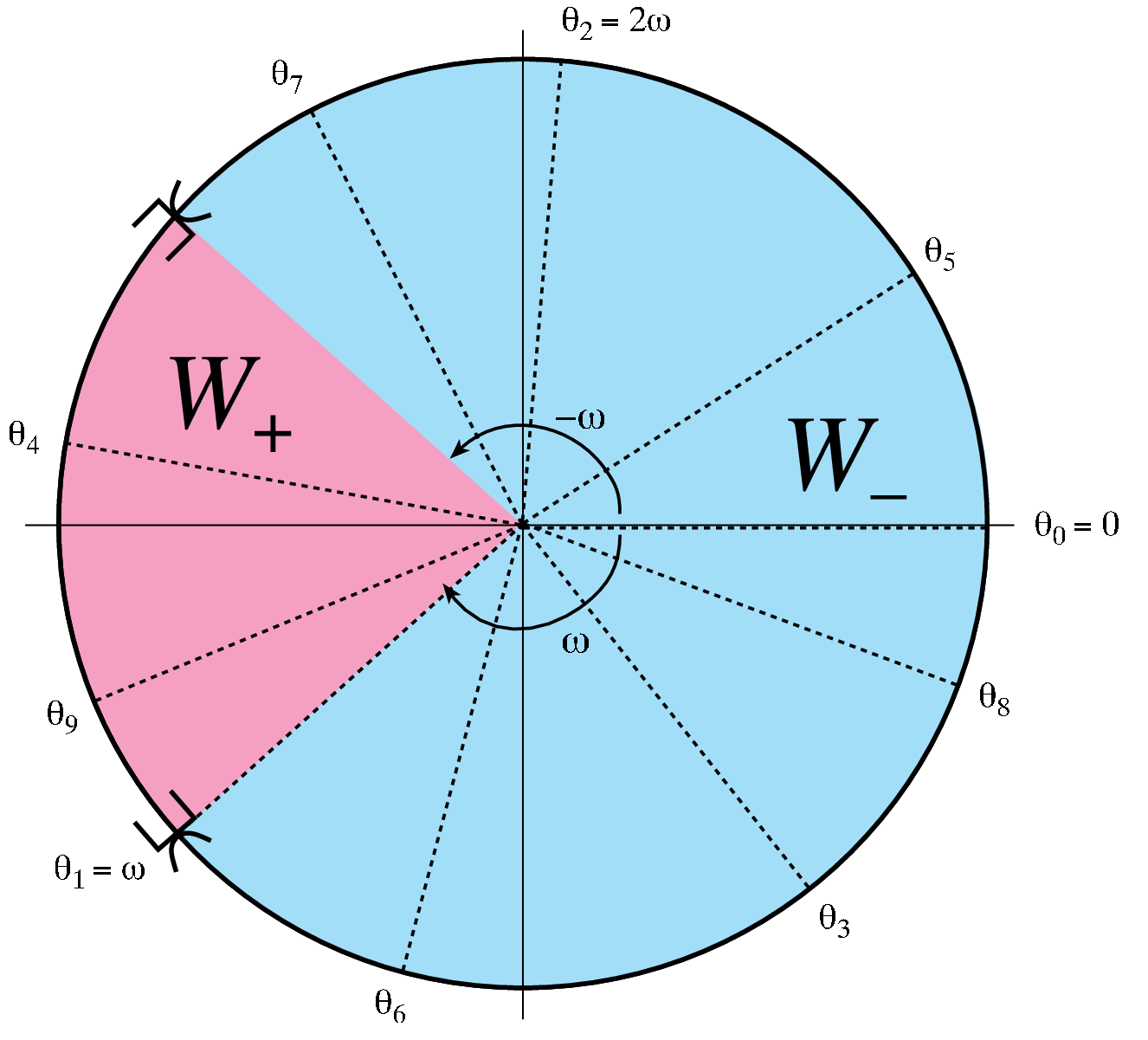} {Construction of the rotational code for
     $\omega = \gamma^{-2}$ (where $\gamma$ is the golden mean) using
     the wedges $W_{\pm}(\omega)$.  The orbit shown corresponds to the
     elliptic code with $\alpha =0$.} {fig:s-code}{3in}

\begin{defn}[Rotational (class-one) Code]\label{def:classOneCodes}
    A point $\theta \in W_s(\omega)$ is defined to have symbol $s$. The
    code for the rotational orbit $\theta_t = \omega t +\alpha$ is
    \begin{equation}\label{eq:scodeRule}
     s_{t} = \pm \quad \mbox{when} \quad \theta_t \in W_{\pm}(\omega) \;.
    \end{equation}
\end{defn}
There are two distinct types of codes that we distinguish by the value
of the initial phase $\alpha$.  When $\alpha = 0$, the first two
symbols of the code are always $-+$ while if $\alpha \in (-\omega, 0)$
then they are $--$.  Indeed, if $\alpha \neq \frc{j\omega}$ for any
integer $j$, then the symbol $-$ always appears doubled in the
code.\footnote
{
   For hyperbolic codes for rational $\omega$ this is Lemma 6.11 in~\cite{SterlingThesis}.
}
To distinguish these two cases, we call the case that $\alpha \ne
\frc{j \omega}$ for some $j \in \Z$ a {\em hyperbolic}-code, and the
alternative case an {\em elliptic}-code.

Note that whenever $\omega \neq \frac12$ there are at least two
distinct codes.  For example if we set $\omega = \frac13$, then the
three possible elliptic codes are $-++$, $+-+$ and $++-$, all of which
represent the same periodic orbit with code, $\per{-++}$.  Similarly
all the hyperbolic codes are cyclic permutations of $\per{--+}$.  In
the exceptional case, $\omega = \frac12$, there is only one period-two
code, the elliptic code, $\per{-+}$.\footnote
{
  This agrees with  the fact that the \hen map also has only one period-$2$ orbit.
}

It is easy to see that a hyperbolic code consists
of blocks of the form $--+^{m-2}$ and $--+^{m-1}$, where $m = \floor{\omega^{-1}}$,
For example, denoting the golden mean by $\gamma \equiv \frac{1+\sqrt{5}}{2}$,
then $\omega = \gamma^{-2} \approx 0.381966011$. Therefore $\gamma^{-2} \in (\frac13,\frac12)$ and
$m = 2$, so that the code is built from
the blocks $--$ and $--+$.  For example if we choose $\alpha = -.1$, we obtain
\[
     (\ldots ----+\pt ----+--+----+----+--+----+--+-\ldots)
\]
For irrational $\omega$ there are infinitely many hyperbolic codes, depending
upon the choice of $\alpha$.
By contrast, there is only one elliptic code, up to cyclic permutations, because any choice
$\alpha = j\omega$ generates the same sequence of symbols.
When $\alpha=0$ the elliptic code for $\omega = \gamma^{-2}$ is
\[
     (\ldots +--+\pt -+--+----+--+----+----+--+----+\ldots)\,,
\]
in agreement with the first few points shown in \Fig{fig:s-code}.
As previously asserted the ``first" two symbols in this code are $-+$, but
all other blocks have an even number of $-$'s.

When $\omega$ is rational, \Def{def:classOneCodes} gives rise to
exactly two distinct codes (up to cyclic permutation).  As we will
see, these correspond to the two class-one orbits with this rotation
number.

\begin{lem}\label{thm:2class1codes}
    For any rational rotation number in $(0 ,\frac12)$, there are exactly
    two codes (the elliptic and hyperbolic codes) up to cyclic permutations.
\end{lem}
\proof
Set $\omega = \frac{p}{q}$, where $p, q \in \Z$ are relatively prime.
The periodic orbit consists of $q$ evenly spaced points on the circle, and
subsequent points are obtained by shifting $p$ points around the circle.
Partition the circle into $q$ half-open sectors
\begin{equation}\label{eq:S-sectors}
       S_j = (\frac{j}{q},\frac{j+1}{q}], \; j = 0,1,\ldots q-1 \;.
\end{equation}
Then each $S_j$ contains exactly one point of the orbit.
Note that $W_{-}$ consists of the interior of the union of $2p$ of the $S_j$
\[
        W_{-} = \Int \displaystyle{\bigcup_{j=-p}^{p-1} S_j} \;,
\]
(where the indices are taken mod $q$)
and the interior of $W_{+}$ is the union of the remaining ones
\[
        \Int W_{+} =  \Int \displaystyle{\bigcup_{j=p}^{q-p-1} S_j} \;.
\]
It is immediately clear that each orbit with $\alpha \in \Int S_0$ has
the same (hyperbolic) code, since the points fall in the interiors of
exactly the same sequence of $S_j$, and therefore in the same sequence
of $W_i$.  Similarly, for any $j$, the codes for orbits that have
$\alpha \in \Int{S_j}$ are all identical.  Finally since an orbit that
starts in the interior of $S_0$ reaches $S_j$ after some number, $t$,
steps, then the codes of orbits that start in these two sectors are
the same up to cyclic permutation.

For the elliptic case we set $\alpha = j/q$ for integer $0 \le j < q$.
Each of these orbits corresponds to the same set of points on the
circle (the boundary points of the $S_j$), differing only by cyclic
permutation.  Thus their codes also differ only by cyclic permutation.

Finally the elliptic code has only $2p-1$ points in $W_{-}(\omega)$,
while the hyperbolic code has $2p$ points.  Thus these two codes
differ.  \qed

We choose a canonical ordering for these two codes by selecting
$\alpha = 0$ for the elliptic case and $\alpha = -\frac{1}{2q}$ for
the hyperbolic case (this mimics the interleaving of the two orbits of
an island chain).  Thus the elliptic code always starts with $-+$ and
the hyperbolic with $--$.  From the third symbol onwards the two codes are identical.
%




%

%
\begin{lem}[Properties of Rotational Codes]\label{thm:class1Prop}
    The canonically ordered elliptic and hyperbolic codes for periodic
    orbits differ only in their second symbol: $s_1 = -$ for
    hyperbolic and $s_1 = +$ for elliptic codes.  Elliptic codes have
    odd parity and hyperbolic codes have even parity.
\end{lem}
\begin{proof}
This follows immediately from the fact that both orbits initially lie in
the sector $S_{q-1}$ defined in the previous lemma.  Thus they visit
exactly the same sectors in the same order.  The only symbolic
difference occurs at $t=1$ where the elliptic point lies at $\theta_1
= \omega \in W_{+}$, while the hyperbolic point is in the interior of
$S_{p-1}$, which is in $W_{-}$.  The elliptic code has odd parity
since it has exactly $2p-1$ points in $W_{-}$.
\end{proof}
A convenient representation for rotational codes is the Farey tree,
\Fig{fig:scodeFarey}.  The tree shows that the canonical hyperbolic
code of a Farey daughter is simply the concatenation of the codes of
its hyperbolic parents (putting the code from the larger rotation
number on the left).  This fact is proved in
\App{sec:accelerationCode}.

     \InsertFig{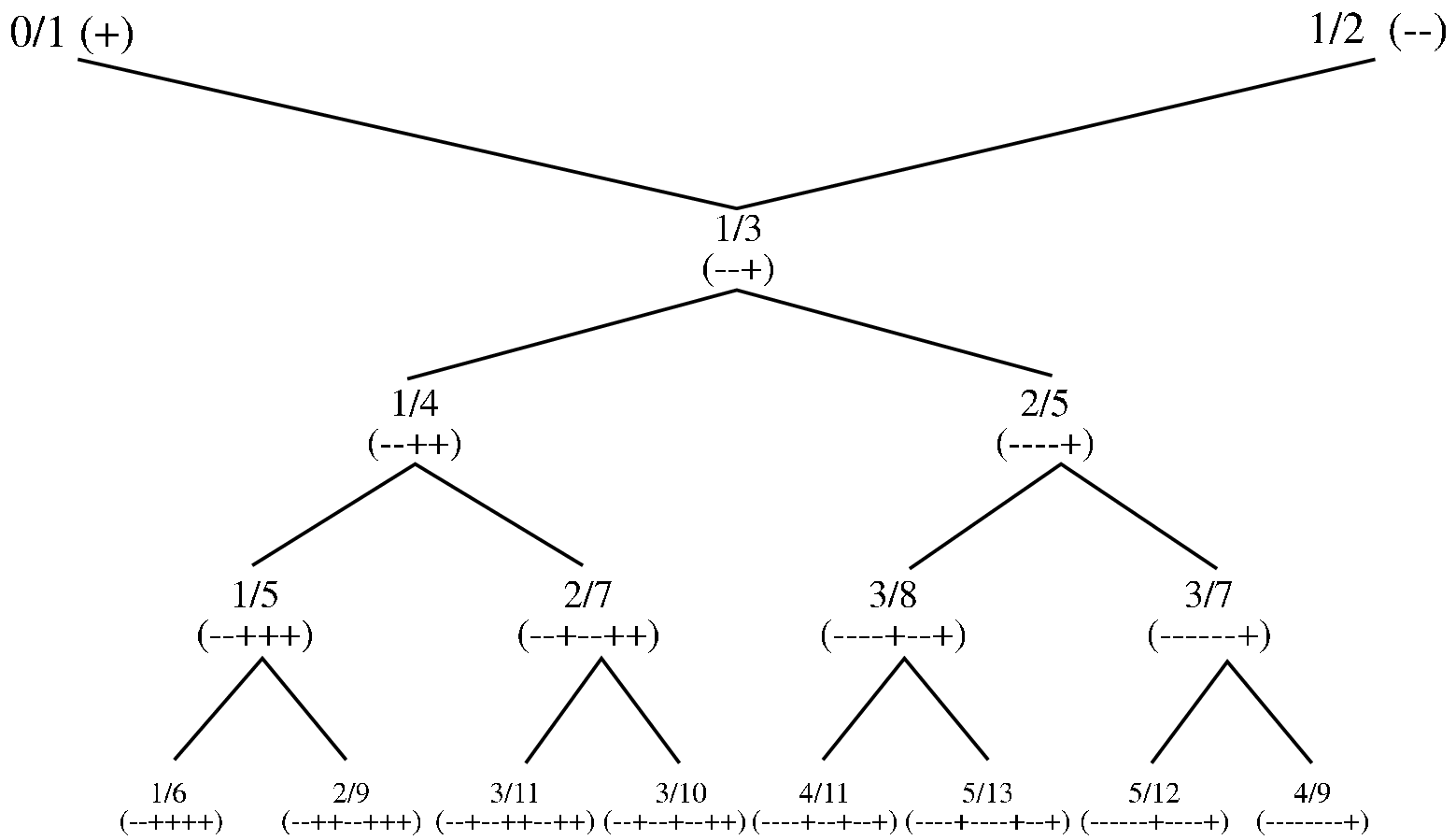} {Farey tree for the base $\tfrac01$ and
     $\tfrac12$ up to level three, and the corresponding codes for
     the hyperbolic rotational orbits.  The code for $\tfrac12$ does
     not correspond to a real orbit, as there is no ``hyperbolic''
     period-two orbit. The codes for the elliptic rotational orbits are
     obtained by flipping the second symbol.} {fig:scodeFarey}{4in}

As we will see in \Sec{sec:classOneSymbol}, the wedge shaped symbol
boundaries in \Fig{fig:s-code}, translate into a similar wedge for the
symbol boundary in the phase space of the \hen map.  However, 
before examining this, we present additional numerical evidence 
that lead to \Def{def:classOneCodes}.

\subsection{Numerical Observations}\label{sec:classOneNumerical}
While there is no proof that the codes for the rotational orbits of the \hen map are generated by 
\Eq{eq:scodeRule}, we conjecture that this is the case:
\begin{con}\label{con:Codes}
      Every class-one rotational orbit of the area-preserving \hen map
      smoothly continues to an orbit of the horseshoe with
      code defined by \Eq{eq:scodeRule}.
\end{con}

As noted in \Sec{sec:Numerical}, our initial observations indicate
that this conjecture is true up to period $20$.  In this section we
show in addition that \Eq{eq:scodeRule} generates the correct
codes for all rotational orbits up to period 99; i.e., that these
orbits are born in a rotational bifurcation of elliptic fixed point at
the value of $k$ given by \Eq{eq:rotBifRes}.  To test this we must
numerically continue each such orbit to parameter value where its
residue vanishes.  We observe that each hyperbolic rotational orbit
has a negative residue near the AI-limit, and that $r$ monotonically
increases as $k$ decreases until it reaches zero at a rotational or
saddle-center bifurcation.  Similarly elliptic rotational orbits have
positive residues near the AI-limit, and $r$ monotonically decreases
to zero as $k$ decreases.  In \Fig{fig:classone-residue} we show the
scaled residue, \Eq{eq:scaledresidue}, as a function $k$ for five
elliptic and hyperbolic class-one orbits.
  \InsertFigTwo{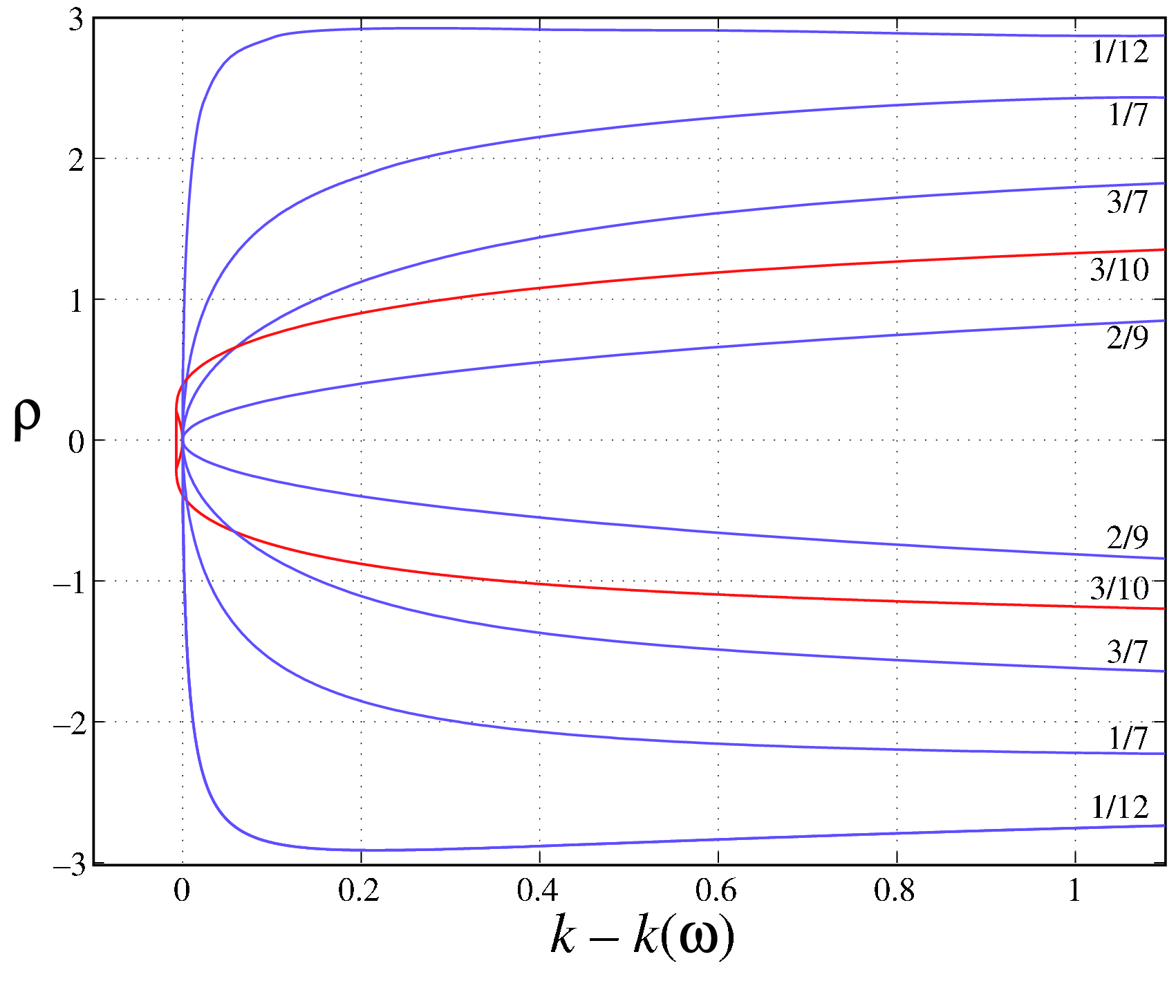}{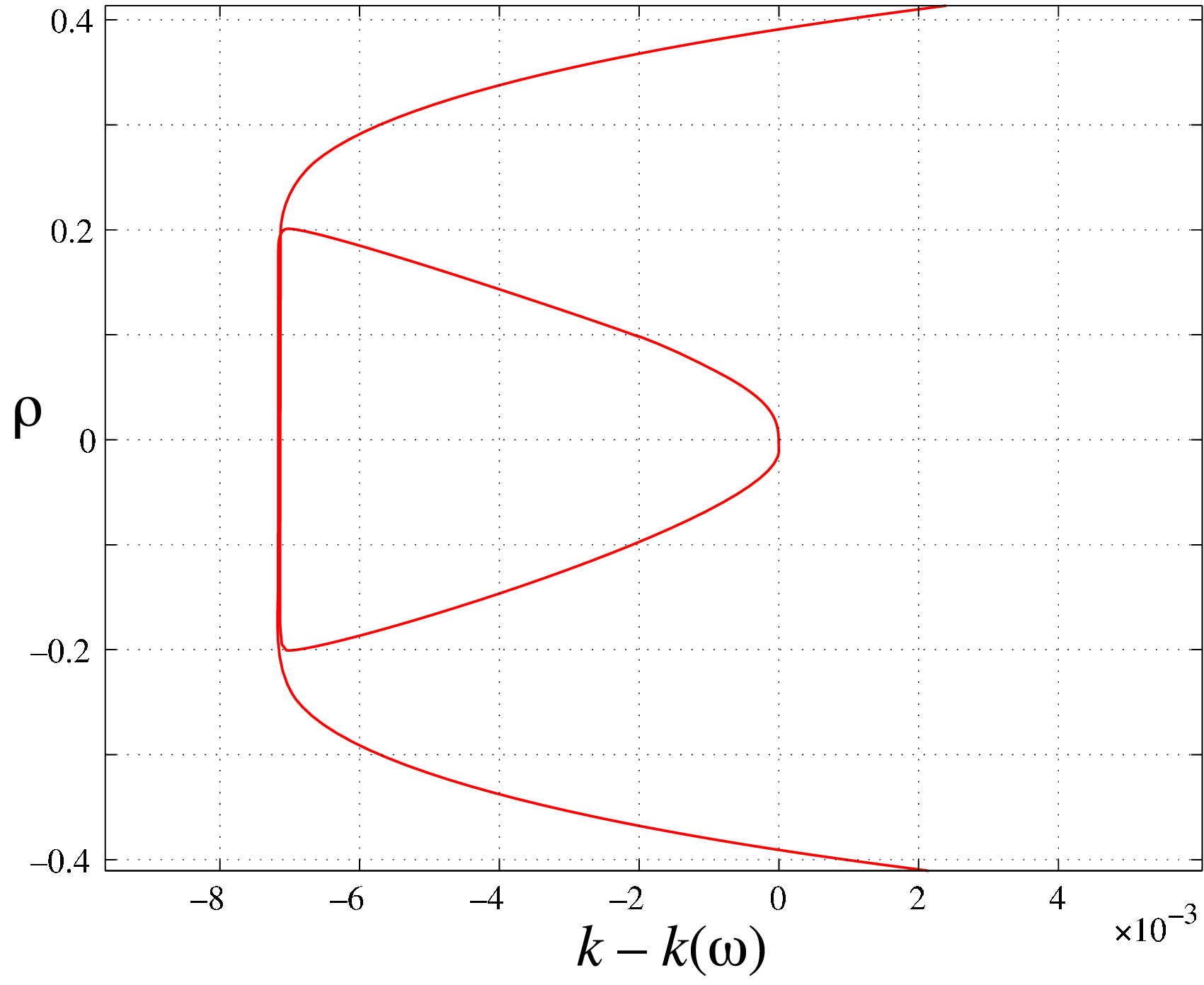} {Left Panel: Scaled residue
  \Eq{eq:scaledresidue} as a function of the deviation $k$ from $k(\omega)$ for $5$
  class-one rotational orbits.  Right panel: an enlargement of the
  residue for the $\omega=\frac{3}{10}$ orbit in the vicinity of the
  rotational and twistless bifurcations.}
  {fig:classone-residue}{3in}

Thus it would appear that a quick method for detecting the rotational
bifurcation would be to find the parameter value, $k_1(\omega)$ for
the first zero of $r$ moving away from the AI limit.  In
\Fig{fig:classone-accuracy} we plot the number of digits of agreement
between $k_1(\omega)$ and $k(\omega)$ for the 1501 class-one
rotational orbits up to period 99 as a function of rotation number.
This shows that the first zero crossing of $r$ corresponds to the
rotational bifurcation for most of the orbits.  The dashed curve in
the right hand pane of the figure shows the cumulative distribution
function for the precision data for the complete set of 1501 orbits.
There are two groups of orbits for which the apparent precision is
lower than $7$ digits.

      \InsertFigTwo{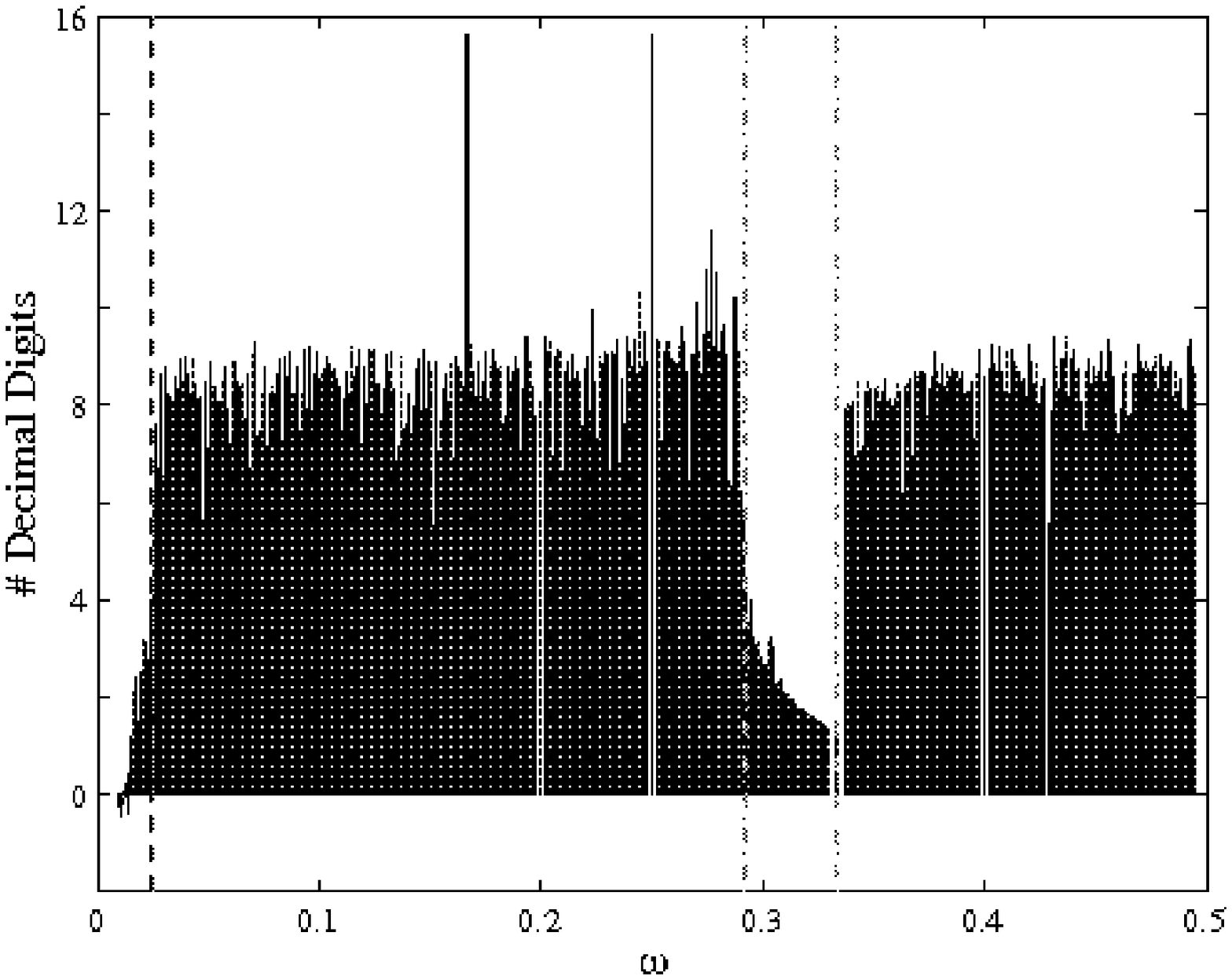}{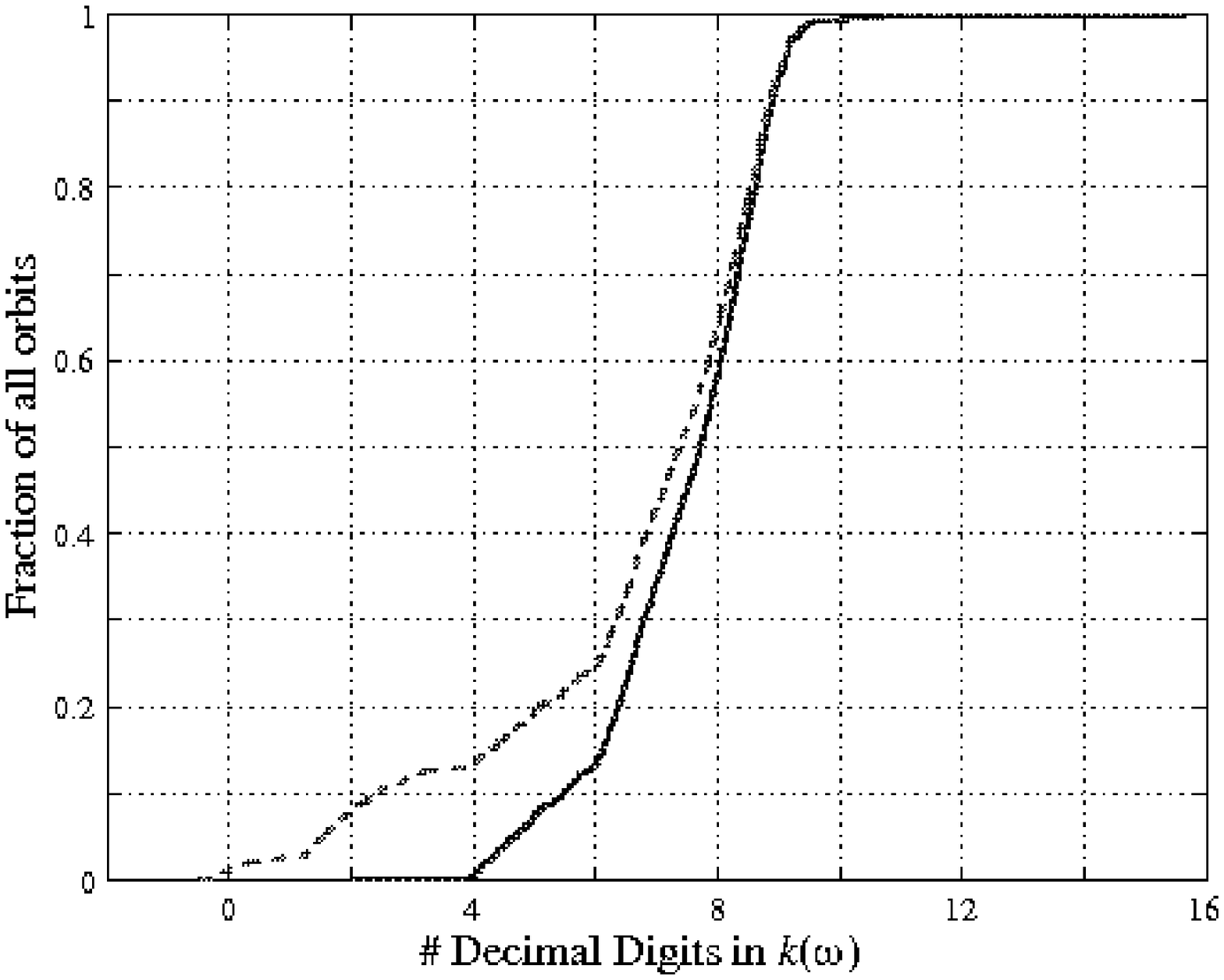} {Left Panel:
      Plot of $-\log_{10}(k_1(\omega) - k(\omega))$ for the 1501 class
      one rotational orbits with periods up to 99.  The right panel
      shows the distribution of digits of precision in the 1501
      numerical bifurcation values (dashed line), and 1305 values for
      orbits with $\omega > 0.025$ and either $k(\omega)
      <\frac{9}{16}$ or $k(\omega) > \frac54$ (solid line).  The
      latter case are the class-one orbits created at parameter values
      where the \hen map has twist and the rotation frequency is large
      enough to avoid numerical
      difficulties.}{fig:classone-accuracy}{3in}

One of these groups corresponds to orbits with $\omega \in [0.29021,
\frac13$], shown by the (green) dashed lines in
\Fig{fig:classone-accuracy}.  This rotation number interval
corresponds to the parameter interval $\frac{9}{16} < k < \frac54$
where the twist of the \hen map is ``anomalous" \cite{Dullin99}.  By
``normal" twist we mean that the rotation number decreases
monotonically moving away from the elliptic fixed point.  In the
anomalous interval the rotation number increases near the elliptic
point before reaching a maximum and eventually decreasing.  For orbits
in this interval the first zero of the residue corresponds to a
saddle-node bifurcation at $k_1(\omega) < k(\omega)$.  For
example, in the right panel of \Fig{fig:classone-residue} we show an
enlargement of the scaled residue plot for the $\frac{3}{10}$ orbits.
For this case, the first zero crossing along the curve from the AI
limit occurs at $k \approx 0.7063832$ for the hyperbolic orbit and $k
\approx 0.7063926$ for the elliptic orbit (these points are almost
indistinguishable in the figure).  
The rotational bifurcation (which is the reference value in \Fig{fig:classone-accuracy})
corresponds to the second zero crossing along each curve;
this occurs at $k(3/10) \approx 0.7135255$.  
This explains the apparent loss in accuracy in the second interval 
(bounded by the green dashed lines) in
\Fig{fig:classone-accuracy}.
%
%

Our continuation method could not find the bifurcation values for
orbits with $\omega < \frac{1}{41}$ (indicated by the (red) dashed
line in \Fig{fig:classone-accuracy}) to more than 4 digits.  This is
due to the ill-conditioning of the equations as $k \rightarrow -1$.
In this region the elliptic and hyperbolic fixed points approach one
another and the map is nearly integrable.  This is reflected in the
fact that periodic orbits very nearly lie on periodic invariant
circles and so there is nearly a null direction in the matrices used
in the continuation method.  Our numerical method fails when the
condition number of the matrices exceeds $10^{12}$, and for low
frequency rotational orbits the onset of this ill-conditioning occurs
much further from the bifurcation value than for higher frequency
orbits.  This causes a premature loss of precision and produces less
accurate estimates of the bifurcation values.

The solid curve in the right hand pane of \Fig{fig:classone-accuracy}
shows the cumulative precision data for the subset of orbits with
$\omega \ge 0.025$ and either $k(\omega) < \frac{9}{16}$ or $k(\omega)
> \frac54$.  The continutation method produced at least 7 digits of
precision for $70\%$ of of the orbits; none of the obits had fewer
than 4 digits of precision once we excluded the orbits with anomalous
twist and low frequencies.  In every case that we investigated, the
precision can be improved by carefully ``hand-tuning" the parameters
of the continuation method (in particular the parameters in the
variable step-size routines).  Thus we believe that
\Fig{fig:classone-accuracy} gives compelling evidence in favor of
\Con{con:Codes}.

%



\subsection{Symmetry}\label{sec:classOneSymmetry}

As is well-known \cite{Lamb98b}, the area-preserving \hen map is
reversible; that is, it is conjugate to its inverse $f \circ S = S
\circ f^{-1}$, where
\begin{equation}\label{eq:sReversor}
      S(x,x') = (x',x) \;.
\end{equation}
Note that the reversor $S$ is orientation reversing, $\det(DS) = -1$,
and is an involution, $S^2 = id$.  Whenever a map is reversible in
this sense, it can be factored into a pair of orientation-reversing
involutions, $f = (fS) \circ S = R \circ S$.  The second reversor for
\Eq{eq:Henon} is
\begin{equation}\label{eq:rReversor}
       R(x,x') = (x,-x'-k+{x}^{2}) \;.
\end{equation}
More generally, any image of a reversor is also a reversor; the full
set of involution pairs $\{ (f^jR, f^jS) , j \in \Z\}$ form a {\em
family} of reversors.  Thus for example,
\begin{equation}
     T(x,x') \equiv f^{-1}S(x,x') = S f(x,x') =  (-x-k+x'^2,x')
\end{equation}
is also a reversor, and since $f^{-1}R = S$, we have $f = ST$.  We
will sometimes find it convenient to use this alternative
factorization.

An orbit $\cO$ is {\em symmetric} if it is invariant under a reversor,
that is if $S\cO = \cO$.  It is easy to see that every symmetric
periodic orbit has two points on the fixed sets of $S$ or $R$
\cite{Lamb98b}; we denote these, for example, as $\fix{S} =
\mbox{fix}(S) = \{(x,y) : S(x,y) = (x,y)\}$:
\begin{align}\label{eq:fixedSets}
     \fix{S} &= \{(x,x'): x=x'\} \;,\nonumber \\
     \fix{R} &= \{(x,x'): 2x' = {x}^{2}-k \} \;.
\end{align}
The fixed set of $T$ is related to \Eq{eq:fixedSets} because
$f^n\fix{R} = \mbox{fix}(f^{2n} R)$, which implies that
\[
   f^{-1}\fix{R} = \fix{T} = \{ (x,x') : 2x = x'^2-k\} \;.
\]

The fixed sets intersect at the elliptic and hyperbolic fixed points.
Following \cite{Meiss86}, we divide the lines at the elliptic fixed
point, labeling the four rays with a subscript $i$ for the {\em
ingoing} half of the line (leading to the hyperbolic point) and
subscript $o$ for the {\em outgoing} half (leading away to infinity).  An example
is shown in \Fig{fig:k275symmetry} for $k=2.75$, where the hyperbolic
fixed point is off-scale (to the upper right), and the elliptic fixed
point is centered.  Shown are a number of class-one invariant circles
and periodic orbits that have class-two invariant circles around them.
Each of the class-one periodic orbits is symmetric; for example, the
$\tfrac37$ elliptic orbit has points on $\fix{R}_o$ and $\fix{S}_i$,
while its hyperbolic partner has points on the other two rays.

     \InsertFig{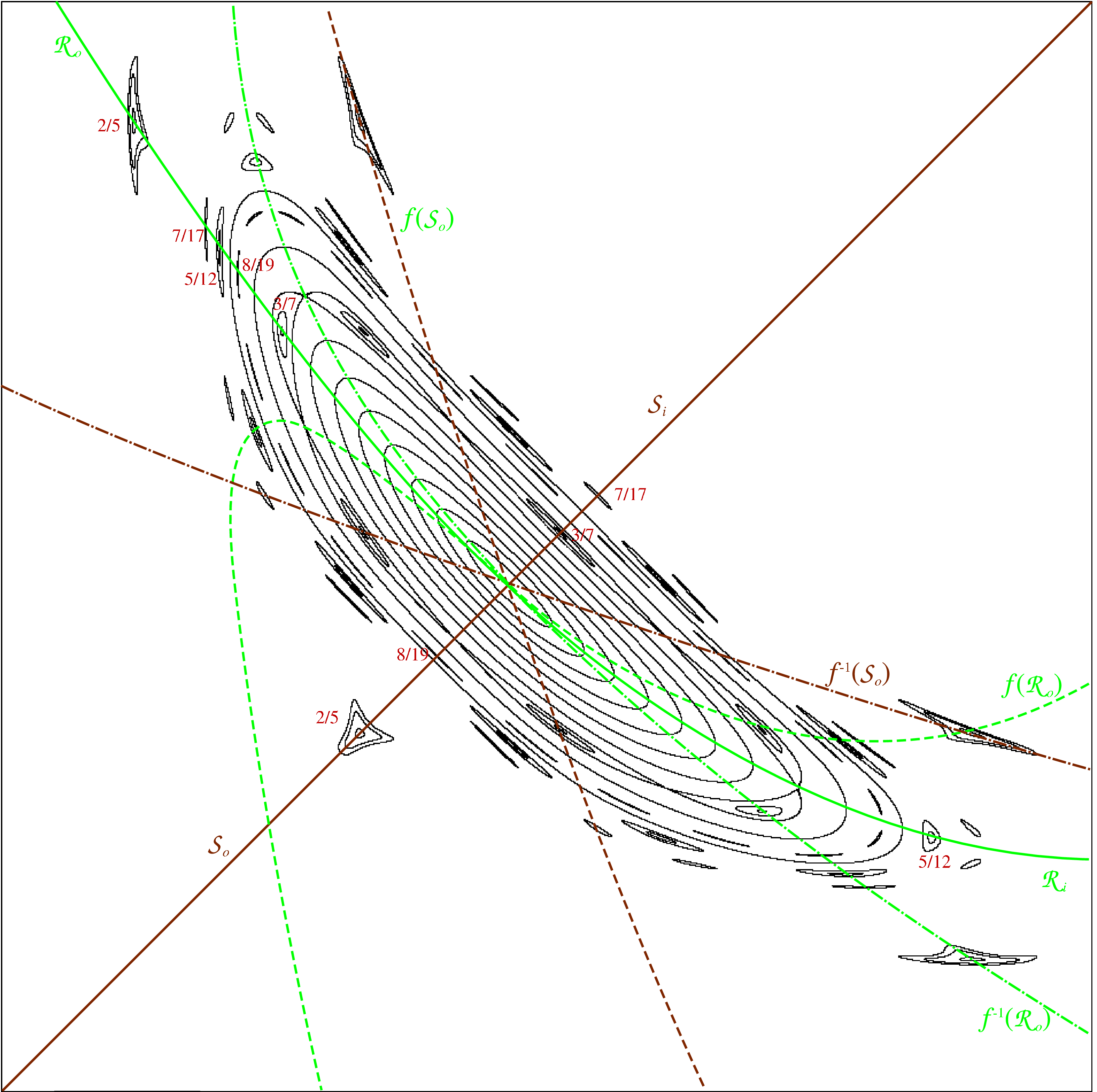} {Phase space of the \hen map for $k = 2.75$ with
     a range $(-1.75,0)$ for $x$ and $x'$. Shown are a number of orbits, including
     invariant circles around the elliptic fixed point, and five class-one island
     chains with rotation numbers $\tfrac25, \tfrac{7}{17},
     \tfrac{5}{12}, \tfrac{8}{19}, \tfrac37$. Also shown are the four
     symmetry rays  (solid curves) and their images (dashed curves) and preimages
     (dash-dot curves).} {fig:k275symmetry}{5in}

It was first noted by Greene that for many reversible examples the
class-one, elliptic, rotational periodic orbits tend to lie on one
particular symmetry ray, called the dominant ray
\cite[\S1.2.3]{MacKay93}.  To our knowledge this has never been proven
in general, though it is known to be true for maps similar to the
standard map for large enough $k$ \cite{Veerman91}.  For the \hen map
the dominant ray appears to be $\fix{R}_o$ for large values of $k$.

However, the identity of the dominant ray depends upon the
monotonicity of the twist of the map.  For the \hen map, there is at
least one anomalous domain, $\frac{9}{16} < k < \frac54$, where the
rotation number is increasing~\cite{Dullin99}.  For orbits in an
anomalous twist region, the elliptic and hyperbolic orbits reverse
roles, and the ray $\fix{R}_o$ contains only hyperbolic orbits.
As a result there is no dominant symmetry line that contains all
elliptic orbits when the twist is reversed.

Under the assumption that $\fix{R}_o$ is dominant, we can easily
determine which of the other symmetry rays contain points on the
elliptic orbit, depending upon whether the numerator and/or the
denominator of the rotation number are even or odd, see
\Tbl{tbl:symmetrylines}.  The labeling of these lines is consistent
with that in \Fig{fig:k275symmetry}.

\begin{table}[htdp]
\begin{center}
\begin{tabular}{c|c|c|c|c|}
       $\omega = \tfrac{p}{q}$  &   $E_d$       & $E_s$  &  $H_d$    &   $H_s$ \\
       \hline
       odd/odd  &   $\fix{R}_{o}$   & $\fix{S}_{i}$& $\fix{R}_{i}$  &$\fix{S}_{o}$\\
       even/odd &   $\fix{R}_{o}$   & $\fix{S}_{o}$& $\fix{R}_{i}$  &$\fix{S}_{i}$ \\
       odd/even &   $\fix{R}_{o}$   & $\fix{R}_{i}$& $\fix{S}_{o}$  &$\fix{S}_{i}$ \\
\end{tabular}
\caption{\footnotesize{The dominant($E_d$, $H_d$) and
subdominant($E_s$, $H_s$) symmetry rays.  Every elliptic (positive
residue) class-one orbit (except when the twist is anomalous) is
observed to have a point on $E_d$.  Each row corresponds to rotation
number $\tfrac{p}{q}$ with $p$ and $q$ even or odd as indicated.}
\label{tbl:symmetrylines}}
\end{center}
\end{table}

The time reversal symmetry operators also have a representation for
the shift map, $\sigma$, \Eq{eq:shiftMap}.  Indeed the interpretation
of $S$ is obvious if we consider its action on a configuration
sequence: $S(x_{t-1},x_{t}) = (x_{t},x_{t-1})$.  Thus as an operator
on symbol sequences, $S$ simply reflects them about the binary point
\[
      S(\ldots s_{-1}\pt s_{0}s_{1} \ldots) = (\ldots s_{1}s_{0}\pt s_{-1} \ldots)
\]
For symbol sequences, it is convenient to use the operator $T = S\sigma$ for the second symmetry (instead of $R$), because this corresponds to a reflection about the symbol $s_0$:
\[
      T(\ldots s_{-1}\pt s_{0}s_{1} \ldots) = (\ldots s_{1}\pt s_{0}s_{-1} \ldots)
\]
Thus the elements of the fixed set of $S$ are sequences symmetric
under a reflection about the binary point, and those of $T$ are
sequences symmetric under a reflection about $s_{0}$.  Using the
geometry of the horseshoe (see \App{sec:horseshoe}) and its implied
ordering relation given in \Lem{thm:ordering}, we can also find the
codes for the symmetry rays.  The $o$ rays correspond to codes whose
symbols are ``smaller" than that for $\per{-}$.  This means that the
parity of sequence $s_0s_1\ldots s_j$, where $s_j$ is the first symbol
that is not $-$, must be negative, i.e., there must be an odd block of
$-$ symbols after the binary point.  Similarly the $i$ rays correspond
to codes that are ``larger" than $\per{-}$, i.e. that have an even
block of $-$'s (or none).  More concretely, define $\kappa(\bs)$ to be
the number of contiguous $-$ symbols following the binary
point:\footnote
{
  $\kappa(\bs)$ does not exist for $\per{-}$, as it should not, since this point is the dividing point
  for the four rays.
}
\begin{equation}\label{eq:kdefine}
     \kappa(\bs) = \min_{j \ge 0} \{j: s_j = +\}
\end{equation}
then we have
\begin{align}\label{eq:symmetryrays}
   \fix{S}_o &= \{\bs : s_{-t} = s_{t-1}, \; \kappa(\bs) \mbox{ odd} \} \;,\nonumber\\
   \fix{S}_i &= \{\bs : s_{-t} = s_{t-1}, \; \kappa(\bs) \mbox{ even}\} \;,\nonumber\\
   \fix{T}_o &= \{\bs : s_{-t} = s_{t},   \; \kappa(\bs) \mbox{ odd}\} \;,\nonumber\\
   \fix{T}_i &= \{\bs : s_{-t} = s_{t},   \; \kappa(\bs) \mbox{ even}\} \;.
\end{align}

It is easy to see that the rotational codes of \Sec{sec:RotationalOrbits} are symmetric. For example, from \Tbl{tbl:classOne}, the code $\bs = (\ldots --+++\pt --+++\ldots)$ for the hyperbolic $\tfrac15$ orbit has two points on symmetry lines,  its first image $ \sigma(\bs) = (\ldots+++-\pt -+++\ldots)$ is a point on $\fix{S}_o$ and $\sigma^3(\bs) = (\ldots --+\pt ++--\ldots)$ is a point on $\fix{T}_i$.

To prove this we start with a lemma about the symmetries of the rigid
rotation \Eq{eq:circleMap}.  Though the reflection point for the rigid
rotation is arbitrary, we choose $\theta = 0$, which conforms to the
symmetry of our $W_\pm$ partitions, and to our selection of $\theta$
for the canonical elliptic code.

\begin{lem}\label{thm:rotSymmetry}
The rigid rotation $F(\theta) = \theta + \omega \mod 1$ can be factored
as $F= ST$, with reversors $T: \theta \mapsto -\theta$, and $S: \theta \mapsto \omega -\theta$.
When $\omega \le \frac12$, the fixed sets of these symmetries divide into the rays
\begin{eqnarray}\label{eq:rotSymmetry}
     \fix{S}_o &= \{\theta : \theta =\frac{\omega}{2}\} \;, \quad
     \fix{S}_i &= \{\theta : \theta =\frac{1+\omega}{2}\} \;,\nonumber\\
     \fix{T}_o &= \{\theta : \theta =0\} \;, \quad
     \fix{T}_i &= \{\theta : \theta =\frac12\} \;.
\end{eqnarray}
Points that start on these rays generate codes with symmetries given by \Eq{eq:symmetryrays}.
\end{lem}

\begin{proof}
    Since the $W_\pm$ partition is symmetric with respect to $\theta =
    0$, if we set $\theta_0 = 0$ then the resulting code is symmetric
    with respect to the binary point.  Since $\theta_0 \in W_-$, and
    $\theta_1 = \omega \in W_+$, the sequence begins $\pt -+\ldots$,
    so that $\kappa(\bs)$ is odd, agreeing with $\fix{T}_o$.
    Similarly, codes generated by $\theta_0 = \frac12$ are also
    invariant under $T$, and start $\pt +\ldots$, so they are on
    $\fix{T}_i$.

    If we set $\theta_0 = \frac{\omega}{2}$, then the symmetry of the
    wedges (recall \Fig{fig:s-code}) implies $s_{-t} = s_{t-1}$, so
    the code is invariant under $S$.  The first symbol is always $s_0
    = -$; we will show that $\kappa(\bs)$ is odd.  Indeed, when
    $\omega \le \frac25$, then $s_1 =+$, since $\theta_1 = \frac32
    \omega \le 1-\omega$.  Otherwise, $1-\omega < \theta_1 < 1$, so
    that $s_1 = -$.  But then $s_2 = -$ too, because $0 < \theta_2 <
    \omega$.  Indeed, a double $-$ must occur whenever the orbit
    enters $W_-$ having skipped $W_+$.  Thus the line $\theta =
    \frac{\omega}{2}$ corresponds to $\kappa(\bs)$ odd, and is
    therefore $\fix{S}_o$.  A similar argument gives the last case.
\end{proof}

Another simple lemma gives relations between the symmetry lines that imply 
the ``dominant-subdominant" pairing in \Tbl{tbl:symmetrylines}.

\begin{lem}\label{thm:subdominant}
    The four symmetry rays can be divided into two pairs $(E_d,E_s)$
    and $(H_d,H_s)$, as given in \Tbl{tbl:symmetrylines} (depending on
    the parity of $\omega$), such that if $\theta$ is a point on a
    dominant ray then $F^{\ceil{\frac{q-1}{2}}}\theta$ is a point on
    its subdominant partner.  Moreover the iterates of lines from
    different pairs are disjoint.
\end{lem}
 \proof We consider only the case $\omega =$ odd/even; the others can
 be done similarly.  First we prove that the stated pair of rays can
 be mapped into each other.  That $E_d$ is mapped into $E_s$ is the
 statement $F^t(\fix{T}_o) = \fix{T}_i$ for some $t$.  This requires
 that $t \omega = \frac12 + j$ for some integer $j$, which is
 equivalent to $\omega = \frac{2j+1}{2t}$.  Thus $\omega$ is of the
 form odd/even, and $t = \frac{q}{2}$.  For the $\fix{S}$ pair we
 similarly have $\frac{\omega}{2} + t\omega = \frac{1+\omega}{2} +j$,
 which again occurs when $t = \frac{q}{2}$.  Since each pair maps to
 itself, it is enough to show that $\fix{T}_o$ does not map to
 $\fix{S}_o$, i.e.\ $F^t(\fix{T}_o) \not = \fix{S}_o$ for all $t$.
 This implies $t \omega \not = \frac{\omega}{2} \mod 1$.  It is clear
 this equation does not have an integer solution for $t$ for the
 odd/even case.  Since $\fix{R}$ is the image of $\fix{T}$, this
 verifies the pairing in \Tbl{tbl:symmetrylines}.

We find that for the odd $q$ case $\fix{T}$ is mapped into $\fix{S}$
after $\frac{q+1}{2}$ iterations.  Since we use $\fix{R}$ in
\Tbl{tbl:symmetrylines}, this decreases the number of iterates by $1$.
\qed

Finally, we use these results to show that the rotational codes have
the symmetries as given in \Tbl{tbl:symmetrylines}.

\begin{teo}\label{thm:symclass1}
     The codes for the rotational periodic orbits are symmetric.
     Moreover the canonically ordered elliptic code is on the dominant ray $\fix{T}_{o}$
     (its image is on $\fix{R}_o$). The other symmetry lines occur as given in \Tbl{tbl:symmetrylines}.
\end{teo}

\begin{proof}
    The canonical code for an elliptic orbit has $\theta_t = \omega t$. Therefore,
    $\theta_0 = 0$, and \Lem{thm:rotSymmetry} implies that this point is on $\fix{T}_o$.
    Since $f(\fix{T}_o) = \fix{R}_o$, the first image of the canonical code is on $\fix{R}_o$.
    Thus all elliptic orbits have points on the dominant symmetry line. By \Lem{thm:subdominant},
    they also must have a point on the subdominant line $E_s$.

    The canonical code for the hyperbolic $\frac{p}{q}$ orbit has $\theta_t= \omega t - \frac{1}{2q}$
    and the orbit covers the $q$ evenly spaced points
    $\frac{2j-1}{2q} \;, j = 0, 1, \ldots q-1$ on the circle.
    Whenever $p$ is odd, there exists a time $t'$ such
    that $2j-1=p$, and then $\theta_{t'} = \frac{p}{2q} = \frac{\omega}{2}$.
    Thus $\theta_{t'} \in \fix{S}_o$. If $q$ is odd, then there exists a $t''$
    such that $2j-1 = q$ so that $\theta_{t''} = \frac12 \in \fix{T}_i$.
    This verifies the line $H_d$. The subdominant line then follows from \Lem{thm:subdominant}.
\end{proof}

In particular, this implies that the dominant symmetry line conjecture
follows from the rotational codes conjecture.

\begin{cor}[Dominant Symmetry Line]
    If \Con{con:Codes} is true, then every elliptic class-one one parameter family of periodic orbits of the
    \hen map starting at the AI limit until residue zero is encountered 
    has a point on $\fix{R}_o$.
\end{cor}
The caveat in this corollary accounts for the domain of anomalous twist. The elliptic orbits born in this region in a rotational bifurcation have points on the $H$ symmetry lines in \Tbl{tbl:symmetrylines}, and connect to the normal-twist hyperbolic orbits in saddle-node bifurcations when they touch the twistless curve. Coming from the AI limit this saddle-node bifurcation is the first encounter of $r = 0$, so the
backwards running branch between $k(\omega)$ and $k_1(\omega)$ is removed, because there
the statement is not true.

\subsection{Symbol Boundary}\label{sec:classOneSymbol}

According to \Th{thm:symclass1}, the boundaries of the wedges
$W_{\pm}$ correspond to the image and preimage of a dominant symmetry
line.  Thus we expect that the symbol partition for class-one orbits
will show this same structure.  Indeed, this is what we observe, see
\Fig{fig:k275boundary}.  The figure shows that the wedge shaped symbol
boundary leaving the elliptic fixed point (small magenta circle) is
delineated by the image and preimage of $\fix{R}_o$, the dominant ray.
This boundary also is valid for the nonrotational orbits shown.

An enlargement is shown in \Fig{fig:k243boundary}, where there is a prominent class-two island around the $\frac25$ rotational orbit. Moving from the elliptic fixed point outward, 
the figure shows class-one periodic orbits with frequencies 
$\frac{16}{39},\; \frac{9}{22},\; \frac{20}{49},\; \frac{11}{27},\; \frac{13}{32},\; \frac{15}{37}$,
and $\frac{17}{42}$.\footnote
{
    It is convenient to use the Farey tree to choose periodic orbits. For $k=2.43$, the
    rotation number of the elliptic point is $\omega \approx 0.4123 < \frac{7}{17}$.
    Starting with the pair of neighboring rationals $\frac25$ and $\frac{7}{17}$, we construct
    the Farey tree for several levels to obtain our orbits.
}
Notice that as expected from the schematic representation above, 
the $-$ region is a wedge with opening angle $2\omega$ centered on the dominant symmetry line. 
Indeed as far as we can tell the symbol boundary for all rotational 
class-one orbits is indeed formed from the image and preimage of $\fix{R}_o$.

Also shown in \Fig{fig:k243boundary} are three orbits that encircle
the $\frac25$ elliptic orbit---orbits of class two.  To the resolution
of this figure, it appears that the wedges formed from images of
$\fix{R}_o$ also provide the symbol boundary for these orbits.
However, as we will see in the sections below, this is not the case.
We start by constructing the rotational codes for higher class orbits.

     \InsertFig{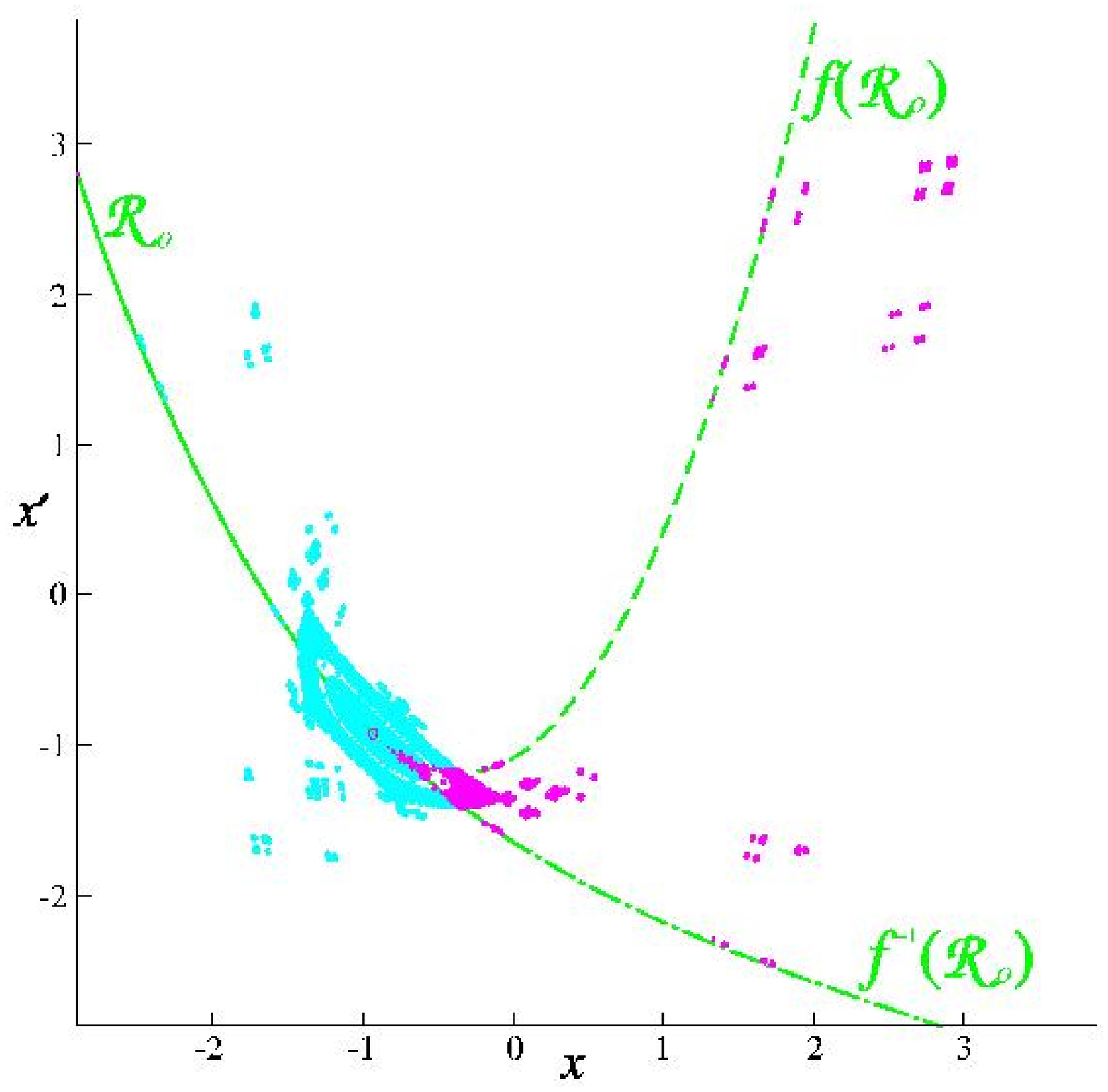} {Symbol boundary for $k = 2.75$ appears
     to coincide with the image and preimage of the dominant symmetry
     line $\fix{R}_o$.  Shown are the same orbits as in
     \Fig{fig:k20boundary}.  Those points with $s_0 = +$ are colored
     magenta and those with $s_0 = -$ are cyan.  The solid green line
     is the dominant symmetry line, $\fix{R}_{o}$, the dashed line is
     its image and the dot-dashed line its preimage.
     }{fig:k275boundary}{4.5in}
     \InsertFig{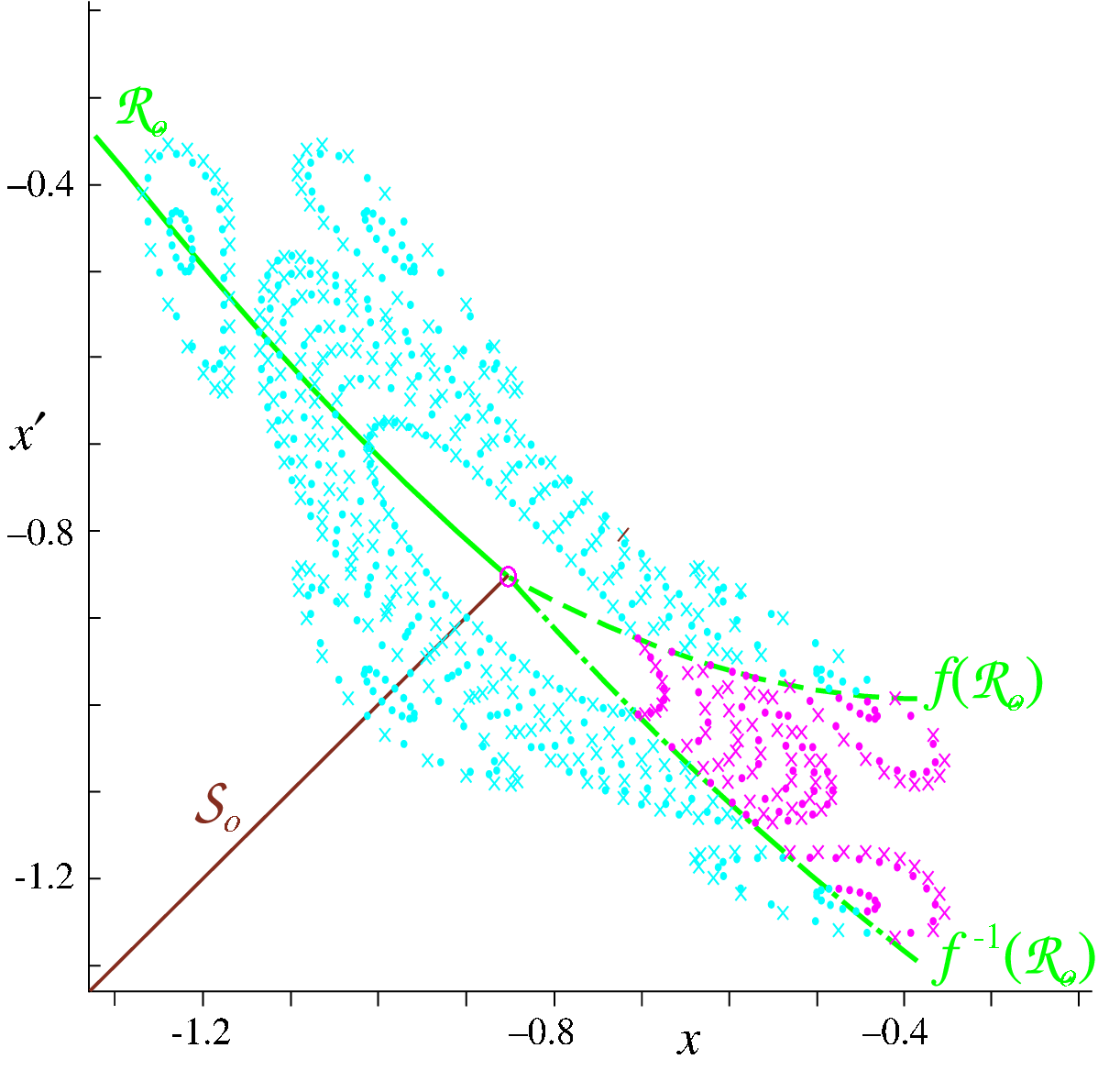} {Class-one symbol boundary for $k
     \approx 2.43$ ($\epsilon = 0.54$) coincides with the image and
     preimage of the dominant symmetry line $\fix{R}_o$.  Shown are
     $13$ class-one rotational orbits about the elliptic fixed point,
     as well as three class-two orbits encircling the $\tfrac25$
     elliptic orbit.  Elliptic (positive residue) orbits are shown as
     dots and hyperbolic as crosses.}{fig:k243boundary}{4.5in}

\section{Class-Two Orbits}\label{sec:classTwo}

As first observed by Birkhoff, a typical elliptic periodic orbit has
satellite periodic orbits in its neigbhorhood, i.e., orbits that
rotate around the elliptic orbit.  Some of these new orbits will in
turn be elliptic and therefore have satellites about them as well.  We
use the term {\em class} to refer to this hierarchy of
islands-around-islands \cite{Meiss86}.  For the \hen map, the elliptic
fixed point is defined to be class-zero, the rotational orbits of 
\Sec{sec:RotationalOrbits} are class one, and a class-two orbit is 
one that rotates around a class-one elliptic periodic orbit, 
see \Fig{fig:k275symmetry} for examples.

Recall that an elliptic/hyperbolic pair of class-one periodic orbits
with rotation number $\omega_1 = \frac{p_1}{q_1}$ is born at the
elliptic fixed point when $k = k(\omega_1)$, \Eq{eq:komega}.  After its birth,
the elliptic class-one orbit is the center of a class-two island chain 
consisting of the invariant circles and cantori that encircle each 
point on the class-one orbit. The island extends to the separatrix formed from stable and
unstable manifolds of the class-one hyperbolic orbit, as schematically
shown in \Fig{fig:islandRotations}.
The class-two orbits have rotation numbers that are measured relative to $f^{q_1}$. 
When the linearized rotation number of the class-one orbit---as determined by its 
residue, \Eq{eq:residue} or \Eq{eq:rotBifRes}---passes through the rational 
frequency $p_2/q_2$, an elliptic/hyperbolic pair of class-two orbits is created in a
$q_2$-tupling bifurcation of $f^{q_1}$.  We denote the rotation number of these
orbits by $\omega_1: \omega_2$; they have period $Q = q_1 q_2$.  Three such 
class-two orbits are shown in \Fig{fig:k243boundary} with rotation
numbers $\frac{2}{5}:\frac{1}{16}, \frac{2}{5}:\frac{1}{20}$ and
$\frac{2}{5}:\frac{1}{22}$.

\InsertFig{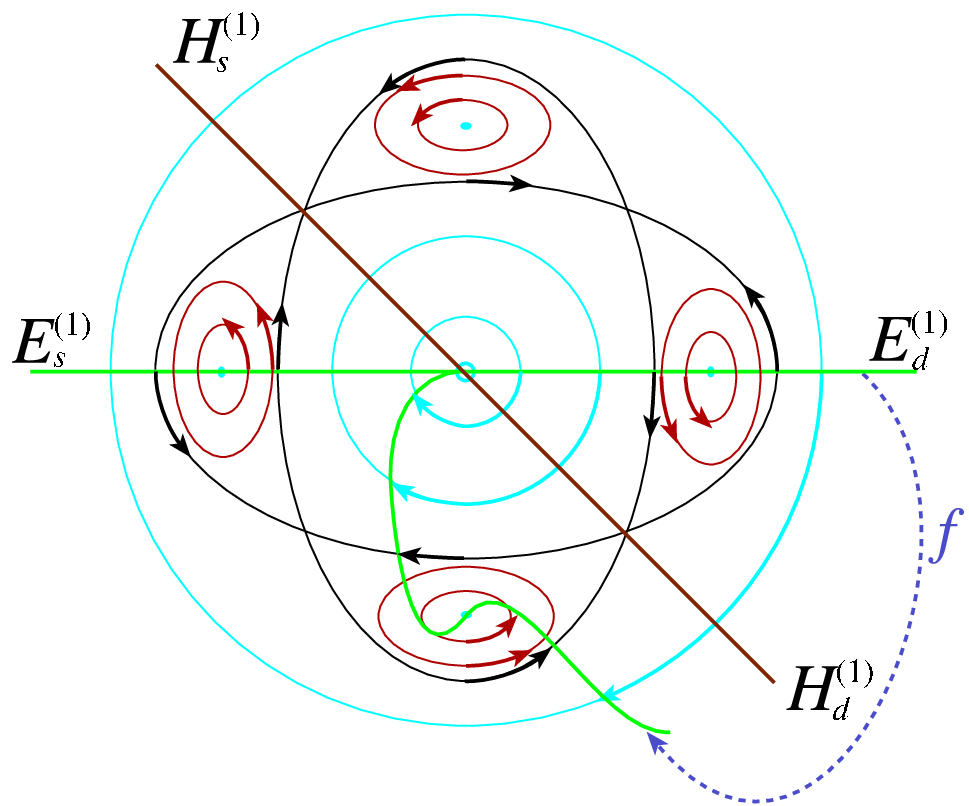} {Schematic class hierarchy: class-one
invariant circles (cyan) and a class-two $\frac14$ island chain (red). 
Since the rotation number of the class-one circles decreases with radius, the image of
the horizontal (green) line shears as shown.  Continuity implies the
direction of rotation reverses as the class is incremented.  Also
sketched are the elliptic (green) and hyperbolic (brown) symmetry rays
for class-one.}{fig:islandRotations}{3in}

For the case of normal twist, the direction of rotation typically reverses 
with each increment of class. This occurs because in an any island the 
rotation number is typically a maximum at the elliptic orbit forming the 
island center and monotonically decreases to zero approaching the separatrix.  
In particular this implies that the inner separatrix of an island advances 
with respect to the center of the island and the outer one retreats, 
see \Fig{fig:islandRotations}.
Continuity then implies that if the class-one invariant
circles have clockwise dynamics, the class-two circles have
counterclockwise dynamics.

Our numerical observations imply that the code for
a class-two $\frac{p_1}{q_1} :\frac{p_2}{q_2}$ orbit is constructed essentially
by repeating the class-one $\frac{p_1}{q_1}$ code $q_2$
times, because most of the class-two points are
deeply buried inside the wedges shown in \Fig{fig:s-code} that
determine the class-one symbols.  However, if the class-one sequence were merely
repeated $q_2$ times, the orbit would not have
the correct minimal period; consequently some of the symbols must be flipped
to obtain the correct code.
Indeed, there are two islands that intersect the class-one symbol boundary
(recall the $\frac25$ island in \Fig{fig:k243boundary}).\footnote
{
 Except for the case $\omega_1 = \frac12$, where there is only one.
}
Such islands are termed {\em ambiguous} because they straddle the symbol
boundary; the determination of the symbols of points in these special islands requires more analysis.
\subsection{Class-Two Codes}\label{sec:classTwoCode}

Schematically, a class-one orbit with rotation number $\omega _1
=\frac{p_1}{q_1}$ can be represented as $q_1$ equally spaced points on
a circle with the dynamics \Eq{eq:circleMap}. In a similar vein, a class-two orbit with rotation number $\frac{p_1}{q_1} : \frac{p_2}{q_2}$ schematically corresponds to points on a set of circles enclosing each point of the class one orbit. To compute these points choose a pair of radii, $r_1$, $r_2$ to represent the sizes of the islands and define
\begin{equation} \label{eq:expClassTwo}
  x + iy = e^{-2\pi i \theta_1} \left( r_1 + r_2 e^{2\pi i \theta_2}\right) \;.
\end{equation}
Here $r_2$ chosen sufficiently smaller than $r_1$ so that none of the class-two islands intersect. This gives an {\em epicycle} view of the orbit.
The angles $(\theta_1,\theta_2)$, representing the island structure, evolve according to the two rotation numbers of the orbit. The schematic dynamics is expressed by using a mixed basis for time:  $t =
t_1 + q_1 t_2$ where  $t_1 = t \mod q_1$
so that $t_i \in [0, q_i)$; this is written in clock-like notation as\footnote
{
Mixed base systems like the above are well known for time measurements, e.g. setting $\omega = \frac{1}{10} : \frac{1}{60} : \frac{1}{60} : \frac{1}{24} : \frac{1}{7} :\frac{1}{52}$ corresponding to tenth of seconds, seconds, minutes, hours, weekdays, weeks, so that in our notation $t = 0:1:30:19:4:5$ is the 4th day in the 5th week, at $19{:}30$ hours plus $1.0$ seconds. Instead of $\frac{1}{24}$ the American system uses $\frac{1}{12}:\frac12$, where the $2$ is represented as AM or PM. Our lives are simpler because all numerators are $1$ for an actual clock, hence $\theta_j$ is not needed; they would have been even more so if the French revolution had succeeded in promulgating a decimal time system.
}
\[
     t \equiv t_1 : t_2 \,.
\]
With this definition, the angles evolve as
\begin{equation}\label{eq:torusmaps}
         \theta_i(t_i) = \omega_i t_i + \alpha_i \;,\; i = 1,2 \;.
\end{equation}
Iteration of this pair of maps produces a set of $Q=q_1q_2$ points on the epicycle.

The signs in the exponentials in \Eq{eq:expClassTwo} are reversed to account for
the reversal of rotation direction that occurs when moving between the
classes. 

Recall that the canonical
elliptic class-one code is obtained by setting $\alpha_1 = 0$ and the
hyperbolic code is obtained when $\alpha_{1}= -1/2q_{1}$.  Since
class-two orbits rotate around elliptic class-one orbits, we set $\alpha_{1}=0$; the phase $\alpha_2$ differentiates between the elliptic and hyperbolic class-two orbits.
The canonical elliptic class-two code has $\alpha_2
= 0$, and the canonical hyperbolic code has $\alpha_2 = -\frac{1}{2q_2}$.

The epicycles for $\omega = \frac27 : \frac29$
are shown in \Fig{fig:classII2729}.  The points of the elliptic
class-one orbit are represented by filled circles evenly spaced along
the large circle.  Surrounding each of the $q_1$ points of this
orbit is a smaller circle containing $q_2$ points of the class-two
periodic orbit.  By convention, we measure $\theta_2$ in a corotating coordinate
system, so that points with $\theta_2 = 0$ are located on radial rays
from the origin.  Notice that a $q_1$-fold iteration maps each small circle to itself,
shifting its points counterclockwise by $\omega_2$.

    \InsertFig{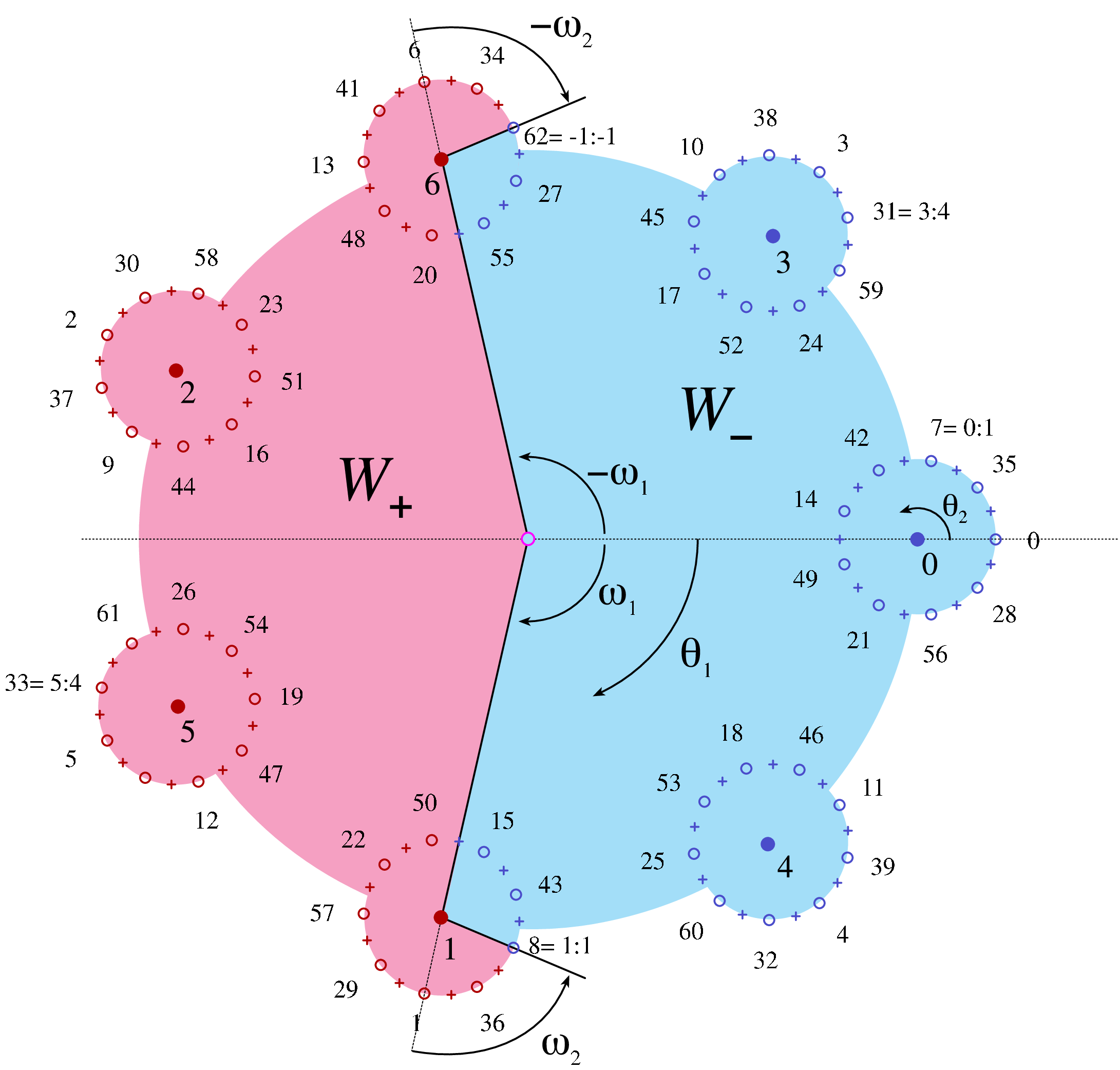}{Epicycle view of a class-two orbit for $\omega = \frac27 :\frac29$.
    The picture is generated by \Eq{eq:expClassTwo} with $r_1 = 1$ and $r_2 = 0.1$.
    The angle $\theta_1$ is measured clockwise, and $\theta_2$ counterclockwise. The elliptic
    class-one $\frac27$ orbit is denoted with solid circles, and the
    class-two orbits are denoted by open circles (elliptic), and
    crosses (hyperbolic).  Points with symbol $-$ are colored blue and
    those with $+$ are colored red. The integers next to each elliptic point
    denote the time $t$ along the orbit.} {fig:classII2729}{4in}

To generate the symbol sequence for a class-two orbit, the epicycle picture must be partitioned into regions corresponding to the $+$ and $-$ symbols. Our observations indicate that this can be done by constructing a pair of ``wedges" $W_\pm(\omega)$ whose boundaries are piecewise linear. The boundaries start as rays from the origin at angles $\theta_1 = \pm \omega_1$---these same rays formed the class-one boundary. The rays end in the center of the two ambiguous class-two islands corresponding to $t_1 = \pm 1$; it is not immediately clear how to continue the symbol boundary in these islands. A cursory look at \Fig{fig:k243boundary} might indicate that the symbol boundary continues straight through the ambiguous islands. However, our numerical observations show that the two rays actually bend at the centers of the $t_1 = \pm 1$ ambiguous islands by angles $\theta_2 = \pm \omega_2$, respectively. This bending is too small to see in \Fig{fig:k243boundary}.

The regions $W_{\pm}(\omega_1:\omega_2)$ bounded by these rays correspond to the $\pm$ symbols. Just as for the class-one case, the region containing $\theta_1 = 0$ corresponds to $s_t = -$. The only remaining point to understand is the symbol for points on the boundary. In contrast to class-one, our observations show that $W_{-}$ is closed and $W_+$ is open.

\begin{defn}[Class-Two Code]\label{def:classTwoCodes}
    $t = t_1 : t_2$ determines a point $z$ in the plane by (\ref{eq:expClassTwo}) 
    and (\ref{eq:torusmaps}). If $z \in W_s(\omega_1:\omega_2)$ then the $t$th symbol is $s$. 
    The region $W_s(\omega_1:\omega_2)$ is bounded by a continuous piecewise linear system 
    of segments. At the origin two  
    segments of length $r_1$ emerge at angles corresponding to $t_1 = \pm 1$.
    Their endpoints are connected to segments of length
    $r_2$ with angles corresponding to $t_1:t_2 = \pm 1: \pm 1$, see, e.g. \Fig{fig:classII2729}.
    The region $W_-(\omega_1:\omega_2)$ is the one that contains $t = 0:0$ and is closed, 
    while $W_+(\omega_1:\omega_2)$ is its complement.
\end{defn}

It is not difficult to translate the epicycle picture into a simple algorithm that generates the codes for any 
$\omega = \frac{p_1}{q_1} : \frac{p_2}{q_2}$.

Continuing the example of \Fig{fig:classII2729}, the orbits are coded so that the points in the red region have symbol $+$ and those in the blue region have symbol $-$. The four points that fall on the symbol boundary (for example $t = 8 = 1:1$ and $t = 62 = 6:1$ for the elliptic orbit) have the symbol $-$.  Note that points in the non-ambiguous islands, corresponding to $t_1=0,2,3,4,5$ have the same code as the elliptic class-one $\frac27$ orbit. Thus many of the symbols of the class-two orbit simply repeat the class-one code.

To see this more clearly, denote the class one code by the row vector $\bs_1$. Since most of the points on the class-two orbit are determined by this code, we first replicate this code $q_c$ times; however, instead of viewing this as a vector of length $q_1 q_2$ we reshape the code as a table with $\bs_1$ as its rows. The columns of this table are labeled by $t_1$, and the rows by $t_2$. In this way each ambiguous island corresponds to a column of the table, i.e. the columns $t_1 = 1$ and $t_1 = -1 \equiv q_1-1$. The codes in the ambiguous columns must now be replaced by the appropriate code for $\omega_2$ using the class-two wedges.

As an example consider again the symbol sequence the $\frac27 : \frac29$ orbit.
The \Tbl{tbl:2729} lists the symbol sequence where the 63 symbols are arranged in 9 rows
of 7 entries each.  The ambiguous islands correspond to the columns
$t_1 = 1$ and $t_1 = 6 = -1 \mod 7$.  The $q_2(q_1 - 2)$ symbols in
the remaining columns are determined by $q_2$-fold repetition of the
elliptic class-one code for $\omega_{1}=2/7$, $\per{-++--++}$.

\begin{table}[htb]
\[
    \begin{array}{rlllllll|l}
        t_1  & 0&1&2&3&4&5&6  & t_2  \\ \hline
        & -&+&+&-&-&+&+ &0 \\
        & -&\fbox{$-$}&+&-&-&+&+ &1\\
        & -&-&+&-&-&+&+ &2 \\
        & -&+&+&-&-&+&- &3 \\
        & -&+&+&-&-&+&+ &4 \\
        & -&+&+&-&-&+&+ &5 \\
        & -&-&+&-&-&+&+ &6 \\
        & -&\fbox{$+$}&+&-&-&+&- &7 \\
        & -&+&+&-&-&+&- &8
    \end{array} \quad
    \qquad
    \begin{array}{c|c@{\;}c}
        \frac{2}{9} & \multicolumn{2}{c}{\frac27 : \frac29} \\ \hline
        s_{t} & s_{1:t_2}  & s_{-1:t_2}  \\ \hline
        -& +  & + \\
        +&  - & +\\
        +&  - & + \\
        +&  + & - \\
        -&  + & + \\
        -&  + & + \\
        +&  - & + \\
        +&  + & - \\
        +&  + & -
    \end{array}
\]
\caption{
    Elliptic class-two code for the $\frac27 : \frac29$ orbit.
    The ambiguous columns are $t_1 = 1$ and $t_1=-1$. The hyperbolic code
    is the same with the exception of the boxed symbols, at $t=8=1:1$ and  $t=50=1:7$,
    which are flipped. The right table compares the code for the elliptic $\frac29$
    class-one orbit with the two ambiguous columns in the $ \frac27:\frac29$ orbit .
    \label{tbl:2729}}
\end{table}

The symbols in the ambiguous columns $t_1 = \pm 1$ are generated with
respect to a wedge formed by the boundary in the class-two island. The codes for the
hyperbolic orbit are obtained by a shift of $\alpha_2 =
-\frac{1}{2q_2}$, (i.e. a clockwise rotation) as shown in
\Fig{fig:classII2729}, this corresponds to flipping the two boxed symbols in \Tbl{tbl:2729}.  As we will see below, when there are two
ambiguous islands, exactly two symbols flip relative to the elliptic case.

\begin{figure}
       \centerline{
         \parbox{3.5in}{ \includegraphics[width=3in]{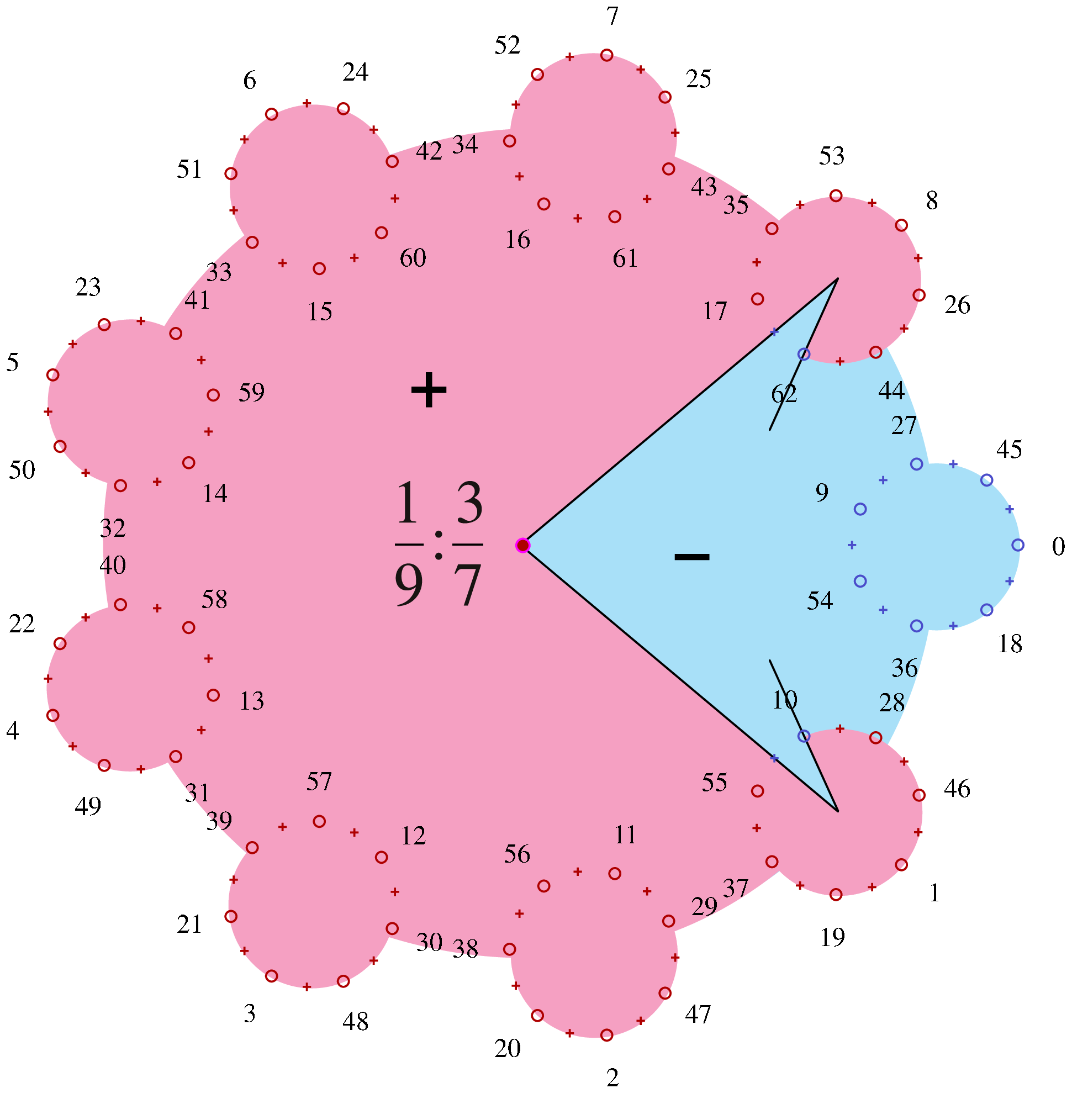}}
         \parbox{3.5in}{  \includegraphics[width=3in]{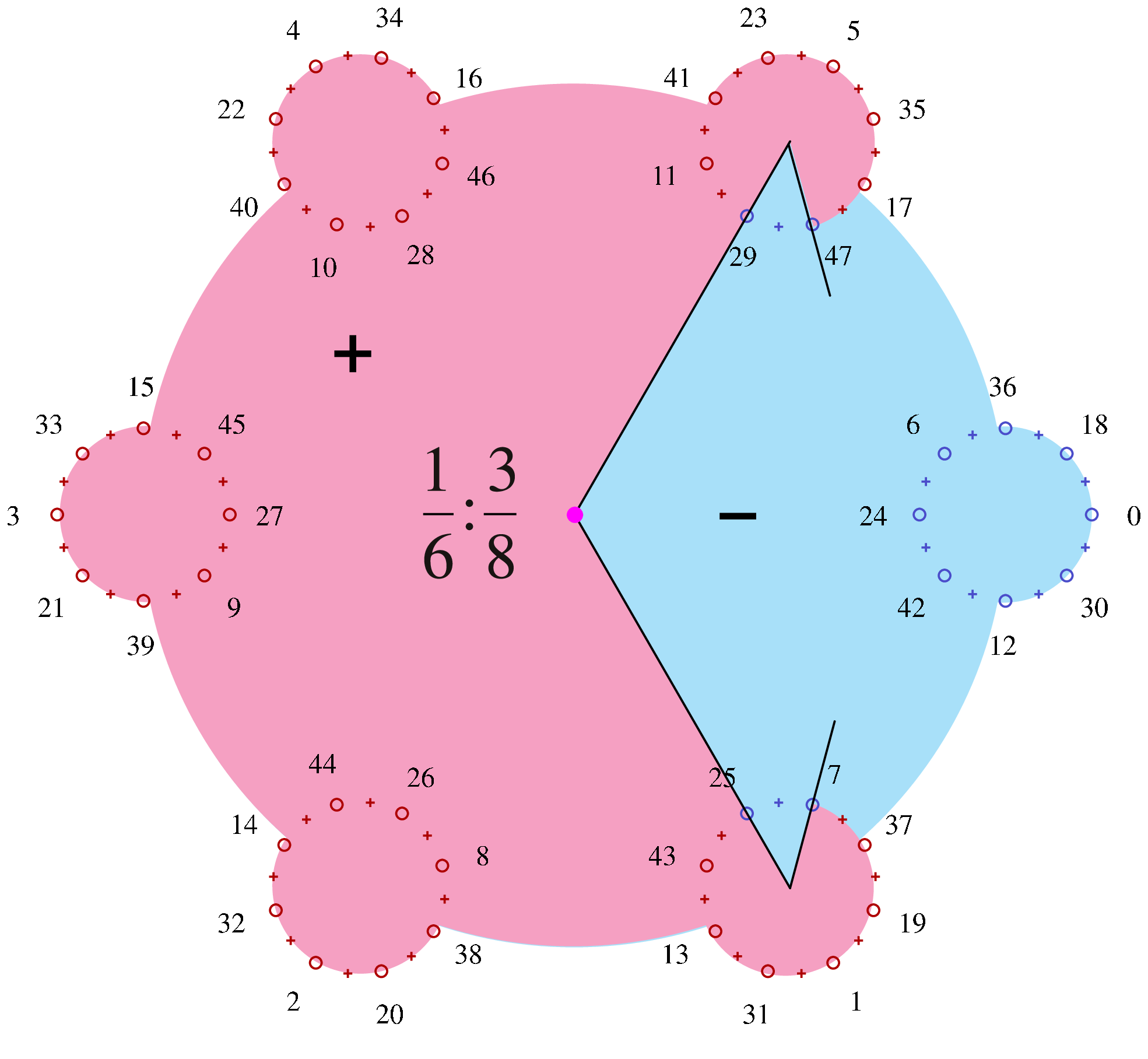}}
     }\centerline{
        \parbox{3.5in}{  \includegraphics[width=3in]{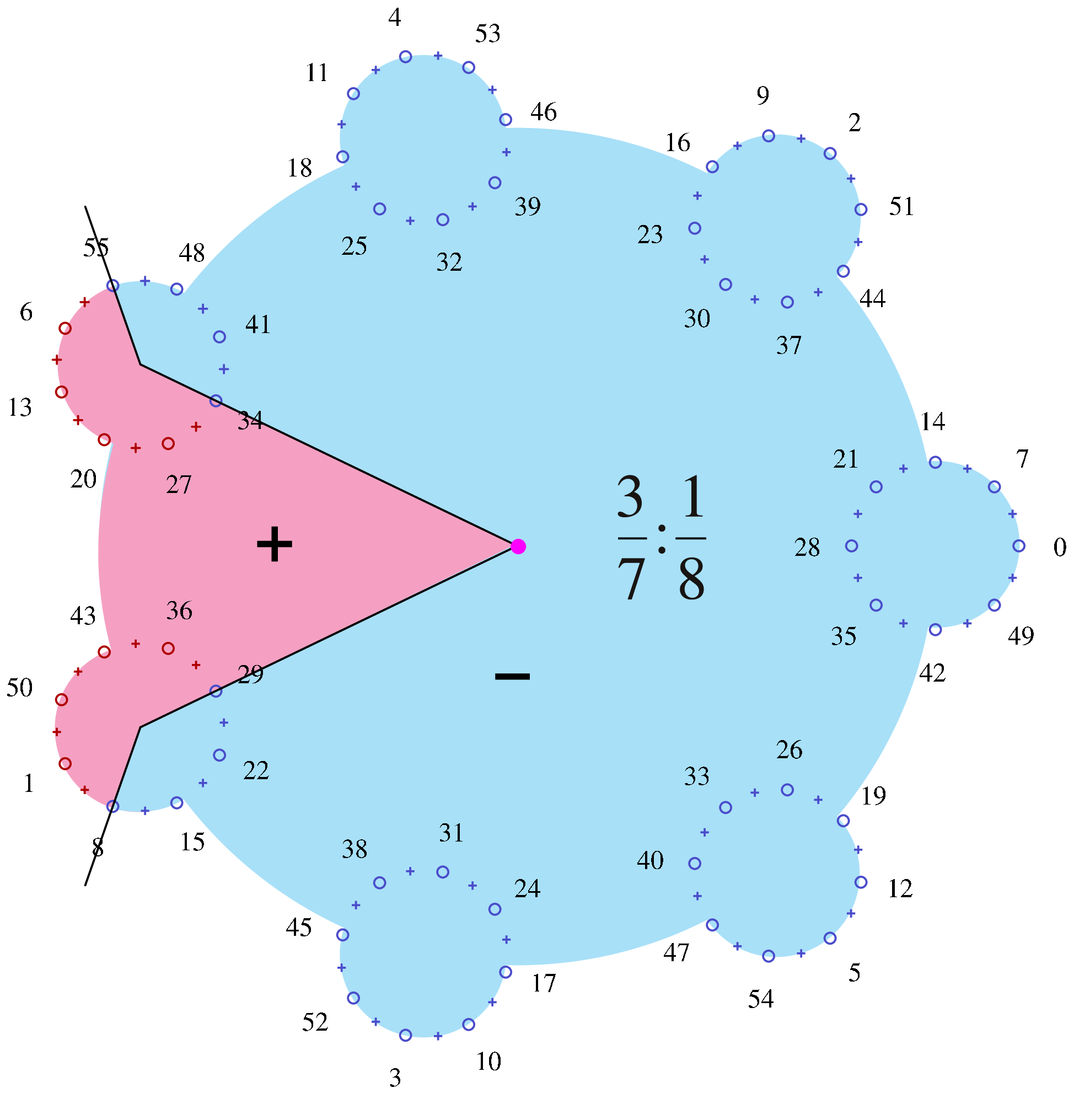}}
        \parbox{3.5in}{  \includegraphics[width=3in]{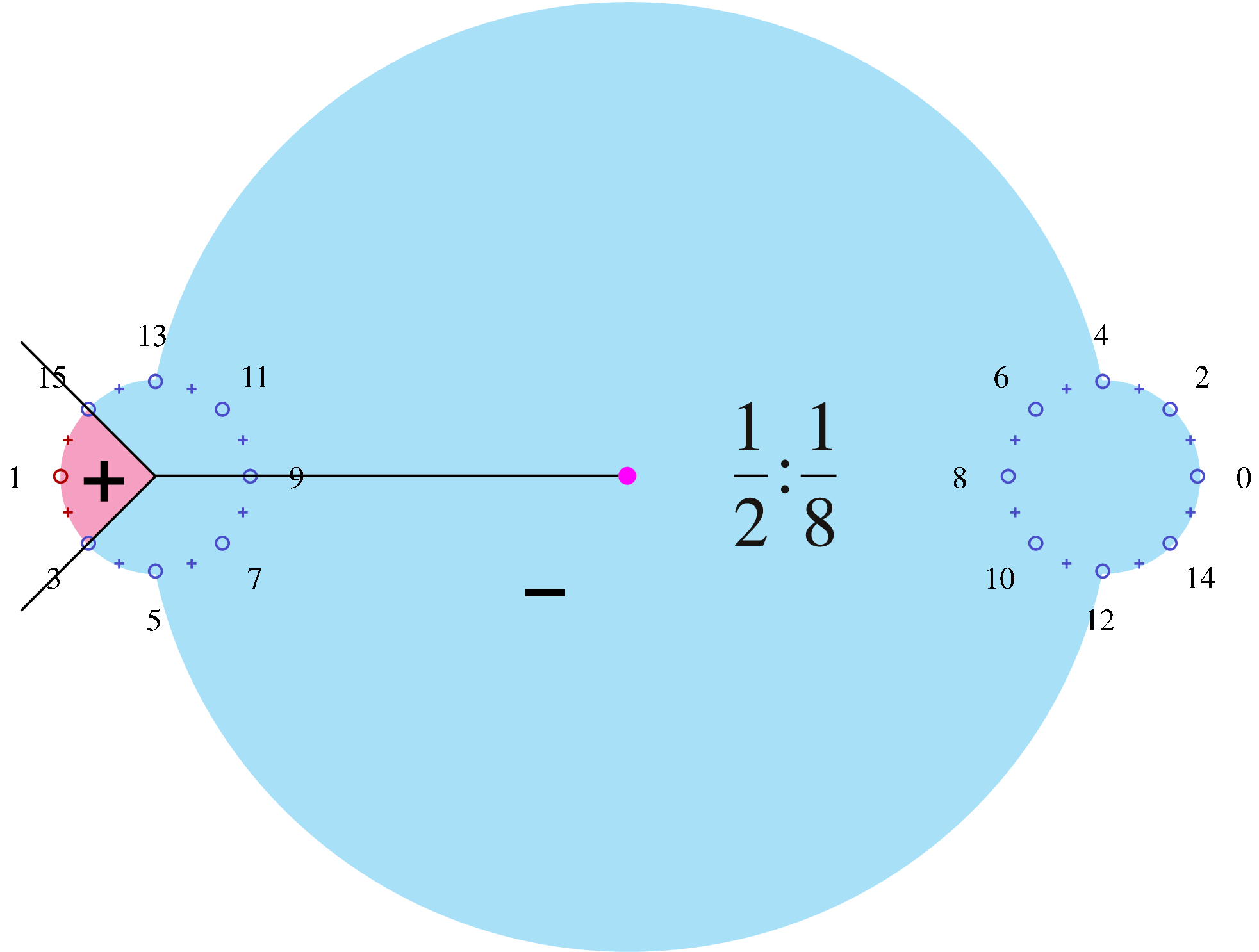}}
       }
       \caption{{\footnotesize  Four examples of schematic class-two orbits. Points are labeled by the time $t$ along the orbit. Symbols in the blue regions are $-$ and in the red regions are $+$.}}
       \label{fig:4wedges}
\end{figure}

Epicyclic pictures for additional class-two orbits
are shown in \Fig{fig:4wedges}.  The codes are easily obtained from the
figures by reading off the symbols according to the
time $t$, using the colors shown.  For example, the code for the
elliptic $\frac19 :\frac37$ orbit is the $7$ fold repetition of the
$\frac19$ elliptic code, $-+^8$, modifying only two symbols in the
columns $t_1 = \pm1$, by setting $s_{10} = s_{62} = -$.  The
hyperbolic code is obtained from the elliptic one by flipping $s_{10}$
to $+$, and $s_{55}$ to $-$.

As $\omega_1$ approaches $\frac12$, the class-one $+$ wedge shrinks to
a single point $\theta_1 = \frac12$.  Orbits with $\omega_1 = \frac12$
are special because the two boundaries of the class-one wedge coincide
so that there is only one ambiguous island with $t_1 = +1 \equiv -1$.
Thus for example, the code for the $\frac12 :
\frac18$ elliptic orbit shown in the last panel of \Fig{fig:4wedges}
is $\per{-+-^{14}}$, while its hyperbolic partner has code
$\per{-+-^{13}+}$.  In this case the ambiguous column of the elliptic
orbit contains $+-^7$ while the unambiguous column simply contains
$-^8$. 

If $\omega_2$ approaches $\frac12$, as in the first two panels of
\Fig{fig:4wedges}, the open class-two wedges grow encompassing all but
$\theta_2 = \frac12$.  The extreme case of this, $\omega_2 = \frac12$
has only one $s=-$ point in the ambiguous islands; for example, the
code for the elliptic $\frac18 : \frac12$ orbit is $\per{-+^7--+^5-}$.

The well-known ``doubling" substitution rule
\begin{equation}\label{eq:doublingSubs}
 - \to -+ \mbox{ and } + \to --
\end{equation}
(see \App{sec:subrule}) can be used to generate the codes for class-two orbits of the form $\frac12:\omega$ from those of the class-one orbit with frequency $\omega$. For example,  applying this substitution rule to the $\frac18$ code, $-+^7$, gives the same code as in \Fig{fig:4wedges}.  That this is true in general follows because the class-two wedge for $s=+$ is $\theta_2 \in (-\omega,\omega)$, which is the same as the class-one wedge for $s=-$.  Hence each symbol flips to create
the ambiguous column; the first, unambiguous column is all $-$.
%

\subsection{Numerical Observations}\label{sec:classTwoNumerical}

In this section we present numerical observations that support the 
class-two code construction for the area-preserving \hen map.  As for the class-one case in \Sec{sec:classOneNumerical}, we checked
that the class-two orbits defined by their codes at the AI limit collide
with the appropriate class-one orbit at the parameter value where the
class-one orbit has the expected residue.  We restricted our attention to class-two orbits with
$q_{1}$ and $q_{2} \le 13$.  \Fig{fig:classtwo-accuracy} shows
the number of digits of precision in the bifurcation value for the class-two rotational orbits with normal twist.\footnote{
There are only eight orbits with fewer than four digits of precision:  $\frac13:\frac{4}{13},
\frac14:\frac{4}{13},\frac15:\frac{4}{13},
\frac25:\frac{4}{13},\frac27:\frac{4}{13},\frac{5}{12}:\frac{3}{13},\frac{5}{13}:\frac{4}{13},
\frac{6}{13}:\frac{6}{13}$. These are probably related to anomalous twist either near 
$\frac13$ or near $\frac14$.
}
      \InsertFig{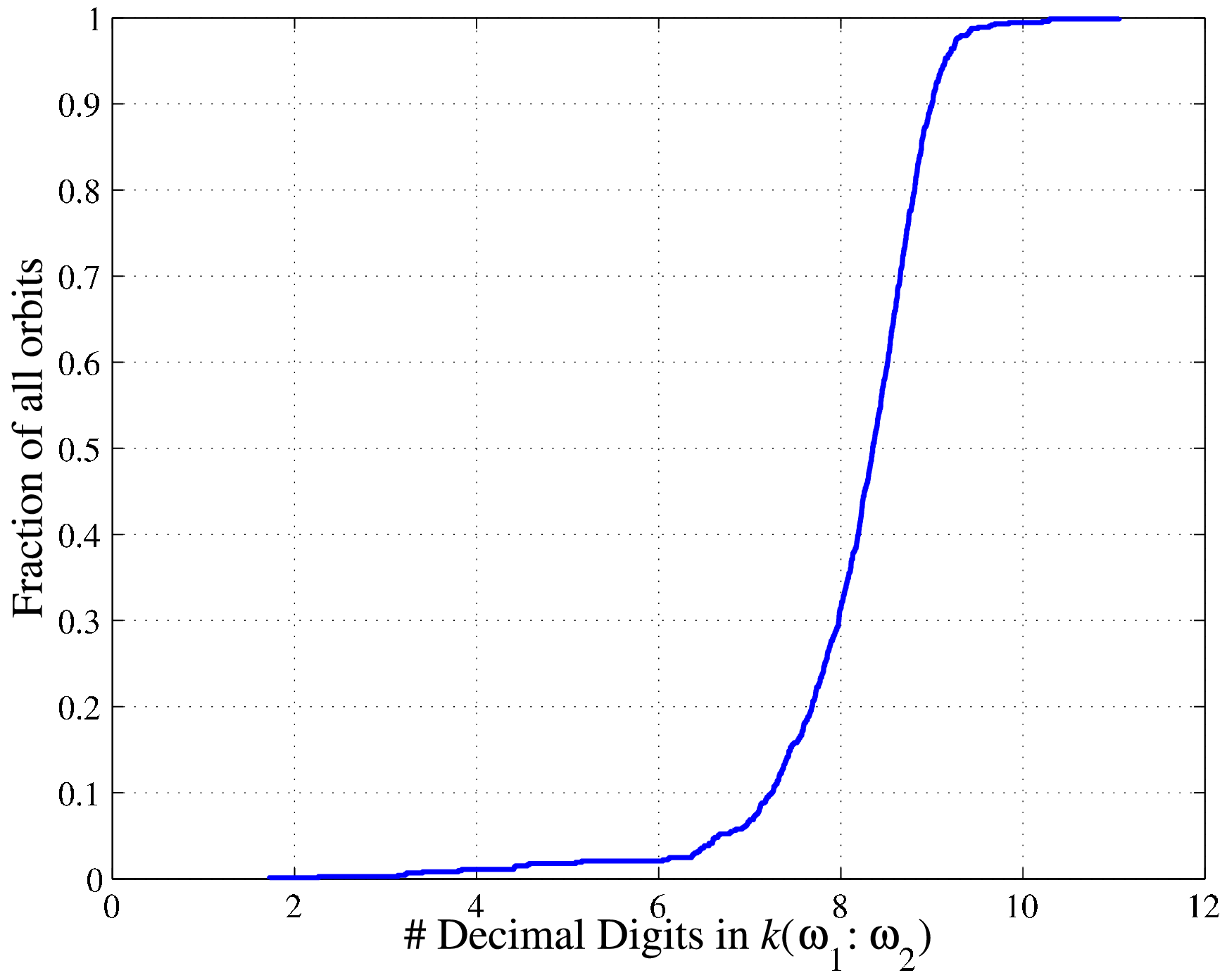} {Number of correct digits in
      parameter value for 727 class-two orbits with $q_{1} \le 13$ and
      $q_{2} \le13$. We excluded frequencies $p_{2}/q_{2} = 1/3, 3/10$ where the class-one twist is 
      anomalous to avoid numerical difficulties with multiple bifurcations.}
      {fig:classtwo-accuracy}{3in}
When $\omega_1=1/2$ there is only a single ambiguous class-two island.
In order to verify the class-two code for this case we compared the 
numerical bifurcation parameters for $236$ class-two
orbits of the form $\frac12:\omega_2$ where $q_2 < 39$ to the exact value \cite{SterlingThesis}
\[
   k(\frac12:\omega_2) = 3 + \sin(\pi \omega_2)^2\,.
\]
Outside of the interval below $\omega_2 = 1/3$ where the twist vanishes at
class-two 
the computed values of the bifurcation parameter were
accurate (to at least 8 significant digits) when compared to the exact value.

The results we have obtained give us confidence to formulate the following 
\begin{con}\label{con:HenCodes}
    The AI code for the class-two orbits of the \hen map is defined by the wedges $W_\pm(\omega_1:\omega_2)$ given in \Sec{sec:classTwoCode}.
\end{con}

\subsection{Parity of Class-Two Codes}
We now show that the elliptic and hyperbolic class-two codes differ by two symbols, but they have the same parity.

\begin{lem}[Properties of class-$2$ codes]\label{lem:class2parity}
Assume $\omega \ne \frac12:\frac12$. Then the elliptic and hyperbolic class-$2$ 
codes differ by flipping two symbols, the symbol at $t = 1:1$ and that at
\begin{align*}
   t &= -1:\frac{q_2}{2}            && \text{$q_2$ even }\\
   t &=  1:\frac{q_2(2j+1)+1}{2p_2} && \text{$q_2$ odd }
\end{align*}
where $j$ is the smallest non-negative integer for which an integer solution for $t_2$ is obtained. The
parity of both orbits is equal to the parity of $q_2$.
\end{lem}
\proof We start by considering the ambiguous island, $t_1 =
+1$.  By definition, the hyperbolic and elliptic codes
differ only when $\omega_2t_2$ and $\omega_2t_2 -\frac{1}{2q_2}$ are
in different wedges.  Let $S_j$ be the sectors defined in
\Eq{eq:S-sectors} relative to $q_2$ for $\theta_2$.  When
$q_2$ is {\em even} the situation is similar to class-one, because the
the wedge boundaries in the ambiguous island coincide with boundaries
of some sectors $S_j$.  Accordingly only the symbol $t =1:1$ changes.  
Alternatively when $q_2$ is {\em odd},
the other wedge boundary at $\frac12$ is not part of the elliptic
orbit since $\omega_2 t_2 \not \equiv \frac12$ for all $t_2$.  Instead
the hyperbolic orbit hits this boundary when $\omega_2 t_2 
-\frac{1}{2q_2} = \frac12 + j$, flipping the symbol at this location.
This gives the above formula.  In order to show that this is the only
other change with $t = 1:t_2$ the argument of
\Lem{thm:2class1codes} can be repeated with sectors $\tilde S_j$ of
half the size.  As a result for all $\alpha_2 \in \tilde S_{-1}$ the
elliptic code is obtained, while for $\alpha_2 \in \tilde S_{-2}$ the
hyperbolic code is obtained.

Now we consider the ambiguous island with $t=-1:t_2$.  The
wedge in this island is a reflection of that in the previous ambiguous island.
However, since the rotation direction is not reversed, the sectors
$S_j$ (or $\tilde S_j$) are not reflected.  This implies that whenever
an open sector end coincides with a closed wedge end for $t =
+1:t_2$, after reflection of the wedge for $t
=-1:t_2$ both endpoints will be open, and vice versa.  Since
a symbol flip only occurs when different types of endpoints coincide
only one of the symbols at $t_2 = \pm 1$ flips, never both.  In
particular when $q_2$ is {\em odd} both symbol flips occur for $t_2 =
+1$, while for {\em even} $q_2$ the symbol half way around the orbit
with $\omega_2 t_2  = \frac12 + j$ flips for $t_2 = -1$.  This gives
the first part of the lemma.

When $q_2$ is even the parity of the unambiguous symbols ($t_1 \not =
\pm 1$) is even, because their number per island is multiplied by
$q_2$.  The parity of the ambiguous symbols is even in the elliptic
case, because the symbol sequences are cyclic permutations of each
other: Starting with $\alpha_2 = 0$ at $t _1=1:1$
gives the same symbol sequence as starting with $\tilde \alpha_2 =
\frac12 - \omega_2$ at $t = -1: \frac{q_2}{2}$
(apply a rotation!).  But $\tilde \alpha_2$ is a point on the orbit
for even $q_2$, hence the elliptic ambiguous columns have the same
numbers of signs.  Therefore their parity is even, and the whole
sequence has even parity when $q_2$ is even.

When $q_2$ is odd the parity of the unambiguous symbols is odd,
because the number per island is odd and they are repeated an odd
number of times.  The number per island is odd because the elliptic
class-one orbits have odd parity, and the two deleted symbols in the
ambiguous column are $+$ signs in the elliptic case.  The number of
signs in the ambiguous columns is the same because the two are time
reversals of each other: upon reflection of the wedge and the
orientation the same sequences are generated.  So their parity is
even, and the total parity is odd.  \qed

Recall from \Lem{thm:class1Prop} that the elliptic/hyperbolic
class-one orbits have opposite parities and this corresponds to their
opposite residue signs.  One important consequence of the previous
lemma is that at class two the elliptic/hyperbolic partners have the
same parity and therefore, according to \Eq{eq:resLimit}, will have
residues with the same sign at the AI limit.  Nevertheless they are
initially born with opposite residues, as this is the normal form of
the $q_2$-tupling bifurcation \cite{MeyerHall92}.  According to
\Eq{eq:resLimit} even parity implies a negative residue at the AI
limit.  Therefore, when $q_2$ is even the residue of the (initially)
elliptic orbit, which is (initially) positive, must cross $r=0$ at
least once as $k$ increases from the bifurcation value in the approach
to the AI limit.  Similarly, when $q_2$ is odd the residue of the
(initially) hyperbolic orbit must cross $r=0$ for some $k$ larger than
the bifurcation value. 

We call bifurcations related to these
additional zeros in the residue function {\em secondary bifurcations}.
In particular when $r=0$ for $k$ larger than the parameter value of
the rotational bifurcation, a secondary pitchfork bifurcation occurs
(super- or subcritical, depending on the parity of the code), and when
$r=1$ a secondary period-doubling bifurcation occurs (again super- or
subcritical, depending on the parity).  Between these two
parameter values, a complete sequence of secondary rotational
bifurcations must take place.  
An example of these bifurcations is shown in \Fig{fig:class2pitch-residue}.

     \InsertFigTwo{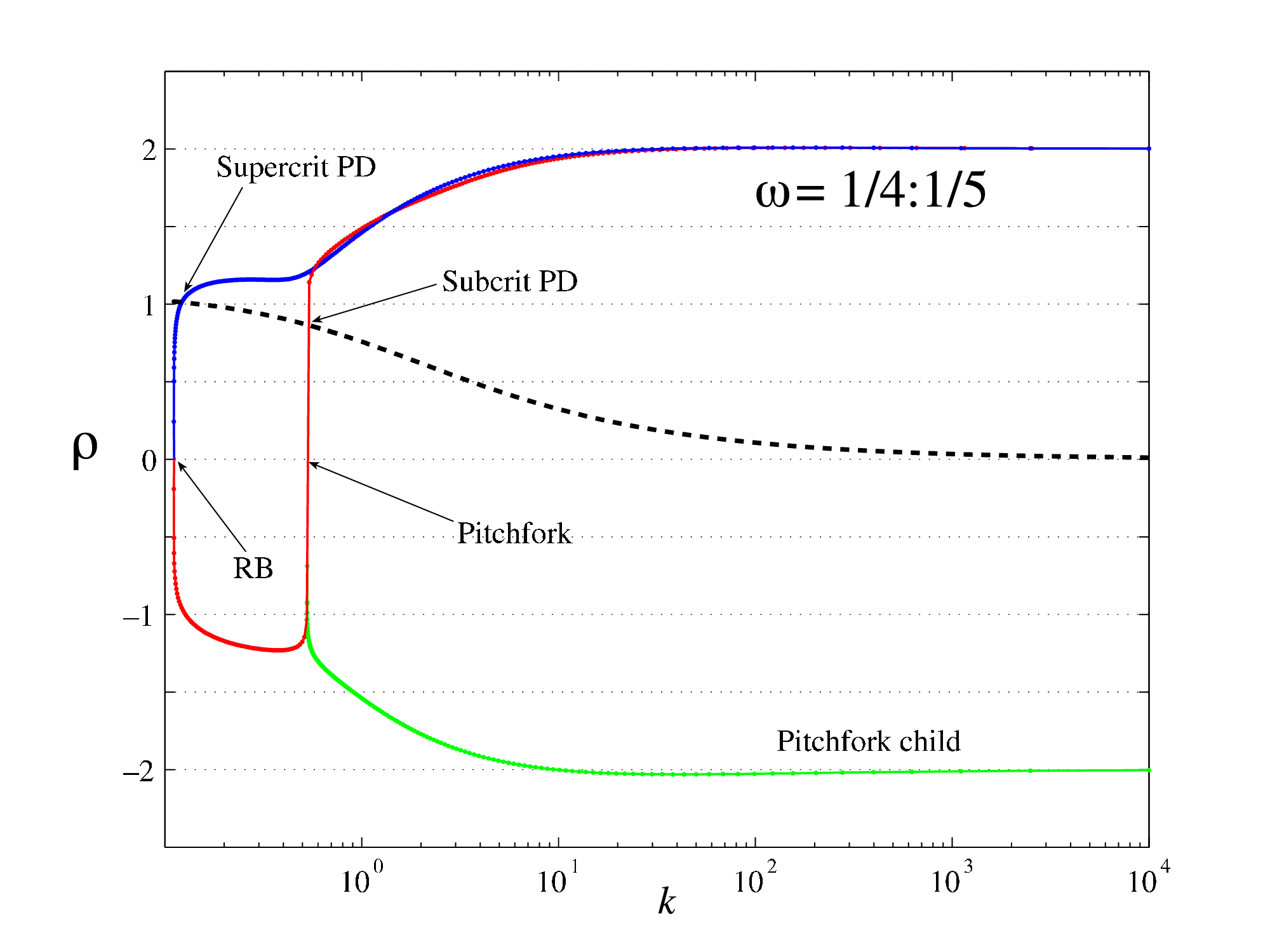}{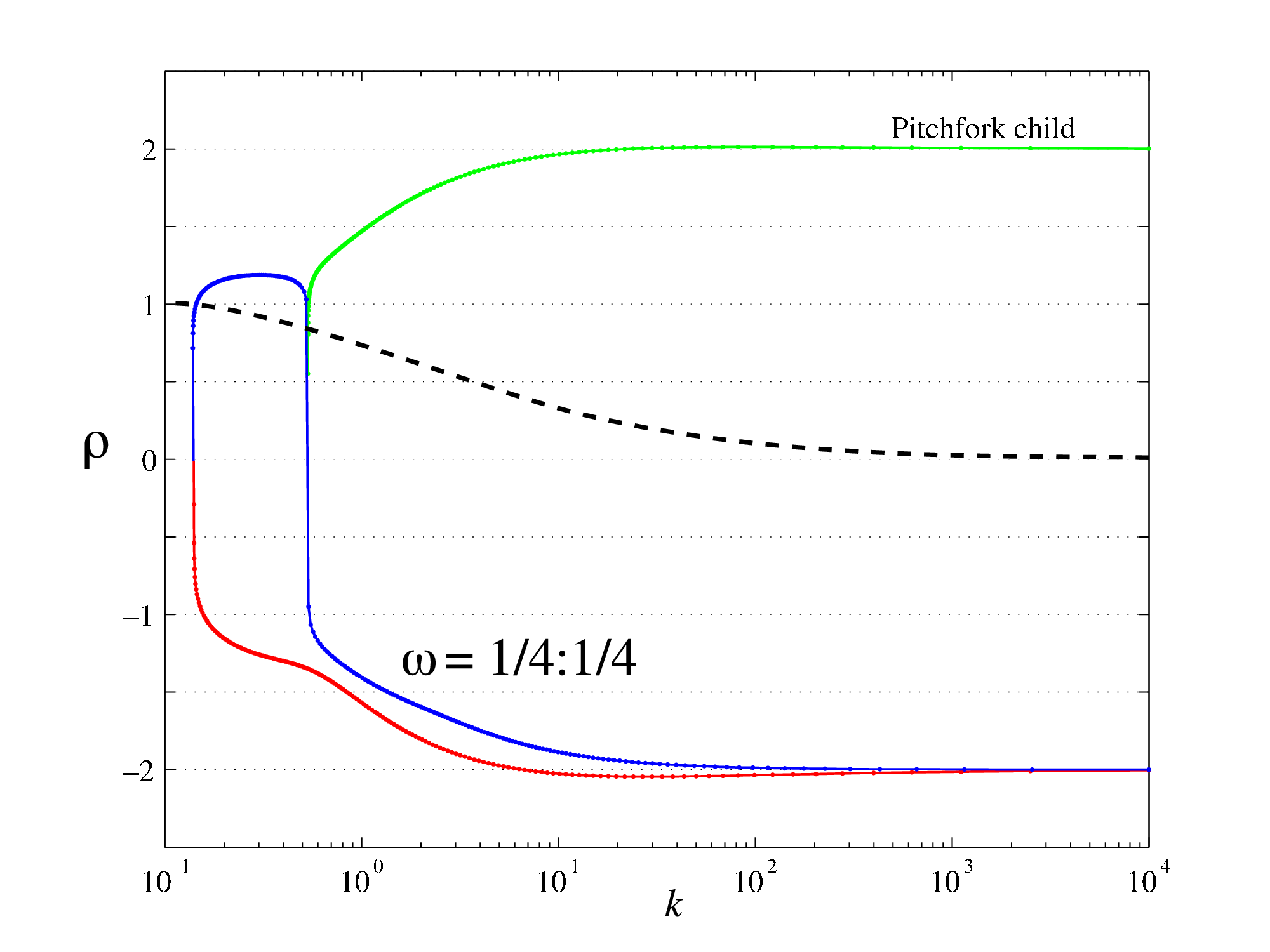}
     {Scaled residue $\rho$, \Eq{eq:scaledresidue}, for typical class-two rotational orbits
     and their pitchfork children with $q_{2}$ odd (left panel) and
     even (right panel).  The (initially) stable class-two orbit is
     rendered in blue, the (initially) unstable orbit in red, and one
     of pitchfork children in green.  The dashed curve is the
     period-doubling curve $r=1$.  The secondary pitchfork bifurcation
     occurs at the rightmost zero crossing on the (initially) stable
     orbit when $q_{2}$ is even and the initially unstable orbits when
     $q_{2}$ is odd.} {fig:class2pitch-residue}{3in}

We will not attempt to completely describe
the symbol sequences associated with these secondary bifurcations, but
only offer an empirical rule based on our numerical observations:

\begin{con}[Pitchfork class-two]\label{con:class2pitch}
    The pitchfork children of class-two orbits only differ in the four
    symbols at the intersection of the ambiguous columns and rows with
    $t_1$ and $t_2$ given in \Lem{lem:class2parity}.  When $q_2$ is
    {\em even} these four symbols ordered by $t$ are $-+--$ for the
    (initially) elliptic orbit and $++-+$ for the (initially)
    hyperbolic orbit.  The secondary bifurcation of the elliptic orbit
    is
    \[
       -+-- \to \mathrm{pf}( -+-+, -++-) \,.
    \]
    When $q_2$ is {\em odd} they are $-++-$ for the elliptic and $++--$ for
    the hyperbolic orbit. The secondary bifurcation of the hyperbolic orbit is
    \[
      ++-- \to \mathrm{pf}(+++-, ++-+) \,.
    \]
\end{con}
It is not hard to show that the four ambiguous symbols in the parent
sequence must be $\mp+--$ as given in \Con{con:class2pitch}; these are
the symbols corresponding to the points on the appropriate class-two
wedge boundaries.

The rule given in \Con{con:class2pitch} generates the symbol sequences
of the daughter orbits created in a secondary pitchfork bifurcation.
These are the orbits whose residues are rendered in green in
\Fig{fig:class2pitch-residue}.  We examined 105 class-two orbits in the
\hen map with frequencies of the form $\frac{1}{j}:\frac{p}{q}$ where
$j=3,4,5$ and $3\le q\le 15$.  The parameter value ($k$) where the
pitchfork children predicted by \Con{con:class2pitch} were born agreed
(to at least eight significant figures\footnote{
With the exception of a six orbits:$\frac13:\frac{4}{13},
\frac14:\frac{4}{11},\frac14:\frac{4}{13},
\frac14:\frac{5}{13},\frac14:\frac{5}{14},\frac15:\frac{4}{13}$ for which
numerical difficulties gave less than six digits of precision.
}) with the value where the residue of the class-two parent orbit crossed
zero in the ``secondary" bifurcation.  These results suggest that the
conjecture is probably valid for the \hen map.

\subsection{Symmetries} \label{sec:classTwoSymmetry}

Numerical observations suggest that the concept of ``dominant"
symmetry line can be extended to class-$2$ orbits (and in fact 
to class-$c$ with $c > 2$, see below); for example, for each
$\omega_1$, there appears to be a particular symmetry ray that
contains all of the elliptic class-two orbits $\omega_1:\omega_2$
\cite{Meiss86}.  Recall from \Tbl{tbl:symmetrylines} that there are
two rays, $E_d$ and $E_s$,  whose identity depends upon the parity of the 
numerator and denominator in $\omega_1$,  that contain elliptic class-one orbits.
Since each class-two orbit rotates around the points on a class-one orbit,
they should have points on the two $E$ rays.  These two rays are divided at the
class-one elliptic points, giving four new rays.  As before, we use $i$ and $o$ to denote
``inward" and ``outward" halves of the rays; the subscript $i$
corresponds to the ray that starts at the elliptic class-$c$ orbit and
heads in toward the elliptic class-$(c-1)$ orbit, while $o$ is the ray
that heads out from the lower class orbit.  Thus, for example, $E_s$
is divided into $E_{so}$ and $E_{si}$.   We denote the
four symmetry rays at class-$c$ by $E^{(c)}_d, E^{(c)}_s,H^{(c)}_d$,
and $H^{(c)}_s$. The observation in
\cite{Meiss86} is that
\begin{equation} \label{eq:JDMrule}
     E_d^{(c)}  = E_{so}^{(c-1)} \,;
\end{equation}
that is, the outward half of the class-$(c-1)$ subdominant ray
becomes the dominant ray for class-$c$, see \Fig{fig:dominantlines}.
This is consistent with \Fig{fig:k243boundary}, as all of the elliptic, class-two
$\frac{2}{5}:\frac{p_2}{q_2}$ orbits have a point on the outward half
of $E^{(1)}_s = \fix{S}_o$.

    \InsertFig{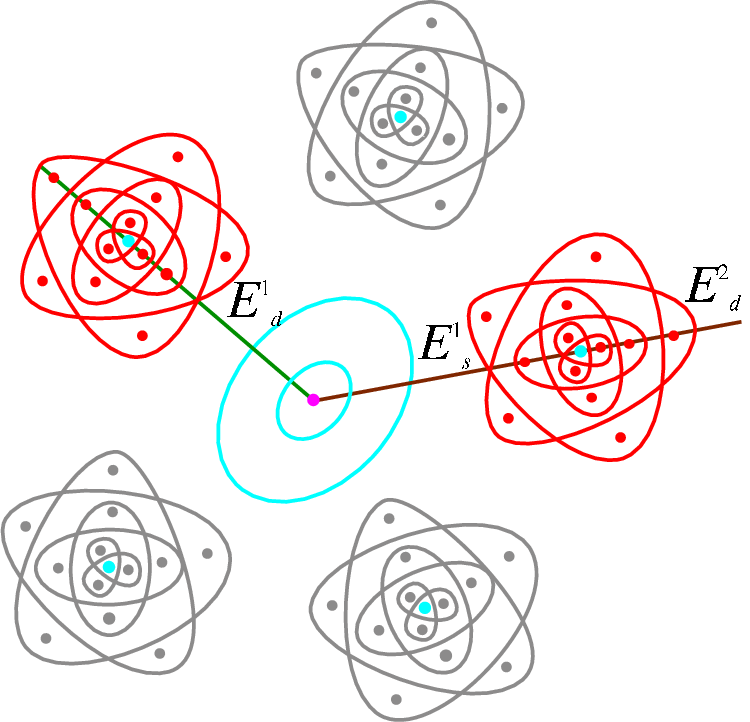} {Sketch of the dominant symmetry line at
    class-two.  Shown are points on a period $5$ elliptic class-one orbit
    (cyan dots), and several class-two orbits (red on the symmetric islands)
    All of the elliptic class-two orbits line up on the outward half of the subdominant
    class-one ray $E_{so}^{(1)}$.} {fig:dominantlines}{3in}

In this section we verify the dominant symmetry conjecture for the
class-two codes.  We also obtain a recipe to describe the
boundary of the class-two wedge in terms of iterates of the
$E^{(2)}_d$, so that the wedge can be computed in the \hen map and
compared to the prediction obtained from numerical continuation.

That the class-two elliptic codes are symmetric is a simple consequence of the
symmetry of the wedges $W_{\pm}$ in the two epicycle picture.

\begin{lem}[Symmetries of class-two codes] \label{thm:symclass2}
Let $\bs$ be the code for the elliptic class-two orbit with rotation
number $\omega = \frac{p_1}{q_1} : \frac{p_2}{q_2}$.
Then $\bs$ has an image on the dominant symmetry
line $E^{(2)}_d = E^{(1)}_s$.  In addition, when $q_2$ is odd, $\bs$
has an image on the subdominant line $E^{(2)}_s = E^{(1)}_d =
\fix{R}_o$, and when $q_2$ is even on $E^{(1)}_s$.  Here
$E^{(1)}_{d,s}$ are the lines in \Tbl{tbl:symmetrylines}.
\end{lem}
\proof Let $\bs$ denote the canonical elliptic symbol sequence, i.e.
starting from the point $\theta = (0,0)$.  Then if $t_+ = 1:1 =
1+q_1$, the point $\sigma^{t_+}\bs$ is on the symbol boundary since it
corresponds to $\theta = (\omega_1, \omega_2)$.  Similarly setting
$t_- = -1:-1 \equiv (q_1-1):(q_2-1) = q_1q_2 -1 \equiv -1$, then
$\sigma^{t_-}\bs$ is also on the symbol boundary, and corresponds to
$\theta = (-\omega_1,-\omega_2)$.  Since the class-two wedge
$W_-$ is closed, these two points have the same first symbols
$s_0 = -$, and by symmetry of the wedges in $\theta \mapsto -\theta$,
the two sequences must be the same when reversed; consequently,
\[
     T \sigma^{t_+}\bs = \sigma^{t_-}\bs \;.
\]
Repeatedly composing this with $\sigma$, and noting that $\sigma T = T
\sigma^{-1}$, we find that when $t_+ + t_-$ is even, $\sigma^a \bs \in
\fix{T}$ where $a = \frac{t_+ + t_-}{2} = \frac{q_1}{2}$.
Alternatively when $t_+ + t_-$ is odd, since $\sigma T = S$ we find
$\sigma^b \bs \in \fix{S}$ where $b = \frac{t_+ + t_- + 1}{2} =
\frac{q_1+1}{2}$.

To decide which half of the symmetry line these points are on, recall
from \Eq{eq:kdefine} that this is determined by $\kappa(\bs)$, the
number of consecutive $-$ symbols starting at $s_0$.  Consider first
$q_1$ even, where the symmetric point is $\sigma^a \bs$.  Since $t = a
= \frac{q_1}{2}:0$, then $\theta = (\frac{\omega_1 q_1}{2} ,0) =
(\frac{p_1}{2},0) = (\frac12,0)$ because $p_1$ is necessarily odd.
Thus the first symbol of this sequence is $s_0 = +$, and so by \Eq{eq:rotSymmetry}
it is on $\fix{T}_i$.  When $q_1$ is odd, the symmetric point is $\sigma^b
\bs$, with $t = \frac{q_1+1}{2}:0$, and so $\theta = (\frac{p_1}{2} +
\frac{\omega_1}{2} , 0)$.  When $p_1$ is even, this is on the line
$\theta_1 = \frac{\omega_1}{2}$, with $s_0 = -$; following the
argument in \Lem{thm:rotSymmetry} this implies the point belongs to
$\fix{S}_o$.  Similarly when $p_1$ is odd, the point is in
$\fix{S}_i$.

Summarizing we obtain the table:
\[
\begin{array}{r|r@{:}lcc|c}
\omega_1         & t_1             &t_2 & \theta_1 & \theta_2 & E^{(2)}_d \\ \hline
\mbox{odd/odd}   & \frac{q_1+1}{2} & 0  & \frac12+ \frac{\omega_1}{2} & 0 & \fix{S}_i \\
\mbox{even/odd}  & \frac{q_1+1}{2} & 0  & \frac{\omega_1}{2}          & 0  & \fix{S}_o \\
\mbox{odd/even}  & \frac{q_1}{2}   & 0  & \frac12                     & 0 & \fix{T}_i
\end{array}
\]
Since this symmetry is independent of $\omega_2$, we are justified in
calling it a ``dominant" symmetry line.

The subdominant symmetry line can by found by noting that since $\bs$
has period $Q = q_1 q_2$, we also have
\[
     T \sigma^{t_+}\bs = \sigma^{t_- + Q}\bs
\]
Iterating this relation we find that if $q_1(q_2+1)$ is even, then
$\sigma^c \bs \in \fix{T}$ when $c = \frac{t_+ +t_- +Q}{2} =
\frac{q_1(q_2+1)}{2}$.  Otherwise, $\sigma^d \bs \in \fix{S}$ where $d
= \frac{q_1(q_2+1)+1}{2}$.

The choice of ray for the subdominant symmetry depends upon the parity
of both $\omega_1$ and $\omega_2$.  When $q_2$ is odd, then the second
line is at $\sigma^c \bs$.  Indeed $t=c = 0:\frac{q_2+1}{2}$, and so
$\theta_1 = 0$ independently of $\omega_1$, and $\theta_2 =
\frac{p_2}{2}+\frac{\omega_2}{2}$.  For any $p_2$ we have
$\sigma^c \bs = \pt -+\ldots$, so that this point is on $\fix{T}_o$.

We now consider $q_2$ even.  If $q_1$ is also even then the symmetry
still occurs at $t = c = \frac{q_1}{2} : \frac{q_2}{2}$, which gives $\theta
= (\frac{p_1}{2}, \frac{p_2}{2}) = (\frac12, \frac12)$ since both
$p$'s are odd.  If $\omega_1 < \frac12$, then $s_o = +$, and the symmetry is $\fix{T}_i$.  A similar argument gives the same result for the frequency $\frac12$ case. The
final case has $q_1$ odd and $q_2$ even, for which the symmetry
corresponds to $t = d = \frac{q_1+1}{2} : \frac{q_2}{2}$, so that
$\theta = (\frac{p_1}{2}+\frac{\omega_1}{2}, \frac{p_2}{2})$.  When
$p_1$ is even, $\theta_1 = \frac{\omega_1}{2}$, so $s_0 = -$.  This
point is on $\fix{S}_o$, as follows from an argument similar to that
in \Lem{thm:rotSymmetry}.  Finally if $p_1$ is even $\kappa(\bs)$ is even, so the
symmetry is $\fix{S}_i$. 

Summarizing, we obtain a table for the subdominant line:
\[
\begin{array}{rr|r@{:}lcc|c}
\omega_1        & \omega_2       & t_1            &  t_2             & \theta_1                   & \theta_2  & E_s^{(2)} \\ \hline
\mbox{any}      &\mbox{even/odd} & 0              & \frac{q_2+1}{2}  &  0                 & \frac{\omega_2}{2}& \fix{T}_o \\
\mbox{any}      &\mbox{odd/odd}  & 0              & \frac{q_2+1}{2}  &  0        &\frac12 + \frac{\omega_2}{2}& \fix{T}_o \\
\mbox{odd/odd}  &\mbox{odd/even} & \frac{q_1+1}{2}&  \frac{q_2}{2}   & \frac12+\frac{\omega_1}{2} & \frac12   & \fix{S}_i\\
\mbox{even/odd} &\mbox{odd/even} & \frac{q_1+1}{2}&  \frac{q_2}{2}   & \frac{\omega_1}{2}         & \frac12   & \fix{S}_o \\
\mbox{odd/even} &\mbox{odd/even} & \frac{q_1}{2}  &  \frac{q_2}{2}   & \frac12                    & \frac12   &  \fix{T}_i
\end{array}
\]
\qed

So far we have shown that the class-two codes are consistent with
\Eq{eq:JDMrule}, but we have not yet shown how to divide $E^{(1)}$
into halves.

\begin{teo}[Class 2 Dominant Symmetry Ray]
The dominant symmetry ray for class-two codes is the outward half of
the subdominant ray at class-one, $E^{(2)}_d = E^{(1)}_{so}$, thus
verifying \Eq{eq:JDMrule} for class-two.
\end{teo}
\proof The division of the symmetry lines for the shift $\sigma$ into
rays at the class-one orbit uses the ordering relation discussed in
\App{sec:horseshoe}.  As shown in \Fig{fig:horseshoeSymmetry}, the
outer halves of $\fix{S}_o$ and $\fix{T}_o$ correspond to codes that
are ``less" than (under this special ordering) those of the class-one
orbit, while the outer halves of $\fix{S}_i$ and $\fix{T}_i$
correspond to codes that are ``greater" than those of the class-one
orbit.  The canonical elliptic
class-two code is the same as that of the class-one orbit up to time
$t = 1:1 = 1+q_1$, where it first hits  the class-two wedge boundary.  
Moreover, according to \Lem{thm:symclass2} and
\Lem{thm:subdominant}, if $\bs$ is the canonical elliptic code, then
its $\floor{\frac{q_1+1}{2}}$th iterate is on $E^{(1)}_s$

Thus the relative ordering of the class-one and class-two codes, following
\Lem{thm:ordering}, depends upon the parity of the symbols of the
class-one orbit starting halfway around the orbit, and going up to
the first differing symbol $s_{q_1+1} = s_1$, i.e. on the sign of
\[
   \rho  =\pi(s_{\floor{\frac{q_1+1}{2}}} \ldots s_{q_1-1} s_0 s_1) \;,
\]
In particular, we must show that $\rho = (-1)^{p_1}$.  This follows
because if $p_1$ is even, $E^{(1)}_s = \fix{S}_o$, and so the code for
the class-one orbit should be greater than that of the class-two
orbit, while if $p_1$ is odd, then the symmetry lines are either
$\fix{S}_i$ and $\fix{R}_i$, so the class-one code should be less than
that for class-two.

As noted in \Lem{thm:class1Prop}, the class-one elliptic code has
$2p_1-1$ symbols that are $-$.  Moreover, the canonical elliptic code
has symmetry $T$, which means it is unchanged under reflection about
$s_0=-$.  This implies that the first half of the orbit $s_1s_2 \ldots
s_{\floor{\frac{q_1-1}{2}}}$ is the same as the second half
$s_{\floor{\frac{q_1+1}{2}}} \ldots s_{q_1-1}$ written backward.  Consequently
they each have $p_1-1$ symbols that are $-$.  Finally, the sequence
$s_{\floor{\frac{q_1+1}{2}}} \ldots s_{q_1-1} s_0 s_1$ has one
additional minus sign, making $p_1$ in total; therefore, $\rho =
(-1)^{p_1}$.  \qed

\subsection{Symbol Boundary}\label{sec:classTwoBoundary}

In order to connect the above findings to the \hen map we now
describe the wedge boundaries in terms of iterates of symmetry lines.
Inside the class-one island the wedge boundary is simply given by the
forward and backward iterate of the dominant symmetry ray $\fix{R}_o$,
as discussed in \Sec{sec:classOneSymbol}.  From the schematic picture it
would seem as if the higher class wedges are also constructed from
iterates of $\fix{R}_o$.  This would, however, contradict the
observation that the dominant symmetry line is different at the next
class \cite{Meiss86}.  Moreover, we expect that a symbol boundary
should be associated with a line containing all the elliptic points.
Thus, as we saw in \Lem{thm:symclass2} it should be associated with
the subdominant elliptic line.

Therefore we now show that appropriate iterates of $\fix{R}_o$ in the
schematic picture can in fact be identified with certain iterates of
the subdominant lines as given by \Eq{eq:JDMrule}.

\begin{lem}[Boundary Symmetry Lines]
The boundaries of the class-two wedge in the ambiguous islands $t_1 =
\pm 1$ for the class-two orbit $\omega = p_1/q_1: p_2/q_2$ are given by
\begin{itemize}
\item the $\lfloor \frac{q_1-1}{2} \rfloor +t_1$ iterate of the subdominant symmetry ray $E_{s}^{(1)}$ and
 \item the $-\lceil \frac{q_1+1}{2}\rceil$ iterate of symmetry ray $E_{do}^{(1)}$. 
\end{itemize}
\end{lem}
\proof Recall the proof of  \Lem{thm:subdominant}.
The schematic boundary of the class-two wedge has two parts.  The inner part
consists of the class-one rays $F^{t_1}(\fix{R}_o)$  with $t_1 = \pm 1$.  
The first part of the lemma is the statement that this ray is an image of $E_s^{(1)} = \fix{S}_o$. 
For example,  when $\omega_1 = e/o$, then
we require that there exists an integer $n$ such that
\[
    \omega_1 = \frac{even}{odd}:  \quad F^{t_1}(\fix{R}_o) = F^n(\fix{S}_o) \quad \Rightarrow \quad
           \omega_1(t_1+1)   + j = \frac{\omega_1}{2} +\omega_1n 
\]
for some integer $j$.  Solving for $\omega_1$ gives
$2j/(2n-2t_1 - 1)$, which is of the form $even/odd$.  When $t_1=1$ the explicit solution
is $n=(q_1+3)/2$, and when $t_1 = -1$ it is $n =
(q_1-1)/2$.  Note that replacing $\fix{S}_o$ by $\fix{S}_i$ or
$\fix{R}_i$ leads to a contradiction.  Hence $\fix{S}_o$ is the unique
symmetry ray that can be mapped to $\fix{R}_o$ for $\omega_1 = even/odd$.
In a similar way the other cases lead to
\[
\omega_1 = \frac{odd}{odd}: \quad F^{t_1}(\fix{R}_o) = F^n(\fix{S}_i) \quad \Rightarrow \quad
   \omega_1(t_1+1) + j = \frac{1+\omega_1}{2} +\omega_1n 
\]
which gives the same values of $n$ as before. Finally, 
\[
\omega_1 = \frac{odd}{even}: \quad F^{t_1}(\fix{R}_o) = F^n(\fix{R}_i)\quad \Rightarrow \quad
  \omega_1(t_1+1) + j = \frac{1}{2} + \omega_1(n+1)
\]
which gives $n = \frac{q_1}{2}+t_1$. Combining theses expressions gives the 
first form.

The second ray of the class-two wedge is formed by taking the $o$ part of
the ray just constructed (i.e.\ the part that goes beyond radius
$r_1$), and turning it by $\pm \omega_2$ for $t = \pm 1$, respectively. Turning by $\pm \omega_2$ is
achieved by iterating one complete cycle of class-one, i.e.\ iterating
$\pm q_1$ times.  For $t_1=-1$ the number of iterates is thus $(q_1-1)/2 -
q_1$ for odd $q_1$ and $q_1/2 -1 - q_1$ for even $q_1$.  Combining
both cases gives $-\lceil(q_1+1)/2\rceil$, and the result follows.
\qed

As a result of the previous Lemma the number of iterates needed to
construct the class-two wedge only depends on $q_1$, the period of the class-one orbit.

It seems pointless to construct the class-two wedge from iterates of
the new dominant symmetry line instead of from iterates of
$\fix{R}_o$, since the two agree in the schematic picture.  
It turns out, however, that the former gives the correct
wedges for the \hen map.  In the schematic pictures the two
constructions coincide because the map has no twist.  At the birth of
the class-two island (when the rotation number of the class-one orbit
is $\omega_2$, and the rotation number of the class-two orbit is 0)
the schematic picture is an accurate description of the
wedge-geometry.
Away from this bifurcation the symbol boundary differs from that in the schematic picture,
and the wedge needs to be constructed from the dominant lines of class 2.

      \InsertFig{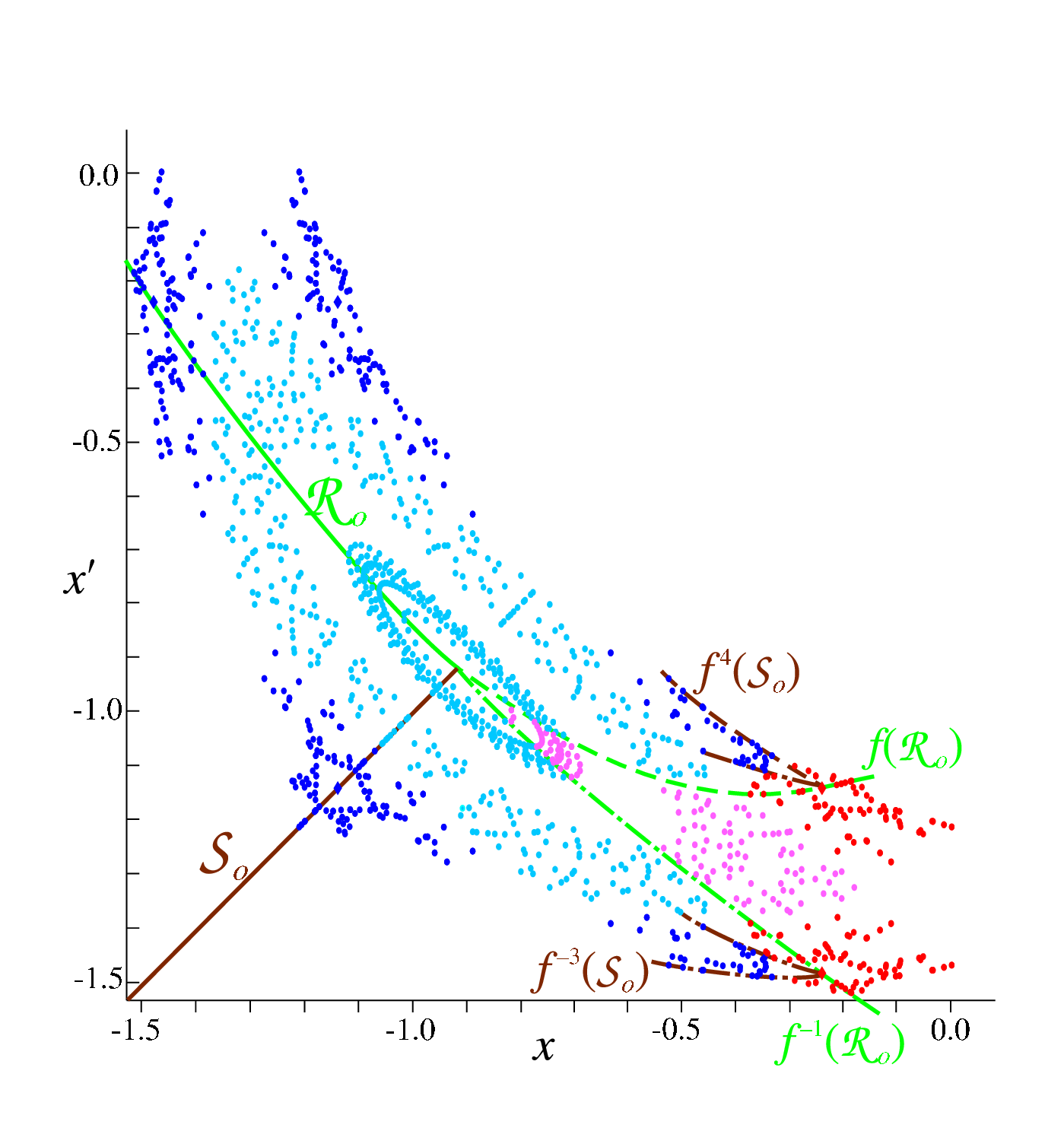} {Class-one (cyan and magenta) and
      class-two (blue and red) rotational orbits for $k=2.684$.  The
      $-$ symbols correspond to the cyan and blue points, and the $+$
      to magenta and red.  Also shown are the $t = \pm 1$ iterates of the symmetry
      line $\fix{R}_o$ (green) that form the class-one boundary, and the iterates
      $t = -1,4$ and $t = -3,2$ of $\fix{S}_o$ (brown) that form the class-two boundary.}
      {fig:k2684boundary}{4in}
In \Fig{fig:k2684boundary}, we show the symbol boundary for $k=2.684$,
where the $\frac25$ island is predominant (the same phase space is
shown in \Fig{fig:enlargements}).  Here the class-one orbits are shown
with colors cyan and magenta for $s = \pm$, respectively.  The
boundary between these colors is given by the image and preimage of
$\fix{R}_o$ as before.  Class-two orbits with $\omega = \frac25
:\frac{p_2}{q_2}$ are shown with colors blue and red for $s=\pm$.
Note that the class-one partition boundary does not work for
these orbits. According to \Lem{thm:subdominant} the boundaries in the $t_1 = -1$ ambiguous island are given by segments of $f^2(\fix{S}_o)$ and $f^{-3}(\fix{S}_o)$; these are shown as the dashed (brown) curves, and clearly delineate the symbol boundary in \Fig{fig:k2684boundary} for the class-two orbits. Similarly segments of $f^{4}(\fix{S}_o)$ and $f^{-1}(\fix{S}_o)$ form the class-two boundary in the $t_1=1$ ambiguous island.

There is apparently a jump in the symbol boundary across the separatrix of the $\frac25$ orbit. We are unable to resolve the behavior near the separatrix numerically as the period of the orbits near the separatrix becomes too large to accurately find the bifurcation values numerically.

The ability to describe the wedge boundary that is obtained by
continuation from the AI limit in terms of iterates of the symmetry
lead to  \Con{con:HenCodes}.


\section{Class-$c$ Codes}\label{sec:classCCode}
Orbits of arbitrarily high classes can also be found in area-preserving maps.
For example, \Fig{fig:enlargements} shows successive enlargements of the phase space of the \hen map and exhibits orbits up to class four.  The enlargements in this figure are not done with the same multiplier; indeed, though there are special parameter values for which 
the class hierarchy exhibits self-similarity \cite{Meiss86}, in general it does not. While there are typically islands of every class, their shape and structure changes with class.

\InsertFig{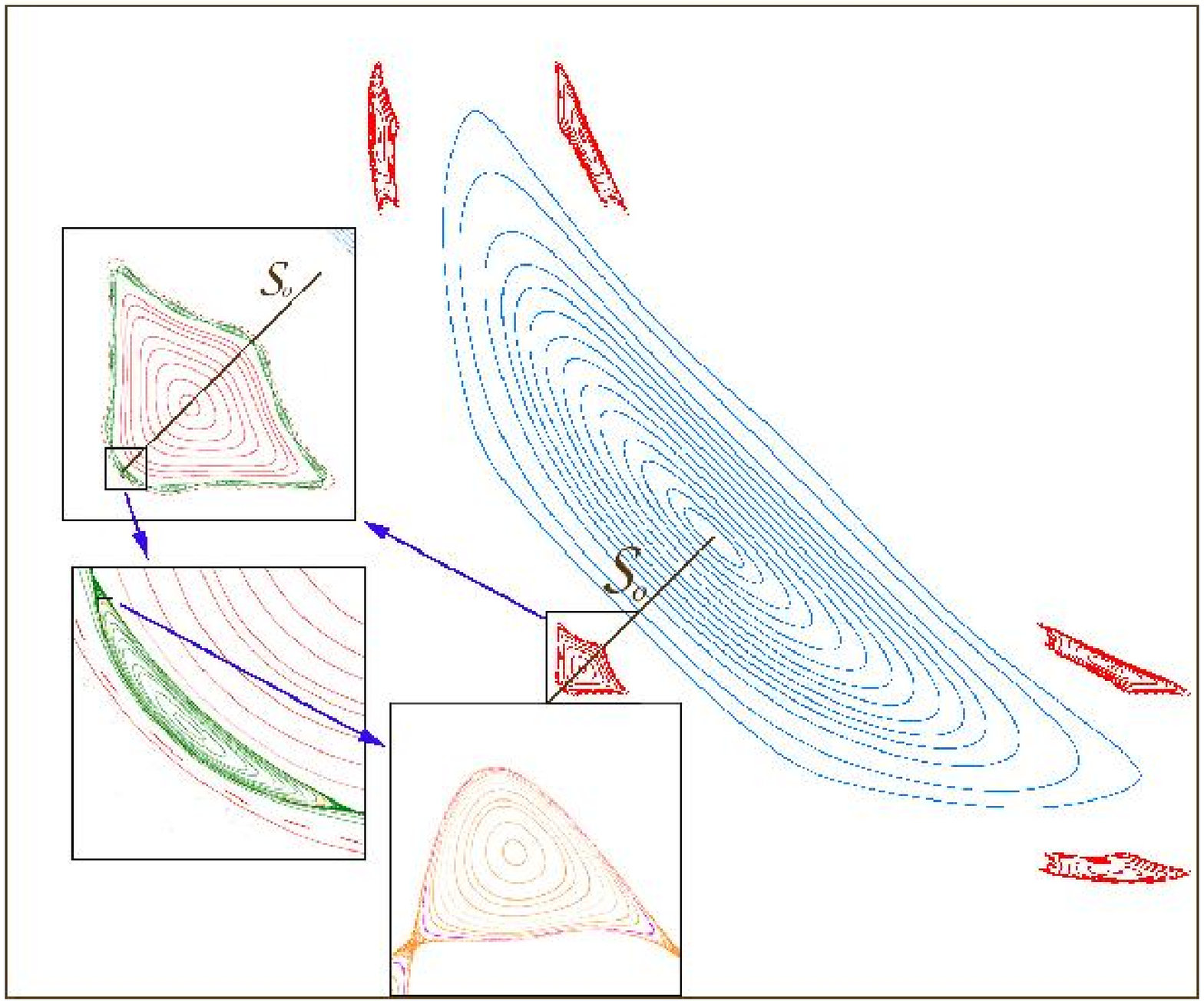}{Phase space of the \hen map for $k = 2.684$.
Enlargements show higher class orbits, ultimately focusing on the
$\tfrac25 : \tfrac{3}{13} : \tfrac{1}{18}: \tfrac{1}{17}$ orbit.  The
class-one invariant circles (cyan) encircle the elliptic fixed point,
class-two (red) the $\tfrac25$ elliptic orbit (red).  Shown in the
enlargements are class-three circles (green), class-four (orange), and
finally a class-five chain (mauve).  The elliptic class-two $\tfrac25
: \tfrac{3}{13}$ orbit has a point on the subdominant elliptic line
$\fix{S}_o$ for the elliptic $\tfrac25$ orbit, as implied by dominant
symmetry line conjecture.} {fig:enlargements}{3.5in}

A class-$c$ orbit rotates around a class-$(c-1)$ elliptic periodic orbit 
with some definite rotation number. We denote the rotation number by a string
\begin{equation}\label{eq:classCRotNo}
   \omega = \omega_1:\omega_2:\ldots:\omega_c \;.
\end{equation}
Each of the numbers $\omega_i= \tfrac{p_i}{q_i}$ is necessarily
rational, except possibly $\omega_c$.  If the class-$c$ orbit is
periodic, it has period $Q = \prod_{i=1}^{c} q_i$.  The rotation
number $\omega_{c+1}$ about a class-$c$ periodic orbit is obtained by
iterating $f^Q$ and considering rotations about one point on the
class-$c$ orbit---if an orbit returns to itself after $q_{c+1}$
iterations of $f^Q$ and has undergone $p_{c+1}$ full rotations, then
$\omega_{c+1} = \frac{p_{c+1}}{q_{c+1}}$.  


Generalizing the epicycle picture in \Fig{fig:classII2729}, we can construct the symbol sequences for periodic orbits of arbitrary class. 
As before, time is written in the mixed $q_i$ basis:
\begin{equation}\label{eq:mixedBasis}
       t = t_1 + q_1\left( t_2 + q_2 \left( t_3 + \ldots + q_{c-1} t_c \right)\right)
         \equiv t_1:t_2:\ldots:t_c \,;
\end{equation}
where the digits $t_i$ are taken modulo $q_i$.  Associated with each $t$ there is a point
$\theta(t) \in \T^{c}$ defined by the mapping
\begin{equation}\label{eq:Torus}
    t \mapsto \theta(t) = (\omega_1 t_1, \omega_2 t_2, \ldots, \omega_c t_c+\alpha_c) \;.
\end{equation}
Here we have set $\alpha_j = 0$ for $j = 1, 2, \ldots c-1$; this
represents the selection of elliptic orbits up through
class-$(c-1)$. The phase $\alpha_c$ selects an elliptic or
hyperbolic class-$c$ orbit.  A class-$j$ island consists of the set of
points $\theta(t)$ with a fixed choice of $t_1:t_2:\ldots :t_{j-1}$.
Incrementing $t_j$ moves the orbit around the island with frequency
$\omega_j$, while incrementing $\theta_j$ by $\frac{1}{q_j}$ moves sequentially through the points in
the island.

The epicycle picture is obtained by generalizing \Eq{eq:expClassTwo}. 
Define a set of radii, $r_i$, $i = 1,\ldots c$, that decrease rapidly enough so that 
the set of circles defined by
\begin{equation}\label{eq:expClassC}
  x + iy = e^{-2\pi i \theta_1} \left( r_1 + e^{2\pi i \theta_2}\left(r_2 + 
       \ldots r_ce^{(-1)^{c} 2\pi i \theta_c}\right) \right) \;,
\end{equation}
do not overlap. As before, the direction of rotation reverses with each class. An example epicycle picture for class three is shown in \Fig{fig:classIII}.

\InsertFig{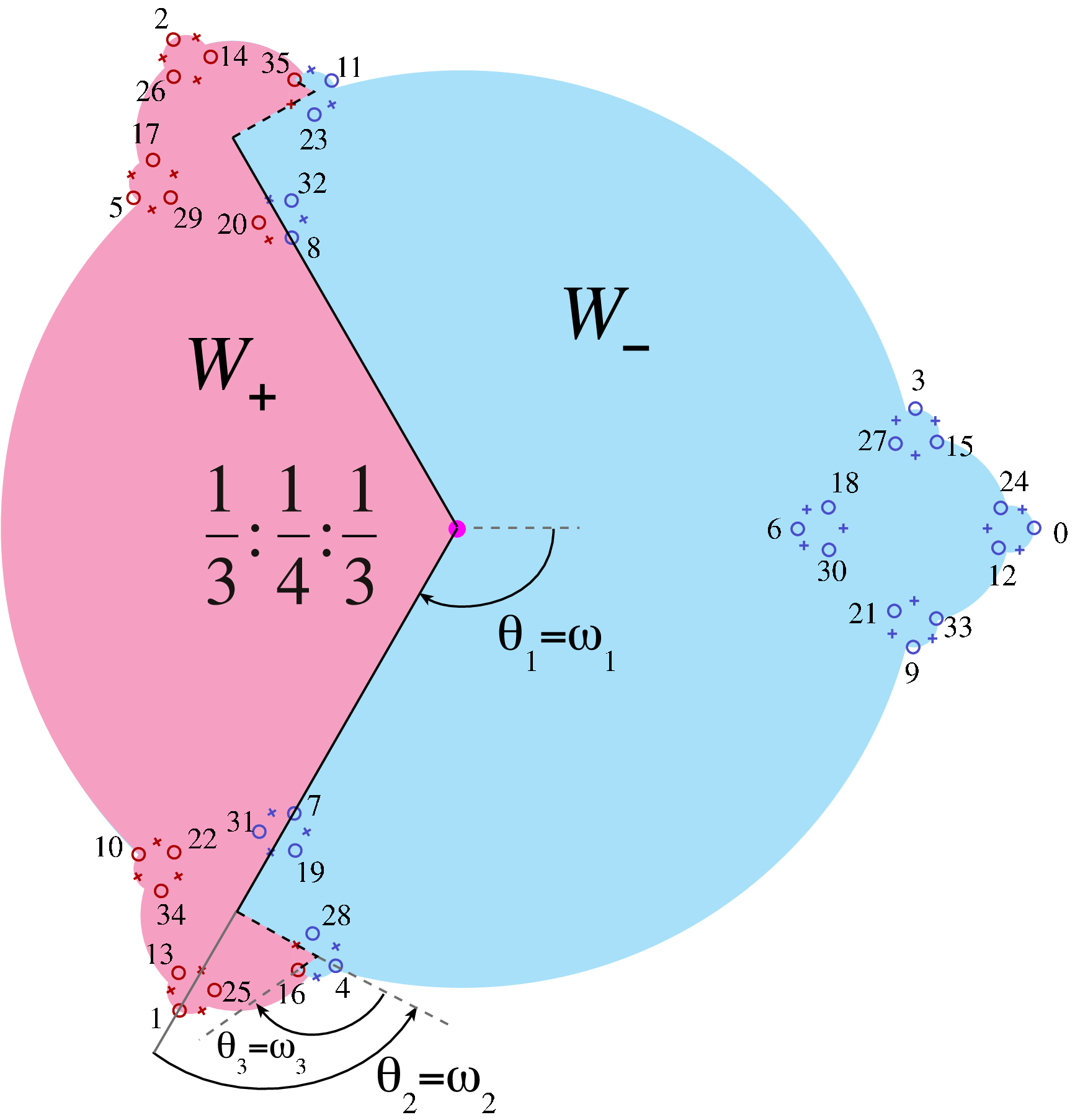}{Epicycle picture for the class-three orbit $\frac13:\frac14:\frac13$. The boundary of $W_{-}$ (blue) is closed on the solid rays and open on the dashed rays.}{fig:classIII}{3.5in} 

The definition of the class-$c$ code is complete once we specify the boundary between the ``wedges" $W_{\pm}$ that define the symbols $s_t = \pm$. As for class-two, our numerical observations indicate that the boundary consists of another wedge.
The rays that bound the wedge leave the origin with angles $\theta_1 = \pm \omega_1$. These extend to the center of the two class-two ambiguous islands at $t_1 = \pm 1$. The rays turn by angles $\theta_2 = \pm \omega_2$, and continue to the center of two class-three ambiguous islands at $t = 1:1$ and $t = -1:-1$, respectively. The rays again turn by angles $\theta_3 = \pm \omega_3$, continuing to the center of class-four ambiguous islands, etc. Note that since the direction of rotation in \Eq{eq:expClassC} reverses, the rays turn in an alternating manner. 

The specification of the symbol boundary is complete once we define the codes for points that fall on the boundary. This is especially important for $c>2$, as there can be more than two ambiguous islands. For example in \Fig{fig:classIII}, since $q_2=4$ is even, there are four ambiguous class-two islands, corresponding to  $t =4 = 1:1$, $7 = 1:2$, $8 = -1:-2$ and $11 = -1:-1$---and the rays defining the boundary of $W_{-}$ intersect all of these islands. By contrast, when $q_2$ is odd, there are only two ambiguous class-two islands, $t=1:1$ and $-1:-1$. 

The openness/closedness state of the boundary of the region $W_{-}$ is most easily described recursively. Recall that at class one, the boundary of $W_{-}$ is open. Given the boundary at class $(c-1)$, the boundary at class $c$ is obtained by reversing the state of the ray at class $c-1$ and appending class-$c$ rays with the same state. Thus for class-two, the previously open class-one ray becomes closed, and the new class-two ray is also closed. For class three, as shown in \Fig{fig:classIII}, the class-two ray and the new class-three ray are now open. Applying this rule to the first few classes generates \Tbl{tbl:classCRays}.

\begin{table}[htdp]
\begin{center}
\begin{tabular}{r|c|c|c|c|c|}
class  & $r_1$ & $r_2$ & $r_3$ & $r_4$ & $r_5$ \\
\hline
1                & - - -&&&&\\
2                &---   & ---    &&& \\
3                &---   & -  - - & - - - &&\\
4                &---   & -  - - & ---   & ---   &\\
5                &---   & -  - - & ---   & - - - & - - -\\
\end{tabular}
\end{center}
\caption{The class-$c$ boundary of $W_{-}$ is open (dashed) or closed (solid) along
the rays, $r_c$, defining the ambiguous islands. }
\label{tbl:classCRays}
\end{table}%

We have checked that this rule works for a number of class-three orbits, and a few orbits of class four. However, the high periods and multiplicity of secondary bifurcations involved prohibit systematic studies of high class orbits. In particular for the class-two orbit $\frac13:\frac14$, this rule generates the correct codes for the elliptic and hyperbolic orbits with $\omega_3 = \frac12$, $\frac13$, $\frac14$, $\frac15$, and $\frac 25$. We have also checked  the case $\frac14:\frac14:\frac13$ and a number of cases with odd $q_2$ including  $\frac13:\frac13:\frac14$.  The correct code is also obtained for $\frac13:\frac14:\frac12:\frac12$.
The class-$c$ codes also work for the period doubling case where it is equivalent to the doubling substitution rule, recall \Eq{eq:doublingSubs}. For example,  the codes for the period-doubling sequence  through class four are
\begin{equation}
\begin{tabular}{ll}
$\frac12:\frac12$                               & $\per{-+--}$ \\
$\frac12 : \frac12 : \frac12 $              &$\per{-+---+-+}$\\
$\frac12 : \frac12 : \frac12:\frac12$   \quad&$\per{-+---+-+-+---+--}$
\end{tabular}
\end{equation}
A simple rule that also generates this sequence is to double the previous entry and flip the last symbol \cite{SterlingThesis}.

Though we have not systematically explored beyond class-two, these few studies give us confidence that the class-$c$ rule works in general.

\section{Conclusion}
We have shown how to systematically generate symbolic codes for rotational orbits and ``islands-around-islands" orbits for the area-preserving quadratic map. The construction of the codes in terms of wedges with opening angles determined by the frequencies $\omega_c$ of the orbits explains numerical observations of the wedge-shaped symbol boundaries for elliptic islands. The wedge boundaries are constructed from segments of the symmetry lines of the mapping.

It is for precisely this case---when there are elliptic orbits---that more traditional, symbolic partitions fail.
It is interesting to note that Christiansen and Politi constructed symbol boundaries in islands of the standard map that have precisely the same structure as ours for class-one rotational orbits \cite{Christiansen96}. However, for this case the identification of symmetry lines is 
complicated by the fact that the standard map has two distinct sets of reversors.

Open questions that we hope will be investigated in the future include:
\begin{itemize}
  \item How does the symbol boundary evolve from the line $x=0$ when there is a horseshoe, to the complex set of epicycle-generated wedges when there are elliptic islands? Note that when an orbit first becomes stable, it typically does so by an inverse period-doubling, so its wedge opening angle will be zero.
  \item Does the epicycle picture for rotational codes also apply to other reversible, area-preserving maps with anti-integrable limits? Examples include the standard mapping, and polynomial automorphisms.  \item How do the wedge boundaries connect across the island separatrices? As was first observed in \cite{Christiansen96}, the symmetry lines seem to naturally connect with the symbol boundary in the chaotic region of phase space that corresponds to primary tangencies. They remark: ``We have no explanation for this nice phenomenon." Neither do we.
  \item What are the codes for rotational orbits generated by bifurcations of the elliptic orbits created in secondary pitchfork and twistless bifurcations?
  \item How are the symbol boundaries organized in maps that are not reversible? Perhaps the simplest example corresponds to some quartic polynomial automorphisms \cite{Gomez04}?
\end{itemize}

\clearpage
\appendix
\section{Codes for Twist Maps}
Our purpose in this appendix is to relate the rotational codes described in
this paper to those used in other papers.  There are three commonly
used codes for rotational orbits of twist maps, the linear, velocity
and acceleration codes, which we denote $a$, $b$, and $c$ respectively
\cite{Percival87a,Percival87b, Chen87,Chen90b}.

Aubry-Mather theory implies that recurrent minimizing orbits for twist
maps have a particularly simple symbolic coding.  Orbits that are
nondegenerate minima of the action are hyperbolic.  Thus the codes we
discuss here will correspond to the hyperbolic codes in
\Sec{sec:RotationalOrbits}.

An area-preserving  map on the cylinder $f: \bS^1 \times \R$
\begin{equation}\label{eq:twistMap}
   (x',y') = f(x,y)
\end{equation}
is a monotone twist map if $\frac{\partial x'}{\partial y} > 0$.  Note
that this hypothesis does not apply to the \hen map, even locally
about $\per{-}$, because it has a twistless bifurcation at $k =
\frac{9}{16}$ leading to a twist reversal \cite{Dullin99}.
Nevertheless, we will see that the $c$-codes for minimizing orbits of
twist maps do correspond to those of the rotational \hen map with the
proper translation.

\subsection{Linear Code}

One way of coding rotational orbits of maps with an angle variable is
to count the number of complete rotations they make.  To do this, we
lift the angle variable to the line, and define the linear
code,
\[
           a_{t} = \ceil{x_{t}} \;.
\]
   Thus $a_{t}$ is the number of complete rotations at time
   $t$.\footnote { The choice of the ceiling function here is
   arbitrary, as are the choices of forward and backward finite
   differences in the next two sections.  Ultimately, these only
   affect the canonical ordering for the codes, but the ordering is
   important for concatenation rules.  }

The linear code of a minimizing orbit of a twist map is determined by its rotation number.

\begin{teo}[Aubry \cite{Aubry83}]\label{thm:aubry}
     Suppose $F: \R^{2}\to \R^{2}$ is a lift of the twist map
     \Eq{eq:twistMap}, and $\{x_{t}, t \in \Z\}$ is the configuration of a
     recurrent, minimizing orbit.  Then there exists a rotation number
     $\omega$ and phase $\alpha$ such that the linear code of $x_{t}$
     is the same as that for the rigid rotation on the circle
     $\theta_{t} = \omega t + \theta_{0}$, i.e.,
     \begin{equation}\label{eq:aCode}
           a_{t} = \ceil{\omega t + \alpha}
     \end{equation}
\end{teo}
Here we take \Eq{eq:aCode} as the definition of the linear code for a
rotational orbit and consider some simple structures that arise.

\subsection{Velocity Code}\label{sec:velocity}

An alternative coding for rotational orbits  is the ``velocity'' code, defined to be the
first difference of the $a$-code:
\begin{equation}\label{eq:bCode}
       b_{t} \equiv a_{t+1}-a_{t}
\end{equation}
Note that knowledge of $\bb$ gives $\ba$ up to an  initial condition,
$a_{0}$, which corresponds to the choice of interval for the
lifted angle $x_{0}$. 

While, the $b$-code is well defined for any $\omega$, if we restrict to
$0 \le \omega < 1$ then we can compute it by dividing the
circumference-one circle into two sectors
\begin{equation}\label{eq:Sectors}
   \begin{split}
     B_{1}(\omega) &\equiv (-\omega,0] \;,\\
     B_{0}(\omega) &\equiv (0,1-\omega] \;.
   \end{split}
\end{equation}
Then we have
\begin{equation}\label{eq:bSector}
      b_{t} = i \quad \mbox{if} \quad \frc{\omega t + \alpha} \in B_{i}(\omega) \;,
\end{equation}
as shown in \Fig{fig:b-code}.

When $\omega$ is irrational, the code changes as
$\alpha$ varies.  We define the ``canonical'' ordering as that obtained
by setting $\alpha = 0$. For example the canonical $b$-code for $\gamma^{-2}$ is
\[
   \bb = (\ldots 0100.10100101001001010010100100101001001010010 \ldots) \;.
\]
The fact that there are uncountably many hyperbolic codes is consistent with
the fact that there are uncountably many orbits on a particular invariant circle
or cantorus.
Note that the velocity code always consists of blocks of the form
$(10^{m})$ and $(10^{m-1})$ where $m = \floor{\omega^{-1}}$.
As we will see below, it is no accident that
these building blocks are the codes for the Farey parents
$\tfrac{1}{m+1}$ and $ \tfrac{1}{m}$ of each frequency in the interval
that they bound.

     \InsertFig{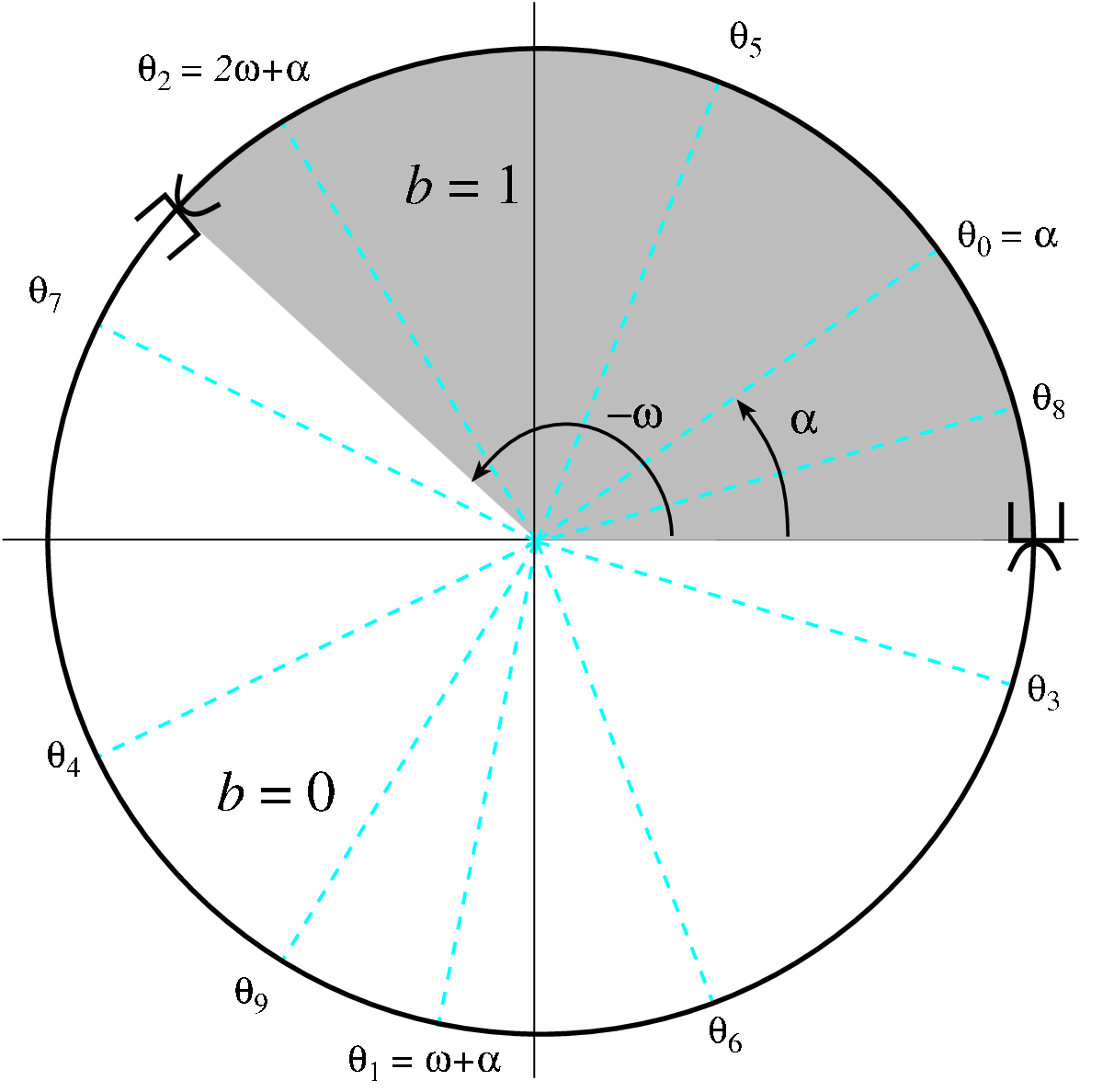} {Construction of the velocity code for $\omega
     = \gamma^{-2}$ showing the sectors $B_{0}$ and
     $B_{1}$ (shaded).  The orbit shown has $\alpha = -0.1$ and its $b$-code is
     $(\ldots 0100.1010010010 \ldots)$.} {fig:b-code}{3in}

When $\omega = p/q$ is rational the velocity code is independent of
choice of initial phase.
\begin{lem}\label{thm:alpha}
     The $b$-code for a rational rotation number $\omega = p/q$
     is independent of the choice of phase $\alpha$ up
     to cyclic permutations.
\end{lem}

\begin{proof}
     The $S_{j}$ defined in \Eq{eq:S-sectors} partition the circle into $q$
     sectors of width $\frac{1}{q}$.
     Note that $B_{0} = \bigcup_{j=0}^{q-p-1} S_{j}$, and
     $B_{1}= \bigcup_{j=q-p}^{q-1} S_{j}$ so the $b$-code
     is determined by these sectors.
     Since for any $\alpha$ the points
     $\theta_{t} = \frc{\omega t + \alpha}$ are spaced
     uniformly on the circle at a distance $\frac{1}{q}$ from each other,
     there is precisely one $\theta_{j}$ in each $S_{j}$.
     Thus the number of $0$ and $1$ symbols
     is independent of $\alpha$. Moreover if $\theta_{t} \in
     S_{j}$, then $\theta_{t+1} \in S_{j+p}$, so that the cyclic order
     of symbols is independent of $\alpha$.
\end{proof}

Note in particular, that all values $-1/q < \alpha \le 0$ give the
canonical velocity code.  Other values of $\alpha$, give a
cyclic permutation of this ``canonical'' ordering for the code.

\begin{cor}
     For $\omega=p/q$ the $b$-code has $p$ $1$'s and $q-p$ $0$'s.
\end{cor}

The velocity codes for various frequencies can be easily constructed
using a Farey tree procedure \cite{Zheng91, Hao98}.  A Farey tree is a binary tree that
generates all numbers in an interval between two ``neighboring''
rationals \cite{Hardy79}.  A pair of rationals $\tfrac{p}{q} < \tfrac{m}{n}$ are neighbors if
\[
         mq-pn = 1
\]
The Farey tree is recursively constructed beginning with a base
defined by a neighboring pair, and recursively applying the Farey sum
operation
\[
        \frac{p}{q} \oplus \frac{m}{n} \equiv \frac {p+m}{q+n} \;,
\]
to each neighboring pair.  Note that $\frac{p}{q} < \frac{p}{q} \oplus
\frac{m}{n} < \frac{m}{n}$, and that the daughter is a neighbor to
each of its parents.

For example, the two rationals $\tfrac01$ and $\tfrac11$ define the
base for a Farey tree that includes all numbers in the unit interval,
see \Fig{fig:bcodeFarey}.  The root of the tree, or level zero, is the
daughter rational defined by the Farey sum of the base rationals.  In
this case the root is $\tfrac12$.  Each subsequent level of the tree
consists of the rationals that are the Farey sum of each number on the
previous level with its two neighbors at earlier levels, thus there
are $2^{j}$ rationals at level $j$.

Every irrational $\omega \in [0,1]$ is uniquely determined by a path defined by an
infinite sequence of left, $L$, and right, $R$, transitions beginning at the
root of the Farey tree. Every rational $\omega \in [0,1]$ is uniquely determined by 
a finite path. For example, the path for $5/13$ is $LRLR$.
There are also paths in the Farey tree that do not correspond
to real numbers: for example $LRLLLLLLL\ldots$ limits to $\frac13$,
but is better thought of as ${\frac13}_{+}$.  Dynamically this sequence 
corresponds to a homoclinic orbit.
The paths of the parents
of any rational are easy to obtain by appropriately truncating its
path; for example, given a path $LRLLL =\tfrac{5}{14}$, then the
direct parent is obtained by simply removing the last symbol, $LRLL
=\tfrac{4}{11}$.  The other parent is obtained by removing all of the
final repeated symbols ($L$ in this case) in the sequence and one more
symbol, yielding the parent $L = \tfrac13$.

     \InsertFig{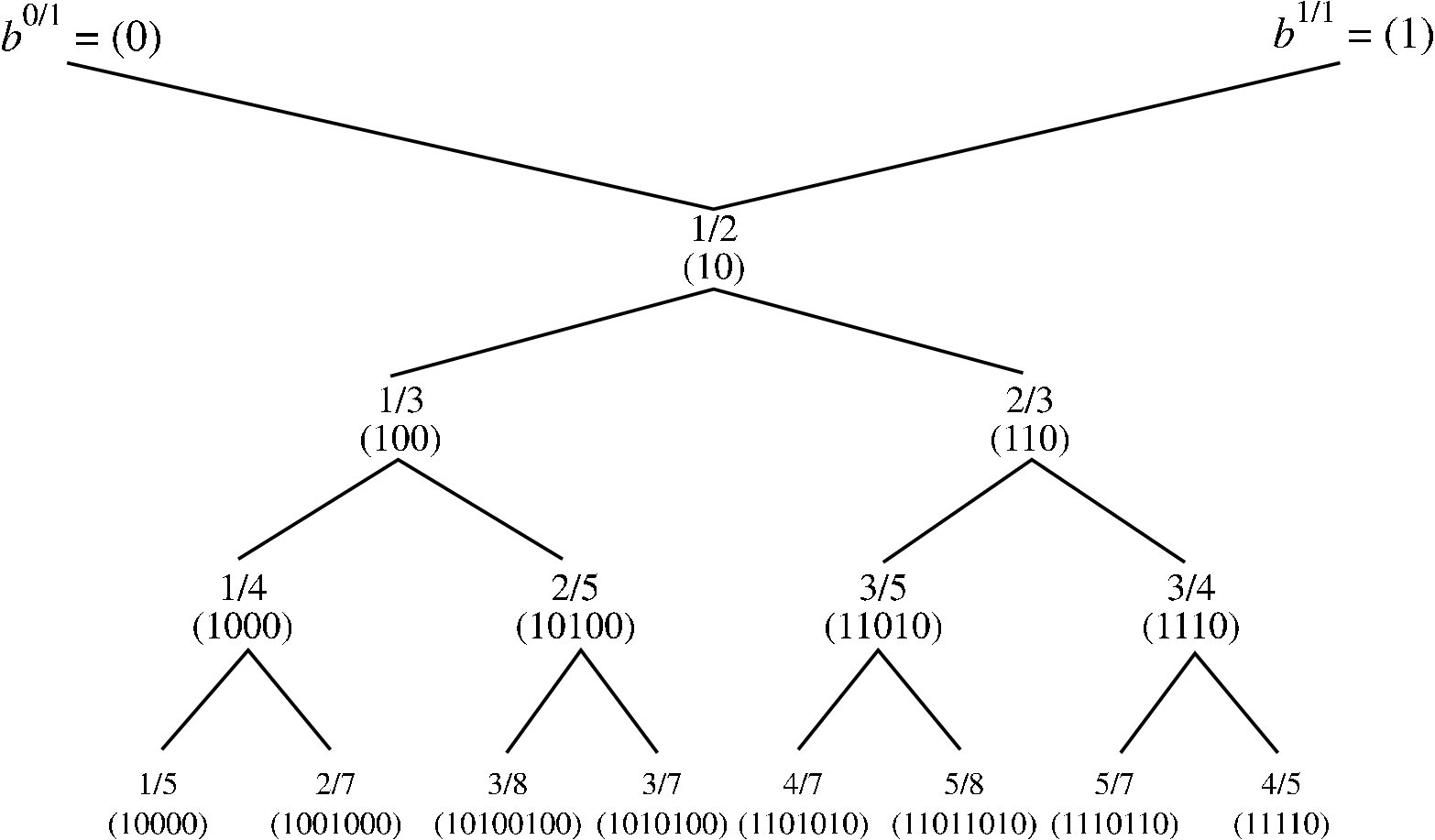} {Farey tree for the base $\tfrac01$ and
     $\tfrac11$ for levels zero to three, and the corresponding $b$-codes.}
     {fig:bcodeFarey}{5in}

\begin{lem} \label{thm:concat}
     The canonical velocity code for the Farey daughter of neighbors
     $\tfrac{p}{q} < \tfrac{m}{n}$ is the concatenation of the codes
     of the parents:     \[
        \bb\left(\frac{p}{q}\oplus \frac{m}{n}\right) =
               \bb\left(\frac{m}{n}\right) \, \bb\left(\frac{p}{q}\right) \;.
     \]
\end{lem}

\proof Let $\omega = \frac{p+m}{q+n}$ denote the daughter, and
$\theta_{t} = \omega t$.  Then $b_{t}(\omega) = i$ if $\frc{\theta_{t}} \in
B_{i}(\omega)$.  We first show that the first $n$ symbols are given
by the code for $\tfrac{m}{n}$, i.e. that $\frc{\tfrac{m}{n} t} \in
B_{i}(\tfrac{m}{n})$ implies that $\frc{\theta_{t}} \in B_{i}(\omega)$ for $0 \le t < n$.

Suppose first that $b_t(\tfrac{m}{n}) = 1$, then there is an integer $j$ such that
\begin{equation}\label{eq:inequalities}
       j- \frac{m}{n} < \frac{m}{n} t \le j \;.
\end{equation}
Since $\omega < \tfrac{m}{n} $, the right inequality implies
$\omega t \le j$; providing $t \ge 0$.
This is the first half of what we desired to show.
Since $t$ is an integer, the left inequality in \Eq{eq:inequalities} implies that
\[
      \frac{m}{n} (t+1) \ge j+\frac{1}{n} \;.
\]
Combining this with the relation
\[
    \omega = \frac{p+m}{q+n} = \frac{m}{n} -\frac{1}{n(q+n)} \;,
\]
implies that
\[
    \omega (t+1) \ge j +\frac{1}{n} - \frac{t+1}{n(q+n)} > j \;,
\]
providing $t < q+n-1$.  Thus we have shown $b_{t}(\tfrac{m}{n}) = 1
\implies b_{t}(\omega) = 1$, for $ 0 \le t < q+n-1$ (this is more than
we needed to prove).

To check that the $0$ symbols agree, we must show that when there
exists an integer $j$ such that
\[
       j < \frac{m}{n} t \le j+1 - \frac{m}{n}
\]
then $\frc{\theta_t} \in B_{0}(\omega)$.  A calculation similar to the previous
one shows that this is true when $0 \le t < q+n$.

So finally we have shown that the symbol sequence $\bb(\omega)$ is
given by $\bb(\tfrac{m}{n})$ repeated, except for the last symbol.

Similar calculations show that all symbols but the first
in  $b_{t}(\omega)$ agree with $b(\tfrac{p}{q})$, repeated from the end.
\qed

It is easy to see that \Lem{thm:concat} is consistent with the
codes in the Farey tree in \Fig{fig:bcodeFarey}.
For example since $b(1/3) = (100)$ and
$\bb(2/5)= (10100)$, then
$\bb(3/8) =  \bb(2/5)\bb(1/3)= (10100100)$, and since
$\tfrac{5}{13} = \tfrac38 \oplus \tfrac25$, then
$\bb(5/13) = \bb(2/5)\bb(3/8) = (1010010100100)$.

Since all numbers can be constructed by Farey paths,
\Lem{thm:concat} extends for irrational numbers as well:
\begin{cor}
     For any $\omega$ whose Farey path includes a parent rational
     $\tfrac{m}{n} > \omega$, the first $n$ symbols of $b(\omega)$
     are those of $b(\tfrac{m}{n})$.
\end{cor}
Thus, for example the first $34$ symbols in the code for $\gamma^{-2}$ are
given by those of its upper Farey neighbor, $\tfrac{13}{34}$, whose code
can be constructed by concatenation:
\begin{align*}
     b(13/34) &= b(5/13)b(8/21) = b(5/13)b(5/13)b(3/8) \\
              &= b(2/5)b(3/8)b(2/5)b(3/8)b(3/8)  = \ldots \\
              &= (1010010100100101001010010010100100)\;.
\end{align*}

\subsection{Acceleration Codes} \label{sec:accelerationCode}
The acceleration code is defined to be the magnitude of the first
difference of the velocity code
\begin{equation}\label{eq:cCode}
      c_{t} = |b_{t} - b_{t-1}| \;.
\end{equation}
This is well defined only for orbits with $0 \le \omega < \tfrac12$.
It could also be called a ``same-different'' or ``exclusive-or'' code, since $c_t = 0$
if the velocities are the same and $1$ if the velocity changes.  The
point is that for $\omega < \tfrac12$, there is a one-to-one
correspondence between allowed $b$ and acceleration codes because the $b$-code
can never have two or more consecutive $1$ symbols, thus we obtain $c_{t} = 0$
only for the case of a double $0$ in the $b$ code.

The acceleration code can be obtained geometrically by defining the sectors
$C_1(\omega) = (-\omega,\omega]$ and $C_1(\omega) = (\omega,1-\omega]$,
so that $c_{t} = i$ when $\frc{\omega t+\alpha} \in C _i(\omega)$. This
follows because we obtain $c_{t}=1$ whenever $b_{t} = 1$ or
$b_{t-1} = 1$. The canonical $b$-code can be reconstructed from the canonical
acceleration code by using the initial condition $b_{0} = 1$.

The geometrical construction shows that the symbol $1$ always appears
doubled in the acceleration code.  Moreover, if $m = \floor{\omega^{-1}}$, then
the acceleration code consists of blocks of the form $110^{m-2}$ and
$110^{m-1}$. The acceleration code inherits  properties of the $b$-code.

\begin{cor}
     The acceleration code for a rational $\omega$ is independent of the phase $\alpha$.
\end{cor}

\begin{cor}
     The canonical acceleration code for a Farey-daughter is the concatenation of the
     codes for its parents.
\end{cor}

\proof This follows from \Lem{thm:concat}, and the fact that the first
symbol in the $b$ code is always $1$ and the last is always $0$.  Thus
concatenation does not disturb the calculation of the acceleration code
symbols.  \qed

The interiors of the sectors $C_{i}(\omega)$ are identical to the
interiors of the wedges $W_{i}(\omega)$ that define the $s$-code in
\Sec{sec:RotationalOrbits}.  Recall that hyperbolic codes were those
in the interior of the $W_{i}$.  Thus if we translate $0$ and $1$ into
$+$ and $-$ appropriately, the acceleration and $s$-codes are identical for
hyperbolic orbits.

\begin{lem}
    If we set $s_{t} =\sgn(2c_{t} -1)$, then the acceleration code for
    an orbit of rotation number $\omega$ becomes the hyperbolic $s$-code.
\end{lem}

For example, since the $\tfrac13$ hyperbolic orbit has acceleration code
$\per{110}$, the corresponding hyperbolic orbit for the \hen map has
$s$-code $\per{--+}$.  The elliptic $\tfrac13$ orbit has code
$\per{-++}$, obtained as usual by flipping the second symbol in the
canonically ordered hyperbolic code; however, the elliptic codes do
not arise directly from the twist map acceleration code.

\section{Substitution Rules}\label{sec:subrule}
An alternative code for rotational orbits is given by the Farey-substitution
rule. This code has been used for monotone circle maps \cite{Hao98}, and turns out to be
identical to the $b$-code.

A substitution rule acts on a symbol sequence $\bs$ by replacing each symbol with a new sequence.
Supposing that $\bs \in \{L,R\}^\infty$, we define two substitution operators for left and right transitions:
\begin{align}
     F_{L}(L) = L  \,&,\quad  F_{L}(R) = RL \nonumber \\
     F_{R}(L) = RL \,&,\quad  F_{R}(R) = R \;.
\end{align}
Then the symbol sequence for a number whose Farey path begins at a root
with symbol sequence $s$, is determined by applying the substitution
operators in the same sequence. Thus, since $2/7$ has Farey path
$LLR$ starting at $1/2$ whose symbol sequence is $(RL)$, the $b$ code
for $2/7$ is
\[
     F_{L}F_{L}F_{R}(RL) = F_{L}F_{L}(RRL) = F_{L}(RLRLL) = RLLRLLL \;.
\]

Note that this operation is associative, i.e. the above could also
be read as
\[
    F_{L}F_{L}F_{R}(RL) = F_{L}F_{L}F_{R}(R) F_{L}F_{L}F_{R}(L) =
   F_{L}F_{L}(R) F_{L}F_{L}(RL) = F_{L}F_{L}(R) F_{L}F_{L}(R) L = \ldots
\]

\begin{lem} \label {thm:path}
     The canonical hyperbolic $b$-code for a frequency $\omega$ is given by
     applying the substitution rule for its Farey path with the translation
     $L \equiv 0$ and $R \equiv 1$.
\end{lem}

\begin{proof}
As shown in \Lem{thm:concat}, the canonical $b$-code for a Farey
daughter is obtained by concatenation of the codes its parents.  We
will show that this concatenation rule is also valid for the
Farey-substitution rule.  Combining this with the fact that the code
for $\tfrac12$ is $RL = 10$, gives the result.

As discussed in \App{sec:velocity}, the Farey paths for the two
parents are obtained by truncating the path of the daughter by
removing the last symbol, and all of the repeated last symbols plus
one more, respectively.  Thus there are two possible cases.  First if
the path for the daughter is $WRL^n$ for an arbitrary sequence $W$ and
$n \ge 1$, then we must show the concatenation rule
\[
   WRL^n = W \oplus WRL^{n-1} \;.
\]
Translating this into substitutions applied to the root $(RL)$, and
using the associative property, we obtain
\begin{align*}
   F_WF_RF_L^n(RL) &= F_WF_RF_L^{n-1}(RLL)  =F_WF_RF_L^{n-1}(RL) WF_RF_L^{n-1}(L) \\
                 &= F_WF_RF_L^{n-1}(RL) F_W(RL)  =  (W \oplus WRL^{n-1}) \,,
\end{align*}
which is what we wanted to show because the $\oplus$ concatenation
writes down the code of the right neighbor first. A similar calculation holds
in the other case, where the daughter is $WLR^n$, and we can show
\[
   WLR^n =  WFLR^{n-1} \oplus W \;.
\]
\end{proof}

There are other substitution rules that apply to period doubling, and
more generally $n$-tupling sequences \cite{Hao98}.  For example, the
doubling substitution rule is
\[
   L \to LR \quad \text{and} \quad R \to LL
\]
Applying this sequence
to the sequence of $\omega$ gives the sequence for $1/2:\omega$.
Applying this recursively gives the $\omega$ bifurcations of any
orbit from the period-doubling sequence.

\section{Symmetries and Codes of the Horseshoe}\label{sec:horseshoe}

In this section we give simple geometric construction of the relation
between the symbol plane and an idealized phase space for the
horseshoe map.  This helps to visualize the relation between the
symmetry lines and the symmetries present in the symbolic codes.
Instead of the standard smooth horseshoe map, we will use a
discontinuous, area-preserving horseshoe map that is defined
everywhere on the square $[-1,1]$ except for the $y$ axis:
\[
     H(x,y) = \left\{\begin{array}{ll}
                  (-2x-1,-\frac12 (1+y))\;, & x < 0  \\
                  (2x-1,\frac12(1+y))\;,    & x > 0  \\
              \end{array}\right. \;,
\]
see \Fig{fig:idealHorseshoe}. The region $x < 0$ corresponds to the symbol $\pt -$,
and  $ x > 0$ to  $+$. Thus the fixed points of $H$ are $\per{+} = (1,1)$, and $\per{-} = (-\frac13, -\frac13)$.

     \InsertFig{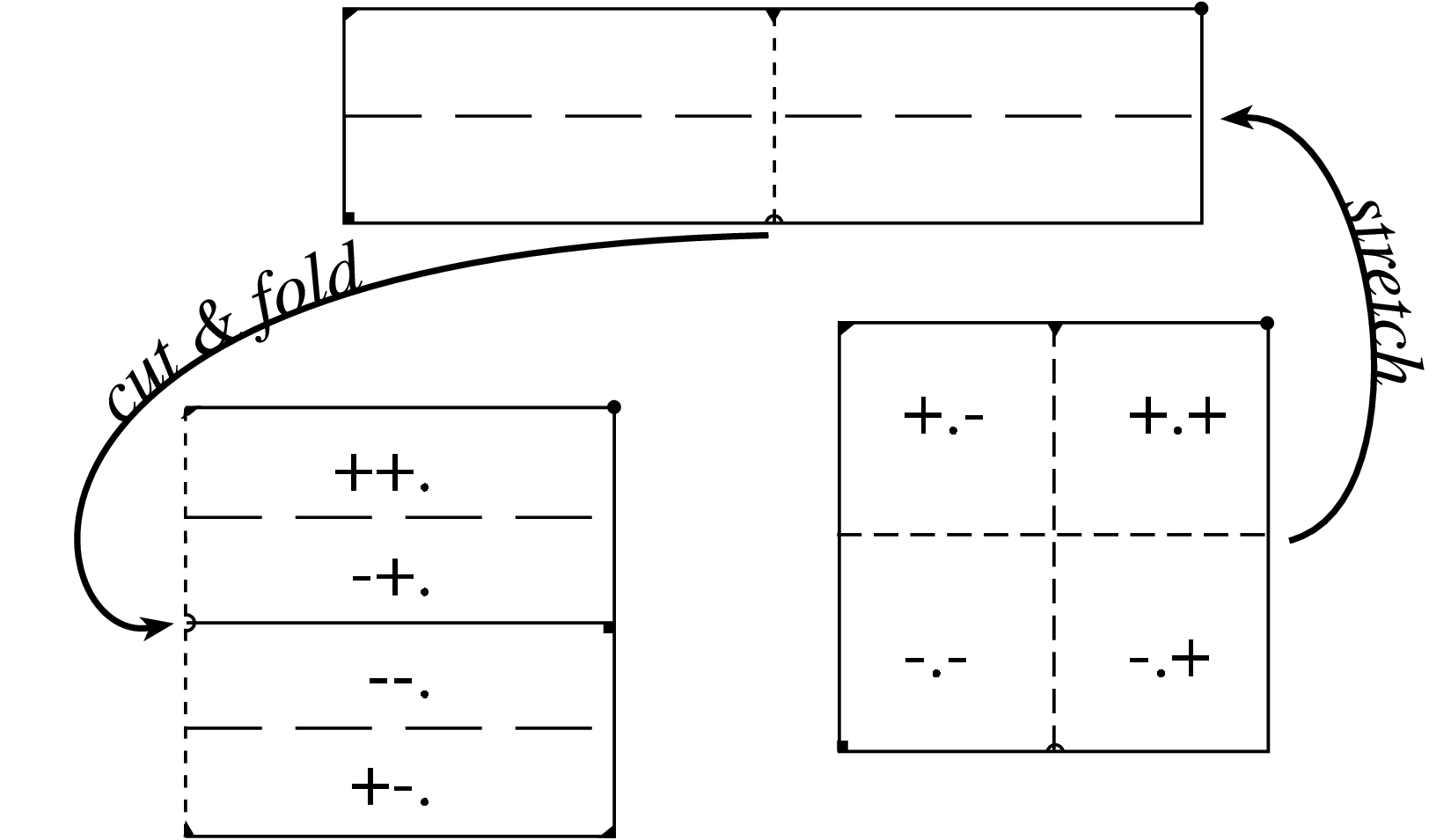} {Idealized, discontinuous horseshoe map.}
     {fig:idealHorseshoe}{3.5in}

This horseshoe map is reversible, with symmetry $S(x,y) = (y,x)$.
As such, it can be factored as $H = RS$, where
\[
   R = HS = \left\{\begin{array}{ll}
                  (-2y-1,-\frac12 (1+x))\;, & y < 0  \\
                  (2y-1,\frac12(1+x))\;,    & y > 0  \\
            \end{array} \right. \;.
\]
Thus the fixed set of $R$ is given by $\fix{R} = \{y=-\frac12 (1+x):
y<0\} \cup \{y=\frac12(1+x): y > 0\}$.  Alternatively, we can factor
$H = ST$ as
\[
   T= SH = \left\{\begin{array}{ll}
                  (-\frac12 (1+y),-2x-1)\;, & x < 0  \\
                  (\frac12(1+y),2x-1)\;,    & x > 0  \\
            \end{array}\right. \;,
\]
with a fixed set $\fix{T} = \{y=-2x-1 : x<0\} \cup \{ y = 2x-1: x >
0\}$.  As usual, we divide the symmetry lines into rays at the
``elliptic'' fixed point $\per{-}$, with the subscript $i$ denoting
the rays that lead to the hyperbolic point and $o$ those that lead
away, see \Fig{fig:horseshoeSymmetry}.

     \InsertFig{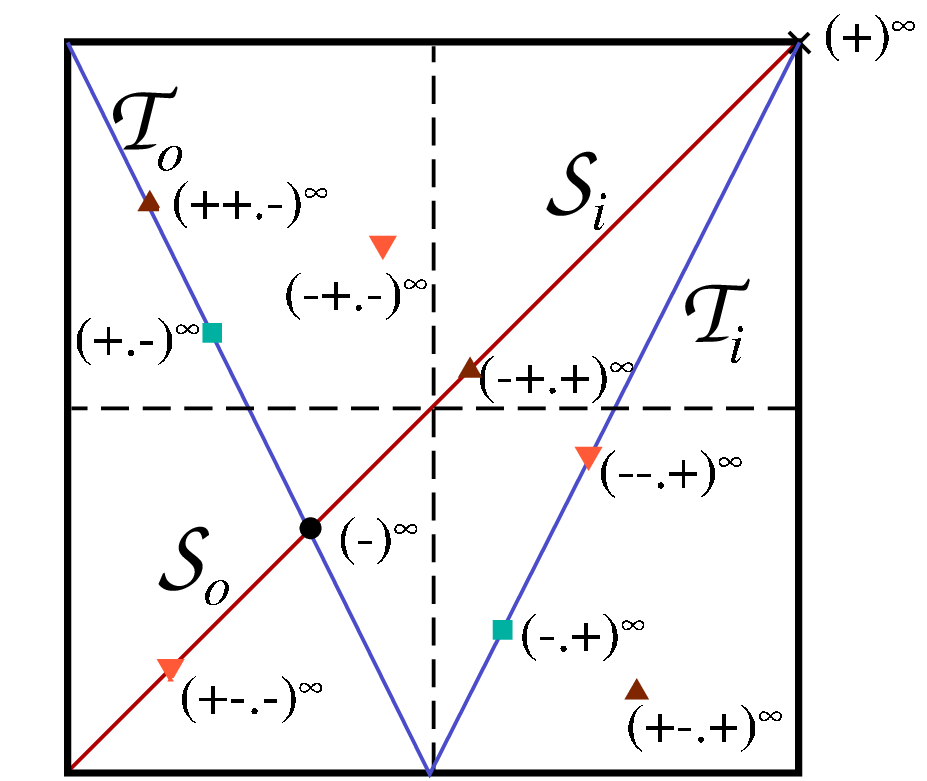} {Symmetry lines and periodic orbits
     of the discontinuous horseshoe.  The two fixed points are shown
     as a circle and $x$, the period-two points as squares, and the
     points on the two period-three orbits as triangles.}
     {fig:horseshoeSymmetry}{3.5in}

We also show in \Fig{fig:horseshoeSymmetry}, the period-two point
$\per{-\pt +} = (\frac15,-\frac35)$ and its image $\per{+\pt -} =
(-\frac35,\frac15)$.  Note that these points lie on $\fix{T}_{i}$ and
$\fix{T}_{o}$, respectively, in accord with \Tbl{tbl:symmetrylines}
since $H \fix{T}_{o} = \fix{R}_{o}$.  Similarly, the elliptic period 3
orbit has points $\per{-+\pt +} = (\frac19,\frac19) \in \fix{S}_{i}$,
$\per{++\pt -} = (-\frac79,\frac59) \in \fix{T}_{o}$, and $\per{+-\pt
+} = (\frac59,-\frac79)$.  The hyperbolic period 3 orbit has points
$\per{--\pt +} = (\frac37,-\frac17) \in \fix{T}_{i}$, $\per{-+\pt -} =
(-\frac17,\frac37)$, and $\per{+-\pt -} = (-\frac57,-\frac57) \in
\fix{S}_{o}$.

We can divide the symmetry lines into rays at the $\per{-}$ fixed
point.  Note that points with $x$ larger than that for the elliptic
fixed point will be on the ray that goes to the hyperbolic point
(labeled $i$), and those with $x$ smaller will be on the ray $o$.  A
standard result gives us the ordering.

\begin{lem}[Ordering (c.f. \protect{\cite[\S2.3.2]{Hao98})}] \label{thm:ordering}

    Suppose two symbol sequences $\bs =\pt s_0s_1s_2\ldots$ and $\bs'$
    agree for their first $j$ symbols, but that $s_j \ne s'_j$.
    Then the corresponding points $x$ and $x'$ on the horseshoe
    are ordered as $x < x'$ if the parity, \Eq{eq:parity}, of $s_0s_1\ldots s_j$
    is odd (and hence that of $s_0s_1\ldots s'_j$ is even).
\end{lem}
\noindent Thus sequences are ordered the same as their parities.  This
implies that symbol sequences for the symmetry rays are distinguished
by ``odd" or ``even" blocks of $-$ symbols after the binary point, see
\Eq{eq:symmetryrays}.

\clearpage

\bibliographystyle{unsrt}
\bibliography{BibFile}

\begin{thebibliography}{10}

\bibitem{Lind95}
D.~A. Lind and B.~Marcus.
\newblock {\em An Introduction to Symbolic Dynamics and Coding}.
\newblock Cambridge University Press, Cambridge, 1995.

\bibitem{Kitchens98}
B.P. Kitchens.
\newblock {\em Symbolic Dynamics}.
\newblock Springer-Verlag, Berlin, 1998.

\bibitem{Hao98}
Bai-Lin Hao and Wei-Mou Zheng.
\newblock {\em Applied Symbolic Dynamics and Chaos}, volume~7.
\newblock World Scientific Publishing Co. Inc., River Edge, {NJ}, 1998.

\bibitem{SDM99}
D.~Sterling, H.R. Dullin, and J.D. Meiss.
\newblock Homoclinic bifurcations for the {H}\'enon map.
\newblock {\em Physica D}, 134:153--184, 1999.

\bibitem{Henon69}
M.~H{\'e}non.
\newblock Numerical study of quadratic area-preserving mappings.
\newblock {\em Quart. Appl. Math.}, 27:291--312, 1969.

\bibitem{Aubry90}
S.~Aubry and G.~Abramovici.
\newblock Chaotic trajectories in the standard map, the concept of
  anti-integrability.
\newblock {\em Physica D}, 43:199--219, 1990.

\bibitem{Sarkovskii64}
A.~N. Sarkovskii.
\newblock Coexistence of cycles of a continuous map of a line into itself.
\newblock {\em Ukr. Mat. Z.}, 16:61--71, 1964.

\bibitem{Metropolis73}
N.~Metropolis, M.~L. Stein, and P.~R. Stein.
\newblock On finite limit sets for transformations of the unit interval.
\newblock {\em J. of Combinatorial Theory}, 15(1):25--44, 1973.

\bibitem{Henon76}
M.~{H}\'enon.
\newblock A two-dimensional mapping with a strange attractor.
\newblock {\em Comm. Math. Phys}, 50(1):69--77, 1976.

\bibitem{Grassberger85}
P.~Grassberger and H.~Kantz.
\newblock Generating partitions for the dissipative {H}\'enon map.
\newblock {\em Phys. Lett. A}, 113(5):235--238, 1985.

\bibitem{Cvitanovic88}
P.~Cvitanovic, G.H. Gunaratne, and I.~Procaccia.
\newblock Topological and metric properties of {H}\'enon-type strange
  attractors.
\newblock {\em Physical Review A}, 38(3):1503--1520, 1988.

\bibitem{Hansen92}
K.T. Hansen.
\newblock Remarks on the symbolic dynamics for the {H}\'enon map.
\newblock {\em Phys. Lett. A}, 165:100--104, 1992.

\bibitem{Giovannini92}
F.~Giovannini and A.~Politi.
\newblock Generating partitions in {H}\'enon-type maps.
\newblock {\em Phys. Lett. A}, 161:332--336, 1992.

\bibitem{Fang94}
H.P. Fang.
\newblock Dynamics for a two-dimensional antisymmetric map.
\newblock {\em J. Phys. A}, 27:5187--5200, 1994.

\bibitem{Biham89}
O.~Biham and W.~Wenzel.
\newblock Characterization of unstable periodic orbits in chaotic attractors
  and repellers.
\newblock {\em Phys. Rev. Lett.}, 63:819--822, 1989.

\bibitem{Sterling98}
D.~Sterling and J.D. Meiss.
\newblock Computing periodic orbits using the anti-integrable limit.
\newblock {\em Physics Letters A}, 241:46--52, 1998.

\bibitem{Grassberger89}
P.~Grassberger, H.~Kantz, and U.~Moenig.
\newblock On the symbolic dynamics of the {H}\'enon map.
\newblock {\em J Phys A}, 22(24):5217--5230, 1989.

\bibitem{Hansen98}
K.~T. Hansen and P.~Cvitanovic.
\newblock Bifurcation structures in maps of {H}\'enon type.
\newblock {\em Nonlinearity}, 11(5):1233--1261, 1998.

\bibitem{Christiansen95}
F.~Christiansen and A.~Politi.
\newblock Generating partition for the standard map.
\newblock {\em Physical Review E}, 51:R3811--R3814, 1995.

\bibitem{Christiansen97}
F.~Christiansen and A.~Politi.
\newblock Guidelines for the construction of a generating parition in the
  standard map.
\newblock {\em Physica D}, 109:32--41, 1997.

\bibitem{Christiansen96}
F.~Christiansen and A.~Politi.
\newblock Symbolic encoding in symplectic maps.
\newblock {\em Nonlinearity}, 9:1623--1640, 1996.

\bibitem{Veerman86}
P.~Veerman.
\newblock Symbolic dynamics and rotation numbers.
\newblock {\em Phys. A}, 134(3):543--576, 1986.

\bibitem{Zheng91}
Wei-Mou Zheng.
\newblock Symbolic dynamics for the circle map.
\newblock {\em Int. J. Mod. Phys. B}, 5:481--495, 1991.

\bibitem{Percival87a}
I.C. Percival and F.~Vivaldi.
\newblock A linear code for the sawtooth and cat maps.
\newblock {\em Physica D}, 27:373--386, 1987.

\bibitem{Meiss86}
J.D. Meiss.
\newblock Class renormalization: Islands around islands.
\newblock {\em Phys. Rev. A}, 34(3):2375--2383, 1986.

\bibitem{Aizawa84}
Y.~Aizawa.
\newblock Symbolic dynamics approach to the two-{D} chaos in area-preserving
  maps.
\newblock {\em Prog. Theor. Phys.}, 71:1419--1421, 1984.

\bibitem{Afraimovich00}
V.~Afraimovich, A.~Maass, and J.~Urías.
\newblock Symbolic dynamics for sticky sets in {H}amiltonian systems.
\newblock {\em Nonlinearity}, 13:617--637, 2000.

\bibitem{Meiss86b}
J.D. Meiss and E.~Ott.
\newblock Markov tree model of transport in area preserving maps.
\newblock {\em Physica D}, 20:387--402, 1986.
\newblock A.P. maps, boundary circles, islands around islands.

\bibitem{MacKay93}
R.S. MacKay.
\newblock {\em Renormalisation in Area-Preserving Maps}, volume~6 of {\em
  Advanced Series in Nonlinear Dynamics}.
\newblock World Scientific, Singapore, 1993.

\bibitem{Meiss92}
J.D. Meiss.
\newblock Symplectic maps, variational principles, and transport.
\newblock {\em Reviews of Modern Physics}, 64(3):795--848, 1992.

\bibitem{Dullin99}
H.R. Dullin, J.D. Meiss, and D.~Sterling.
\newblock Generic twistless bifurcations.
\newblock {\em Nonlinearity}, 13:203--224, 1999.

\bibitem{Devaney79}
R.L. Devaney and Z.~Nitecki.
\newblock Shift automorphisms in the {H}\'enon mapping.
\newblock {\em Commun. Math. Phys.}, 67:137--146, 1979.

\bibitem{Aubry92}
S.~J. Aubry.
\newblock The concept of anti-integrability: Definition, theorems and
  applications to the standard map.
\newblock {\em Twist Mappings and Their Applications, Ed Richard McGehee,
  Kenneth R. Meyer}, pages 7--54, 1992.

\bibitem{MeyerHall92}
K.R. Meyer and G.R. Hall.
\newblock {\em Introduction to the Theory of {H}amiltonian Systems}, volume~90
  of {\em Applied Mathematical Sciences}.
\newblock Springer-Verlag, New York, 1992.

\bibitem{Franks90}
J.~Franks.
\newblock Periodic points and rotation numbers for area-preserving
  diffeomorphisms of the plane.
\newblock {\em Publications Math\'ematiques De l'IH\'ES}, 71:105--120, 1990.

\bibitem{Moser94}
J.K. Moser.
\newblock On quadratic symplectic mappings.
\newblock {\em Math. Zeitschrift}, 216:417--430, 1994.

\bibitem{Aubry83}
S.~Aubry and P.Y. Le~Daeron.
\newblock The discrete {F}renkel-{K}ontorova model and its extensions.
\newblock {\em Physica D}, 8:381--422, 1983.

\bibitem{Backer97}
A.~Backer and H.R. Dullin.
\newblock Symbolic dynamics and periodic orbits for the cardioid billiard.
\newblock {\em Journal of Physics A}, 30(6):1991--2020, 1997.

\bibitem{DSM00}
H.R. Dullin, D.~Sterling, and J.D. Meiss.
\newblock Self-rotation number using the turning angle.
\newblock {\em Physica D}, 145(1-2):25--46, 2000.

\bibitem{SterlingThesis}
D.~Sterling.
\newblock {\em Anti-Integrable Continuation and the Destruction of Chaos}.
\newblock Ph{D} {T}hesis, University of Colorado, 1999.

\bibitem{Lamb98b}
J.W.S. Lamb and J.A.G. Roberts.
\newblock Time-reversal symmetry in dynamical systems: A survey.
\newblock {\em Physica D}, 112:1--39, 1998.

\bibitem{Veerman91}
J.J.P. Veerman and F.M. Tangerman.
\newblock Intersection properties of invariant manifolds in certain twist maps.
\newblock {\em Communications in Mathematical Physics}, 139:245--265, 1991.

\bibitem{Gomez04}
A.~G\'omez and J.D. Meiss.
\newblock Reversors and symmetries for polynomial automorphisms of the complex
  plane.
\newblock {\em Nonlinearity}, 17:975--1000, 2004.

\bibitem{Percival87b}
I.C. Percival and F.~Vivaldi.
\newblock Arithmetical properties of strongly chaotic motion.
\newblock {\em Physica D}, 25:105--130, 1987.

\bibitem{Chen87}
Q.~Chen, J.D Meiss, and I.C Percival.
\newblock Orbit extension methods for finding unstable orbits.
\newblock {\em Physica D}, 29:143--154, 1987.

\bibitem{Chen90b}
Q.~Chen, I.~Dana, J.D. Meiss, N.~Murray, and I.C. Percival.
\newblock Resonances and transport in the sawtooth map.
\newblock {\em Physica D}, 46:217--240, 1990.

\bibitem{Hardy79}
G.H. Hardy and E.M. Wright.
\newblock {\em An Introduction to the Theory of Numbers}.
\newblock Oxford Univ. Press, Oxford, 1979.
\newblock Theorem 25.

\end{thebibliography}

\end{document}